\documentclass[12pt]{article}
\usepackage{amsfonts}
\usepackage{amsmath}
\usepackage{epsfig}
\usepackage{latexsym}
\usepackage{graphicx}
\usepackage{pdflscape}
\numberwithin{equation}{section}
\allowdisplaybreaks

\DeclareMathOperator{\Tr}{Tr}
\DeclareMathOperator{\Ai}{Ai}
\renewcommand{\thefootnote}{\fnsymbol{footnote}}
\newcounter{aff}
\renewcommand{\theaff}{\fnsymbol{aff}}
\newcommand{\affiliation}[1]{
\setcounter{aff}{#1} $\rule{0em}{1.2ex}^\theaff\hspace{-.4em}$}
\newcommand{\nn}{\nonumber \\}
\def\til#1{\widetilde{#1}}
\newcommand{\bra}{\langle}
\newcommand{\ket}{\rangle}
\def\({\left(}
\def\){\right)}
\def\cO{{\mathcal O}}
\newcommand{\pd}{\partial}
\def\l{\ell}
\def\Res{\mathop {\rm Res} \limits}

\begin{document}
\begin{titlepage}
\hfill\hfill
\begin{minipage}{1.2in}
DESY 12-196\\
TIT/HEP-624 
\end{minipage}

\bigskip\bigskip
\begin{center}
{\LARGE\bf Instanton Effects in ABJM Theory}\\
\vskip3mm
{\LARGE\bf from Fermi Gas Approach}\\
\bigskip\bigskip
{\large Yasuyuki Hatsuda\footnote[1]{\tt yasuyuki.hatsuda@desy.de},
Sanefumi Moriyama\footnote[2]{\tt moriyama@math.nagoya-u.ac.jp} and
Kazumi Okuyama\footnote[3]{\tt kazumi@azusa.shinshu-u.ac.jp}
}\\
\bigskip\bigskip
\affiliation{1}
{\normalsize\it DESY Theory Group, DESY Hamburg\\ 
Notkestrasse 85, D-22603 Hamburg, Germany\\
and\\
Department of Physics, Tokyo Institute of Technology\\
Tokyo 152-8551, Japan} \bigskip\\
\affiliation{2} {\normalsize\it Kobayashi Maskawa Institute 
\& Graduate School of Mathematics, Nagoya University\\
Nagoya 464-8602, Japan} \bigskip\\
\affiliation{3} {\normalsize\it Department of Physics, Shinshu University\\
Matsumoto 390-8621, Japan} \bigskip\\
\end{center}

\begin{abstract}
We study the instanton effects of the ABJM partition function using
the Fermi gas formalism.
We compute the exact values of the partition function at the Chern-Simons levels
$k=1,2,3,4,6$ up to $N=44,20,18,16,14$ respectively, and extract
non-perturbative corrections from these exact results.
Fitting the resulting non-perturbative corrections by their expected
forms from the Fermi gas, we determine unknown parameters in them.
After separating the oscillating behavior of the grand potential,
which originates in the periodicity of the grand partition function,
and the worldsheet instanton contribution, which is computed from the
topological string theory, we succeed in proposing an analytical
expression for the leading D2-instanton correction.
Just as the perturbative result, the instanton corrections to the
partition function are expressed in terms of the Airy function.
\end{abstract}

\end{titlepage}

\renewcommand{\thefootnote}{\arabic{footnote}}
\setcounter{footnote}{0}
\setcounter{section}{0}

\section{Introduction}

Partition functions are the most fundamental quantities in both quantum
systems and statistical systems.
Although it typically serves only the role of normalization in the
physical interpretation, it encodes important thermodynamic
information such as the free energy.
In field theories, the partition function, in general, contains the
information of not only perturbation but also instanton effects as
well as some global analytical structures.
Therefore it is natural to study it as a first step.

M-theory is supposed to unify all our understandings of
non-perturbative effects in the string theory.
In the seminal paper \cite{ABJM}, the low-energy effective theory on
$N$ multiple M2-branes on the geometry
${\mathbb C}^4/{\mathbb Z}_k$ was proposed to be the
${\mathcal N}=6$ supersymmetric Chern-Simons theory with gauge group
$U(N)_k\times U(N)_{-k}$ and bifundamental matters $A_{1,2}, B_{1,2}$
forming the superpotential
$W=\frac{2\pi}{k}(A_1B_1A_2B_2-A_1B_2A_2B_1)$.

It was found in \cite{KWY,J,HHL} that the infinite-dimensional path
integral of the partition function on $S^3$ and the expectation values
of BPS Wilson loops are reduced to a finite dimensional matrix model
by the localization techniques \cite{P}.

There is much progress in this matrix model including the derivation
of the degree of freedom $N^{3/2}$ \cite{DMP1} expected from the
AdS/CFT correspondence \cite{KT} and its perturbative completion by
the Airy function \cite{FHM}. 
Recently, the ABJM partition function was rewritten as that of a Fermi gas
system \cite{MP,MPinteract,KMSS}, where the standard method in
statistical mechanics is applicable and the above properties are
reproduced easily.

The Airy function and the Fermi gas formalism imply a deep structure
of M-theory.
The Airy function has the integral representation of the exponentiated
cubic term, which looks similar to the partition function of the
Chern-Simons theory, and appears also in other contexts of M-theory
\cite{OVV}.
Therefore we would like to study the partition function further and
to extract some physical information out of it.

In our previous work \cite{HMO}, we have proposed a method to compute
the exact partition function $Z_k(N)$ at $k=1$ and indeed computed it up to $N=9$.
Here with a generalization of our method and a technique from
\cite{PY}, we perform the computation of the exact partition function at $k=1, 2, 3, 4, 6$ up
to $N=44, 20, 18, 16, 14$ respectively.

After the computation of the exact values, we proceed to study
the non-perturbative effects.
We can extract the non-perturbative corrections from the exact values
of the partition function by subtracting the perturbative result.
The main motivation of this paper is to explore an analytic form of
the non-perturbative correction.
It is known that there are two kinds of instantons that induce the
non-perturbative corrections to the ABJM partition function.
One is called the worldsheet instanton, and the other the D2-instanton
(membrane instanton).
From the viewpoint of the gravity dual, the Type IIA string theory on
$AdS_4\times\mathbb{CP}^3$, the worldsheet instanton comes from the
fundamental string wrapping the holomorphic cycle
$\mathbb{CP}^1\subset\mathbb{CP}^3$ \cite{DMP1,Cagnazzo:2009zh}, while the D2-instanton comes from
the D2-brane wrapping the Lagrangian submanifold
$\mathbb{RP}^3\subset\mathbb{CP}^3$ \cite{DMP2}.
The worldsheet instanton behaves as $e^{-2\pi\sqrt{2N/k}}$ while the
D2-instanton as $e^{-\pi\sqrt{2kN}}$ in the large $N$ limit.

Let us summarize our results here.
Using the Fermi gas formalism (and some results of the topological
string), we can know the expected forms of the instanton corrections
to the partition function.
In the Fermi gas formalism, it is useful to consider the grand
partition function.
The grand potential (the logarithm of the grand partition function),
in general, receives the non-perturbative corrections as
\begin{align}
J_k^{\rm (np)}(\mu)\equiv J_k(\mu)-J_k^{\rm (pert)}(\mu)
=J_k^{\rm D2}(\mu)+J_k^{\rm WS}(\mu)+\cdots,
\label{eq:Jnp0}
\end{align}
where $\mu$ is the chemical potential, and $J_k^{\rm (pert)}(\mu)$,
$J_k^{\rm D2}(\mu)$ and $J_k^{\rm WS}(\mu)$ are the perturbative,
D2-instanton and worldsheet instanton contributions, respectively.
We note that the dots in \eqref{eq:Jnp0} represent an additional
oscillatory contribution and, if any, bound states of the worldsheet
instanton and D2-instanton.
The oscillatory behavior will be studied in subsection \ref{sec:osc}
in detail, which we ignore here.
The D2-instanton and the worldsheet instanton corrections are expected
to take the following forms:
\begin{align}
J_k^{\rm D2}(\mu)&=\sum_{m=1}^\infty J^{{\rm D2}(m)}_k(\mu)
=\sum_{m=1}^\infty\Big[a_k^{(m)}\mu^2+b_k^{(m)}\mu+c_k^{(m)}\Big]e^{-2m\mu},\nn
J_k^{\rm WS}(\mu)&=\sum_{n=1}^\infty J^{{\rm WS}(n)}_k(\mu)
=\sum_{n=1}^\infty d_k^{(n)}e^{-\frac{4n\mu}{k}}.
\label{eq:JWS}
\end{align}
We would like to determine the prefactors of those instanton
corrections.
The prefactors of the worldsheet instanton corrections are determined
by the results of the topological string on local $\mathbb{F}_0$,
where the first few prefactors are
\begin{align}
d_k^{(1)}=\frac{1}{\sin^2(\frac{2\pi}{k})},\quad
d_k^{(2)}=-\frac{1}{2\sin^2(\frac{4\pi}{k})}
-\frac{1}{\sin^2(\frac{2\pi}{k})},\quad 
d_k^{(3)}=\frac{1}{3\sin^2(\frac{6\pi}{k})}
+\frac{3}{\sin^2(\frac{2\pi}{k})}.
\label{eq:d_n}
\end{align}
On the other hand, there are no systematic ways to determine the
prefactors of the D2-instanton corrections.%
\footnote{One approach is to consider the `WKB' (small $k$) expansions
of the prefactors as in \cite{MP}.
However, it is hard to determine the higher order corrections in this
way.}
In this paper, we propose the analytic form of the leading
D2-instanton correction,
\begin{align}
a_k^{(1)}&=-\frac{4}{\pi^2 k}\cos\(\frac{\pi k}{2}\),\quad
b_k^{(1)}=\frac{2}{\pi\tan(\frac{\pi k}{2})}\cos\(\frac{\pi k}{2}\),\nn
c_k^{(1)}&=\biggl[-\frac{2}{3k}+\frac{5k}{12}+
\frac{k}{2\sin^2(\frac{\pi k}{2})}
+\frac{1}{\pi\tan(\frac{\pi k}{2})}\biggr]\cos\(\frac{\pi k}{2}\).
\label{eq:a1b1c1}
\end{align}
We arrive at these expressions from many constraints (by matching the
exact data, the numerical data from TBA, pole structure, small $k$
behavior etc.) and a few assumptions.
We believe that these results are valid for any $k$. 
We also find the higher instanton corrections at $k=1,2,3,4,6$ from
the numerical fitting.
Here is the summary of our results:
\begin{align}
J_1^{\rm (np)}&=\biggl[\frac{4\mu^2+\mu+1/4}{\pi^2}\biggr]e^{-4\mu}
+\biggl[-\frac{52\mu^2+\mu/2+9/16}{2\pi^2}+2\biggr]e^{-8\mu}
\nonumber\\
&\quad+\biggl[\frac{736\mu^2-152\mu/3+77/18}{3\pi^2}-32\biggr]e^{-12\mu}\nn
&\quad+\biggl[-\frac{2701\mu^2-13949\mu/48+11291/768}{\pi^2}+466\biggr]e^{-16\mu}
+\cO(e^{-20\mu}), \nonumber\\
J_2^{\rm (np)}&=\biggl[\frac{4\mu^2+2\mu+1}{\pi^2}\biggr]e^{-2\mu}
+\biggl[-\frac{52\mu^2+\mu+9/4}{2\pi^2}+2\biggr]e^{-4\mu}
\nonumber\\
&\quad+\biggl[\frac{736\mu^2-304\mu/3+154/9}{3\pi^2}-32\biggr]e^{-6\mu} \nn
&\quad+\biggl[-\frac{2701\mu^2-13949\mu/24+11291/192}{\pi^2}+466\biggr]e^{-8\mu}
+\cO(e^{-10\mu}), \nonumber\\
J_3^{\rm (np)}&=\frac{4}{3}e^{-\frac{4}{3}\mu}
-2e^{-\frac{8}{3}\mu}
+\biggl[\frac{4\mu^2+\mu+1/4}{3\pi^2}+\frac{20}{9}\biggr]e^{-4\mu}
-\frac{88}{9}e^{-\frac{16}{3}\mu}+\cO(e^{-\frac{20}{3}\mu}), \nonumber\\
J_4^{\rm (np)}&=e^{-\mu}+\biggl[
-\frac{4\mu^2+2\mu+1}{2\pi^2}\biggr]e^{-2\mu}
+\frac{16}{3}e^{-3\mu}+\left[-\frac{52\mu^2+\mu+9/4}{4\pi^2}+2\right]e^{-4\mu}
+\cO(e^{-5\mu}), \nonumber\\
J_6^{\rm (np)}&=\frac{4}{3}e^{-\frac{2}{3}\mu}
-2e^{-\frac{4}{3}\mu}
+\biggl[\frac{4\mu^2+2\mu+1}{3\pi^2}+\frac{20}{9}\biggr]e^{-2\mu}
-\frac{88}{9}e^{-\frac{8}{3}\mu}+\cO(e^{-\frac{10}{3}\mu}).
\label{Jnp}
\end{align}

Once the instanton corrections to the grand potential are fixed, we
can easily translate them into the instanton corrections to the
partition function,
\begin{align}
Z_k^{\rm (np)}(N)
\equiv\frac{Z_k^{\rm (exact)}(N)}{Z_k^{\rm (pert)}(N)}-1,\label{eq:Znp}
\end{align}
where $Z_k^{\rm (pert)}(N)$ is the perturbative result in \cite{FHM}, whose
explicit form is given by \eqref{eq:Zpert}.
The leading corrections of both instantons are written by the Airy
function (and its derivative):
\begin{align}
Z_k^{\rm D2}(N)&=\Bigl(C_k^{-1}(N+2-B_k)a_k^{(1)}+c_k^{(1)}\Bigr)
\frac{\Ai[C_k^{-1/3}(N+2-B_k)]}{\Ai[C_k^{-1/3}(N-B_k)]}\notag \\
&\quad 
-C_k^{-1/3}b_k^{(1)}
\frac{\Ai'[C_k^{-1/3}(N+2-B_k)]}{\Ai[C_k^{-1/3}(N-B_k)]},
\label{eq:Zinst_D2}\\
Z_k^{\rm WS}(N)&=d_k^{(1)}
\frac{\Ai[C_k^{-1/3}(N+\frac{4}{k}-B_k)]}{\Ai[C_k^{-1/3}(N-B_k)]},
\label{eq:Zinst_ws}
\end{align}
with
\begin{align}
B_k=\frac{k}{24}+\frac{1}{3k},\quad C_k=\frac{2}{\pi^2 k}.
\label{eq:BC}
\end{align}
Some remarks on our results are mentioned in order.

Firstly, we emphasize that our results include all the $1/N$
corrections.
One can check that in the large $N$ limit, both instanton corrections
show the correct exponentially suppressed behavior as mentioned
above.

Secondly, the worldsheet instanton correction \eqref{eq:Zinst_ws} is
divergent at $k=1,2$, and the D2-instanton  correction
\eqref{eq:Zinst_D2} is also divergent for even $k$.%
\footnote{For odd $k$, the D2-instanton correction \eqref{eq:Zinst_D2}
trivially vanishes, and the leading correction starts from
$\cO(e^{-2\pi\sqrt{2kN}})$.}
Therefore we cannot apply the above result to these values of $k$.
In these cases, the D2-instanton and worldsheet instanton corrections
are mixed, and the sum of them totally become finite.
At $k=1,2$, for example, such mixed instanton corrections are given by
\begin{align}
Z_1^{\rm D2+WS}(N)
&=\Bigl(2N+\frac{C_1}{8}+\frac{29}{4}\Bigr)
\frac{\Ai[C_1^{-1/3}(N+\frac{29}{8})]}{\Ai[C_1^{-1/3}(N-\frac{3}{8})]}
-\frac{C_1^{2/3}}{2}
\frac{\Ai'[C_1^{-1/3}(N+\frac{29}{8})]}
{\Ai[ C_1^{-1/3}(N-\frac{3}{8})]},\label{eq:Zinst_k=1} \\
Z_2^{\rm D2+WS}(N)
&=\Bigl(4N+C_2+7\Bigr)
\frac{\Ai[C_2^{-1/3}(N+\frac{7}{4})]}{\Ai[C_2^{-1/3}(N-\frac{1}{4})]}
-2C_2^{2/3}
\frac{\Ai'[C_2^{-1/3}(N+\frac{7}{4})]}{\Ai[ C_2^{-1/3}(N-\frac{1}{4})]}.
\end{align}
One can check that the leading large $N$ behavior of
\eqref{eq:Zinst_k=1} precisely reproduces the result in \cite{PY}.
We note again that our results include all the $1/N$ corrections.
 
Finally, for $k>2$, the leading non-perturbative contribution comes
from the worldsheet instanton.
Since our result is valid for any $N$ and $k$, we can take the 't
Hooft limit.
Our worldsheet instanton correction \eqref{eq:Zinst_ws} must give the
correct answer in this limit.
We check that the genus-zero and genus-one worldsheet instanton corrections
to the free energy studied in \cite{DMP1}
are indeed reproduced from our result \eqref{eq:Zinst_ws} after taking
the 't Hooft limit.
This is a non-trivial check of our result.

The paper is organized as follows.
In section~2, we review the method to compute the partition function exactly.
We generalize our previous method in \cite{HMO} to arbitrary level $k$.
In section~3, using the exact partition function obtained by this method, we numerically study the behavior of the instanton corrections.
We focus on the grand potential and the partition function.
We find that the grand potential shows an oscillatory behavior and that it is explained by the periodicity of the grand partition function.
By the numerical fitting, we determine the instanton corrections to the grand potential at $k=1,2,3,4,6$.
In section~4, we compute the worldsheet instanton corrections by using
the results in the topological string on local $\mathbb{F}_0$.
As a non-trivial test, we take the 't Hooft limit of our results, and confirm that our results correctly reproduce the worldsheet instanton
corrections to the genus-zero and genus-one free energies up to 3-instanton.
In section~5, we propose an analytic form of the D2-instanton correction, and check its validity.
Section~6 is devoted to conclusions and discussions.
In appendix~A, we prove a new relation found in this paper.
In appendix~B, we summarize the exact values of the partition function at $k=1,2,3,4,6$.

\section{Exact Results of Partition Function}\label{sec:exactZ}
In \cite{MP} the ABJM partition function
\begin{align}
Z_k(N)=\frac{1}{(N!)^2}\int\frac{d^Nx_i}{(2\pi)^N}\frac{d^Ny_i}{(2\pi)^N}
\frac{\prod_{i<j}[2\sinh\frac{\mu_i-\mu_j}{2}]^2
[2\sinh\frac{\nu_i-\nu_j}{2}]^2}
{\prod_{i,j}[2\cosh\frac{\mu_i-\nu_j}{2}]^2}
\exp\biggl[\frac{ik}{4\pi}\sum_i(\mu_i^2-\nu_i^2)\biggr],
\label{ABJMmatrix}
\end{align}
is rewritten in terms of the partition function of a Fermi gas system
\begin{align}
Z_k(N)=\frac{1}{N!}\sum_{\sigma\in S_N}
(-1)^{\epsilon(\sigma)}\int\frac{d^Nq}{(2\pi )^N}
\prod_i\rho(q_i,q_{\sigma(i)}),
\end{align}
with the density matrix given by
\begin{align}
\rho(q_1,q_2)=\frac{1}{k}\frac{1}{\sqrt{2\cosh\frac{q_1}{2}}}
\frac{1}{2\cosh\frac{q_1-q_2}{2k}}
\frac{1}{\sqrt{2\cosh\frac{q_2}{2}}}.
\end{align}
After introducing the chemical potential $\mu$ or the fugacity
$z=e^\mu$, the grand partition function
\begin{align}
\Xi_k(\mu)=1+\sum_{N=1}^\infty Z_k(N)e^{\mu N},
\label{grand}
\end{align}
is given by
\begin{align}
\Xi_k(\mu)=\det(1+z\rho)
=\exp\left[-\sum_{n=1}^\infty\frac{(-z)^n}{n}\Tr\rho^n\right].
\end{align}
This means that if we want to know the partition functions for various
$N$, all we have to do is to compute $\Tr\rho^n$ for $n=1, \cdots, N$.

As explained previously in \cite{HMO}, it is difficult to compute
$\Tr\rho^n$ because of the complexity in convoluting the density
matrix.
This difficulty was, however, overcome by a lemma from \cite{TW}
stating that when a kernel $K(q_1,q_2)$ takes the form
\begin{align}
K(q_1,q_2)=\frac{E(q_1)E(q_2)}{M(q_1)+M(q_2)},
\label{eq:K-form}
\end{align}
we can compute its power $K^n(q_1,q_2)$ by
\begin{align}
K^n(q_1,q_2)=\frac{E(q_1)E(q_2)}{M(q_1)+(-1)^{n-1}M(q_2)}
\sum_{\ell=0}^{n-1}(-1)^\ell\phi^\ell(q_1)\phi^{n-1-\ell}(q_2),
\end{align}
where $\phi^\ell(q)$ is defined by
\begin{align}
\phi^\ell(q)=\frac{1}{E(q)}\int\frac{dq'}{2\pi}K^\ell(q,q')E(q')~.
\end{align}
We can compute $\phi^\ell(q)$ recursively
\begin{align}
\phi^\ell(q)=\frac{1}{E(q)}\int\frac{dq'}{2\pi}K(q,q')E(q')\phi^{\ell-1}(q'),
\label{phirecursive}
\end{align}
with the initial condition $\phi^0(q)=1$.
Though we can apply this method to the density matrix $\rho$ directly,
there is a more convenient way to compute the grand partition function
as explained in \cite{HMO}.

Let us briefly review the method found in \cite{HMO}.
As our density matrix is invariant under the
parity
\begin{align}
\rho(-q_1,-q_2)=\rho(q_1,q_2),
\end{align}
we can decompose it into two parts:
\begin{align}
\rho(q_1,q_2)=\rho_+(q_1,q_2)+\rho_-(q_1,q_2),\quad
\rho_\pm(q_1,q_2)\equiv\frac{\rho(q_1,q_2)\pm\rho(q_1,-q_2)}{2}.
\end{align}
Since $\rho_+$ (and also $\rho_-$) is written as the form
\eqref{eq:K-form},
\begin{align}
\rho_+(q_1,q_2)
=\frac{E(q_1)E(q_2)}{\cosh(\frac{q_1}{k})+\cosh(\frac{q_2}{k})},
\end{align}
with
\begin{align}
E(q)=\frac{\cosh(\frac{q}{2k})}{\sqrt{2k\cosh(\frac{q}{2})}},
\end{align}
we can compute $\rho_+^n$ from a series of functions $\phi^\ell(q)$ as
\begin{align}
\rho_+^n(q_1,q_2)=\frac{E(q_1)E(q_2)}
{\cosh(\frac{q_1}{k})+(-1)^{n-1}\cosh(\frac{q_2}{k})}
\sum_{\ell=0}^{n-1}(-1)^\ell\phi^\ell(q_1)\phi^{n-1-\ell}(q_2).
\end{align}
The important fact is that using \eqref{phirecursive} the functions
$\phi^\l(q)$ are determined by the following recursive relation:
\begin{align}
\phi^\l(q)=\frac{1}{2k \cosh(\frac{q}{2k})}
\int_{-\infty}^\infty\frac{dq'}{2\pi}
\frac{1}{2\cosh (\frac{q-q'}{2k})}
\frac{\cosh(\frac{q'}{2k})}{\cosh (\frac{q'}{2})}
\phi^{\l-1}(q').
\label{eq:int_eq_phi}
\end{align}
One can show that the grand partition function is factorized as
\begin{align}
\Xi(\mu)=\det(1+z \rho_+)\det(1+z\rho_-).
\end{align}
In \cite{HMO}, we found that the density matrices enjoy an interesting
property.
\begin{align}
\frac{\det(1+z\rho_-)}{\det(1-z\rho_+)}=\sum_{\l=0}^\infty\phi^\ell(0)z^\l.
\label{rhoidentity}
\end{align}
Thus the grand partition function is rewritten as
\begin{align}
\Xi(\mu)=\det(1-z^2\rho_+^2)\cdot\sum_{\l=0}^\infty\phi^\ell(0)z^\l
\label{Xi},
\end{align}
and determined only by the even parity functions $\phi^\l(q)$.

In \cite{HMO}, we further found that the eigenvalue problem of the density matrix $\rho$ is
mapped to that of an infinite-dimensional parity-preserving Hankel matrix, whose form is given by
\begin{align}
H=\begin{pmatrix}
h_0&0&h_1&0&h_2&0&\\
0&h_1&0&h_2&0&h_3&\\
h_1&0&h_2&0&h_3&0&\cdots\\
0&h_2&0&h_3&0&h_4&\\
h_2&0&h_3&0&h_4&0&\\
0&h_3&0&h_4&0&h_5&\\
&&\vdots&&&&\ddots
\end{pmatrix}.
\end{align}
Thus the grand partition function is also expressed as
\begin{align}
\Xi(\mu)=\det (1+z\rho)=\det(1+zH).
\end{align}
In this paper, we find a general formula for arbitrary parity-preserving Hankel matrices, which reduces to \eqref{rhoidentity} for our current case.
Let $H_\pm$ be the even/odd parity block of the original Hankel
matrix,
\begin{align}
H_+=\begin{pmatrix}
h_0&h_1&h_2&\\
h_1&h_2&h_3&\cdots\\
h_2&h_3&h_4&\\
&\vdots&&\ddots
\end{pmatrix},
\quad
H_-=\begin{pmatrix}
h_1&h_2&h_3&\\
h_2&h_3&h_4&\cdots\\
h_3&h_4&h_5&\\
&\vdots&&\ddots
\end{pmatrix}.
\end{align}
Then it satisfies the following relation
\begin{align}
\frac{\det(1+zH_-)}{\det(1-zH_+)}=\langle e_0|\frac{1}{1-zH_+}|v\rangle,
\label{Hankelidentity}
\end{align}
with
\begin{align}
|e_0\rangle=(1,0,0,\cdots)^{\rm T},\qquad
|v\rangle=(1,1,1,\cdots)^{\rm T}.
\label{evdef}
\end{align}
In appendix A, we shall give a proof of \eqref{Hankelidentity}  for a
certain class of Hankel matrices.

Using these methods we computed the partition functions at
$k=1,2,3,4,6$.
The exact values are given in appendix B.
As can be imagined from \cite{O}, the partition functions have close
relations with $\tan(\frac{\pi}{k})$.
Therefore it is especially clean for these values of $k$.

\section{Numerical Study}
As summarized in appendix \ref{sec:exactvalue}, we obtain the exact
values of the partition functions at $k=1,2,3,4,6$ up to
$N_{\rm max}=44,20,18,16,14$, respectively, following the procedure in
section \ref{sec:exactZ}.
After computing the exact partition function, let us extract important information from it.
Here we numerically study the non-perturbative corrections to the partition function.
We shall plot various physical quantities computed by our exact values
and fit them by the theoretically expected forms
\cite{MP,DMP1,DMP2}.

\subsection{Grand Potential}\label{sec:J}

First, we start with the grand potential $J_k(\mu)$ defined by
\begin{align}
J_k(\mu)=\log\Xi_k(\mu)~.
\label{J}
\end{align}
From the theoretical expectation the prefactor of the instanton
corrections are at most quadratic functions of $\mu$
(see \eqref{eq:JWS}).
In fact, we find numerically that the grand potential $J_k(\mu)$ ($k=1,2,3,4,6$)
generally can be expanded as%
\footnote{We expect that this property holds for $k \in \mathbb{Z}$. However, we cannot establish it at present.}
\begin{align}
J_k(\mu)=J_k^{\rm (pert)}(\mu)
+\sum_{n=1}^\infty J_k^{(n)}
e^{-\frac{4n\mu}{k}}~,\quad
J_k^{(n)}=\alpha_k^{(n)}\mu^2+\beta_k^{(n)}\mu+\gamma_k^{(n)}~.
\label{Jexpand}
\end{align} 
In the following we shall plot the grand potential and their
instanton corrections, fit them by the quadratic functions and read
off the coefficients numerically.

However, not all of the coefficients are read off from this analysis
of $J_k(\mu)$.
As we shall explain below we will encounter some
oscillations in $\mu$ and it is difficult to study the exact values of
the coefficients due to large errors caused by the oscillations.
Some of the values are actually first read off from the analysis of
the partition function $Z_k(N)$ as in subsection \ref{sec:Z} which does
not have the oscillating behavior, while some of them are found after
we understand the oscillating behavior well as in subsection
\ref{sec:osc} and subtract its contribution.
It is simpler, however, to present our result as if we read all of the
exact values from the grand potential and explain various behaviors
later.
Hence, we shall take this path in presenting our results.

\begin{figure}[tb]
\begin{center}
\begin{tabular}{cc}
\resizebox{60mm}{!}{\includegraphics{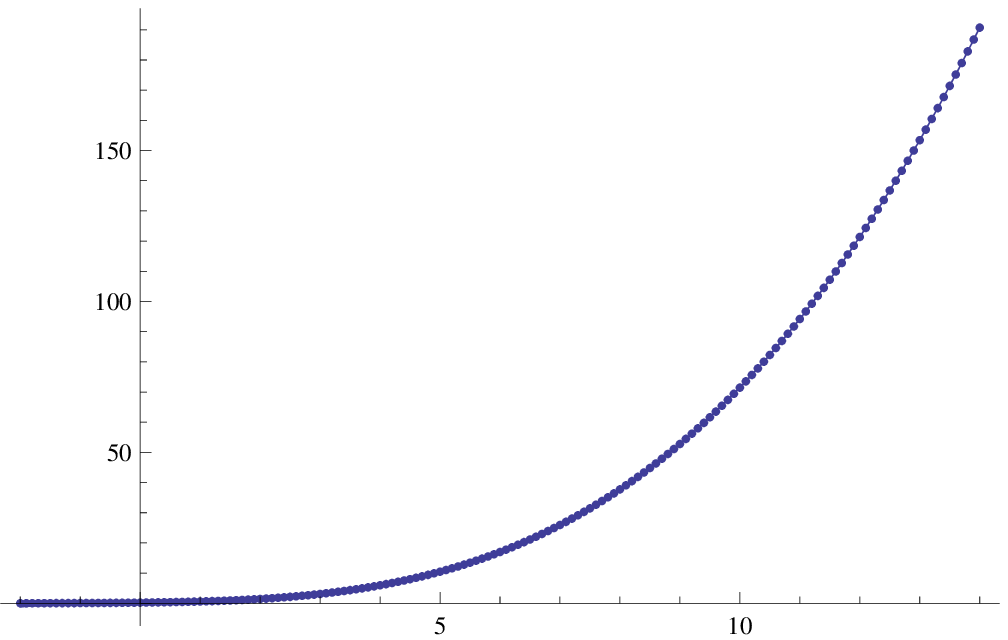}}
&
\resizebox{60mm}{!}{\includegraphics{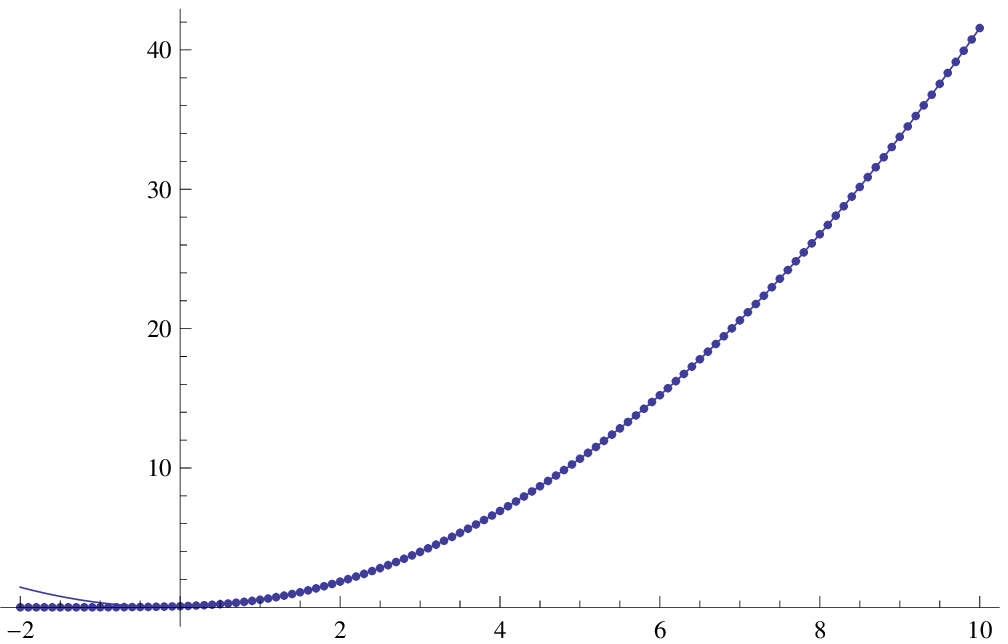}}
\\(a)&(b)\\[10pt]
\resizebox{60mm}{!}{\includegraphics{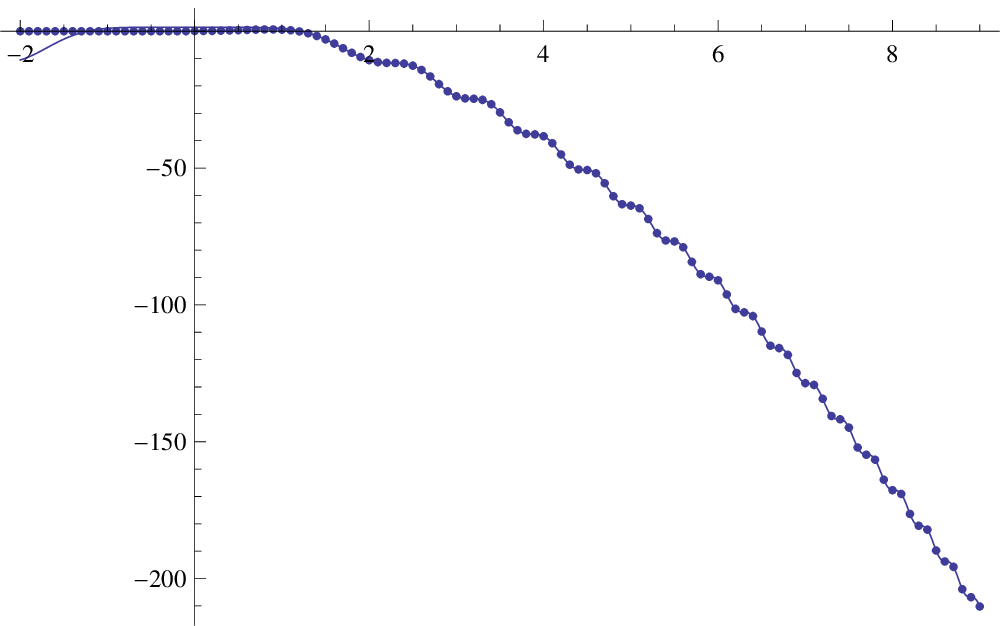}}
&
\resizebox{60mm}{!}{\includegraphics{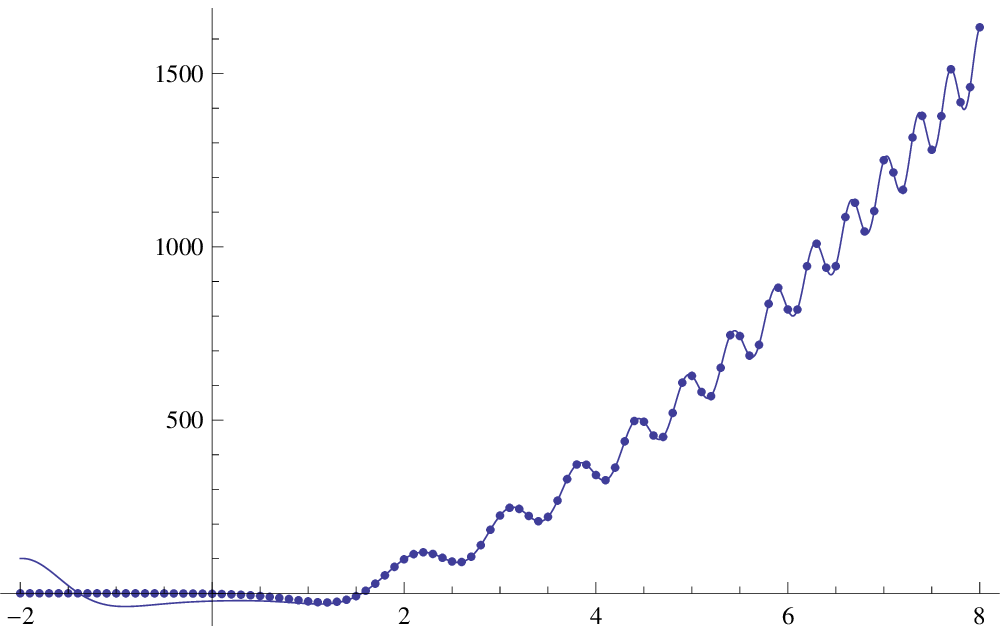}}
\\(c)&(d)
\end{tabular}
\end{center}
  \caption{(a) The perturbative, (b) 1-instanton, (c) 2-instanton and
    (d) 3-instanton corrections of the grand potential $J_k(\mu)$ at
    $k=1$. The dots represent the grand potential obtained by the exact partition function while the solid lines represent
    the fitted functions.}
  \label{fig:J1}
\end{figure}

Let us start to study the grand potential at $k=1$.
In figure~\ref{fig:J1}(a), we have plotted the
grand potential $J_1(\mu)$ (for selected values of $\mu$) in dots together with its perturbative
expression $J_1^{\rm (pert)}(\mu)$ in a solid line.
The grand potential $J_k(\mu)$ \eqref{J} is given by the grand
partition function $\Xi_k(\mu)$ \eqref{Xi} with the summation of $N$
truncated at our maximal values $N_{\rm max}$, while the perturbative
expression is given by
\begin{align}
J_k^{\rm (pert)}(\mu)=\frac{C_k}{3}\mu^3+B_k\mu+A_k.
\label{Jpert}
\end{align}
Here the constants $B_k$ and $C_k$ are given by \eqref{eq:BC}, and
$A_k$ was conjectured in \cite{MP,KEK} as
\begin{align}
A_k=-\frac{\zeta(3)}{8\pi^2}k^2
+\frac{1}{6}\log\frac{4\pi}{k}+2\zeta'(-1)
-\frac{1}{3}\int_0^\infty dx\,
\frac{1}{e^{kx}-1}
\left(\frac{3}{x^3}-\frac{1}{x}-\frac{3}{x \sinh^2x}\right).
\end{align}
We find a perfect match, especially for large $\mu$.
Therefore, we can plot
$J_1^{(1)}(\mu)=(J_1(\mu)-J_1^{\rm (pert)}(\mu))/e^{-4\mu}$
next in figure~\ref{fig:J1}(b) and fit the result by a quadratic
function.
We find the behavior matches well with
\begin{align}
J_1^{(1)}(\mu)=\frac{4\mu^2+\mu+1/4}{\pi^2}.
\label{k1Jfit}
\end{align}
We continue this process further and plot
$J_1^{(2)}(\mu)=(J(\mu)-J_1^{\rm (pert)}(\mu)-e^{-4\mu}J_1^{(1)}(\mu))/e^{-8\mu}$
in figure~\ref{fig:J1}(c).
Now the plot starts to oscillate.
We can fit the result by the sum of a quadratic function with the
oscillations $\cos(\frac{4\mu^2}{\pi})$ and $\sin(\frac{4\mu^2}{\pi})$,
\begin{align}
J_1^{(2)}(\mu)&=-\frac{52\mu^2+\mu/2+9/16}{2\pi^2}+2
+J_1^{(2){\rm osc}}(\mu),\nonumber\\
J_1^{(2){\rm osc}}(\mu)&=2\cos\Bigl(\frac{4\mu^2}{\pi}-\frac{7\pi}{12}\Bigr).
\label{eq:J_2}
\end{align}
Similarly, the 3-instanton
$J_1^{(3)}(\mu)=(J_1(\mu)-J_1^{\rm (pert)}(\mu)
-e^{-4\mu}J_1^{(1)}(\mu)-e^{-8\mu}J_1^{(2)}(\mu))/e^{-12\mu}$
is plotted in figure~\ref{fig:J1}(d), and the fitted function results in
\begin{align}
J_1^{(3)}(\mu)&=\frac{736\mu^2-152\mu/3+77/18}{3\pi^2}-32
+J_1^{(3){\rm osc}}(\mu),\nonumber\\
J_1^{(3){\rm osc}}(\mu)
&=-\frac{4}{\pi}(8\mu+1)\sin\Bigl(\frac{4\mu^2}{\pi}-\frac{7\pi}{12}\Bigr)
-32\cos\Bigl(\frac{4\mu^2}{\pi}-\frac{7\pi}{12}\Bigr).
\label{J13}
\end{align}

Note that our plot range of $\mu$ changes slightly.
This is because in going to higher and higher instanton effects, we
have more and more numerical errors and it is difficult to find a
sensible plot for large $\mu$.

\begin{figure}[tb]
\begin{center}
\begin{tabular}{cc}
\resizebox{60mm}{!}{\includegraphics{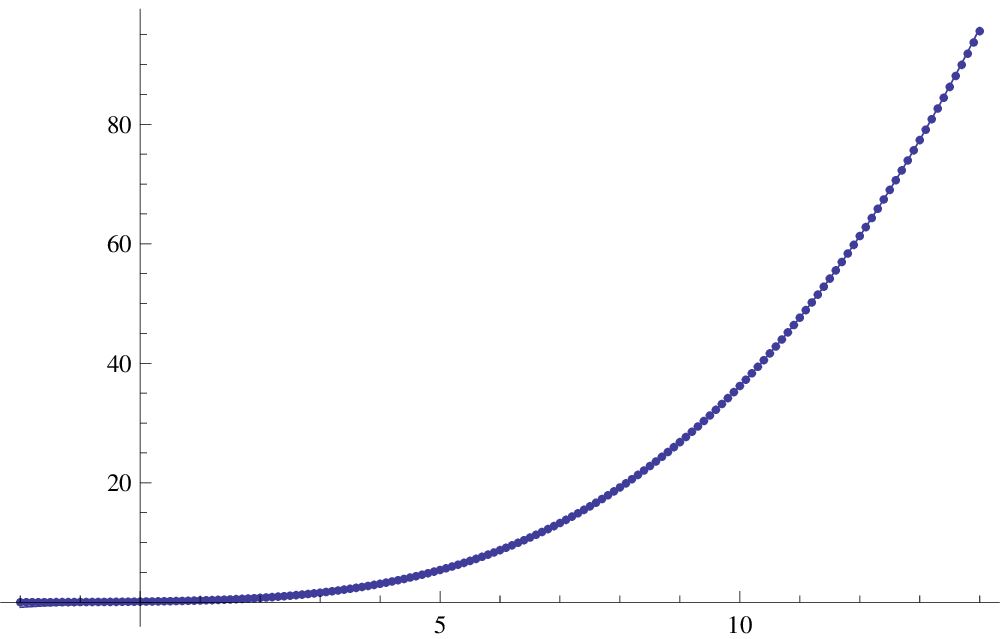}}
&
\resizebox{60mm}{!}{\includegraphics{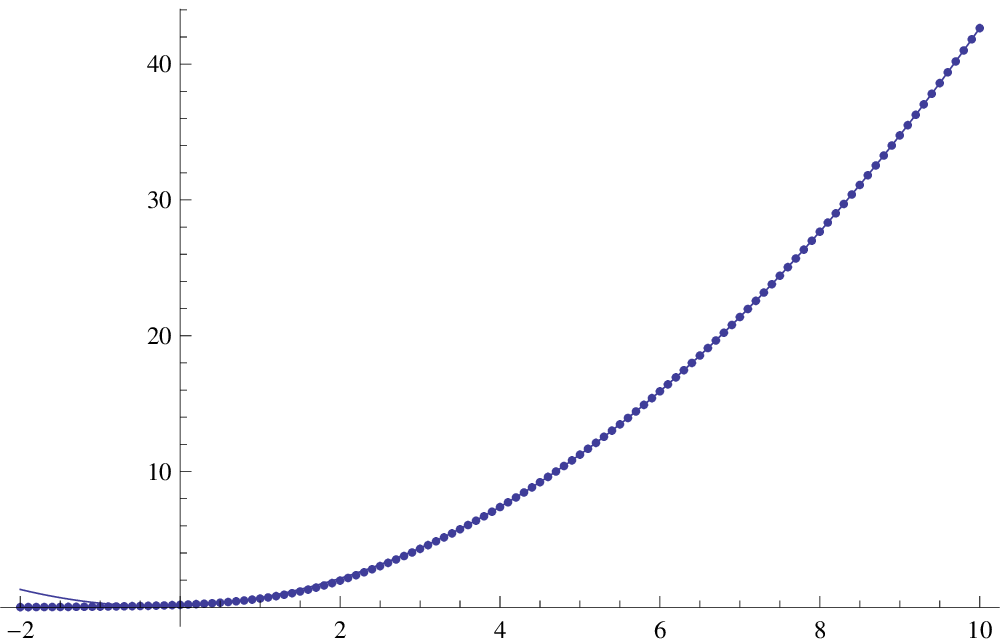}}
\\ (a) & (b)\\[10pt]
\resizebox{60mm}{!}{\includegraphics{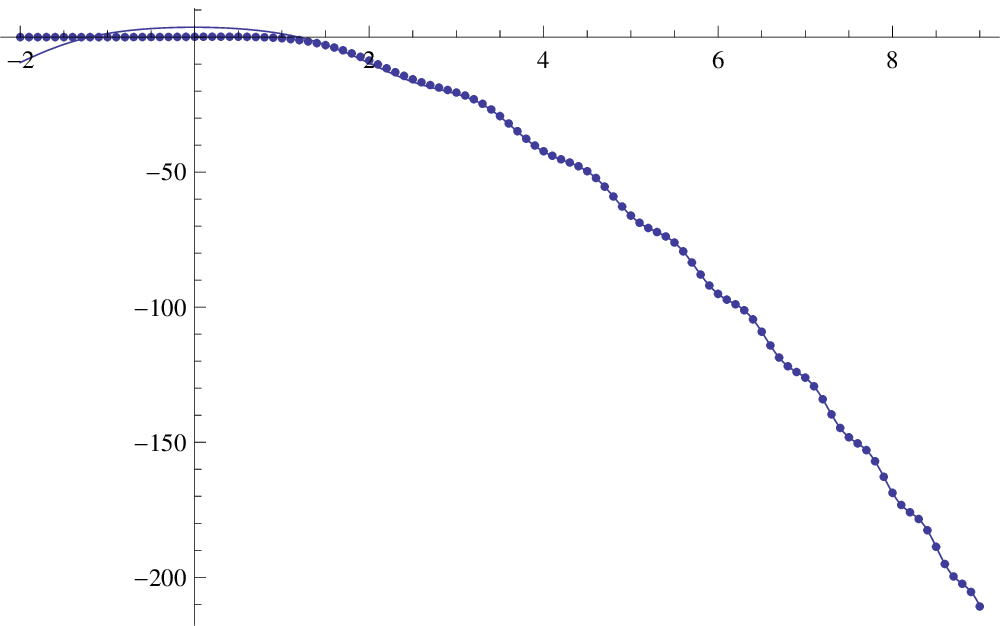}}
&
\resizebox{60mm}{!}{\includegraphics{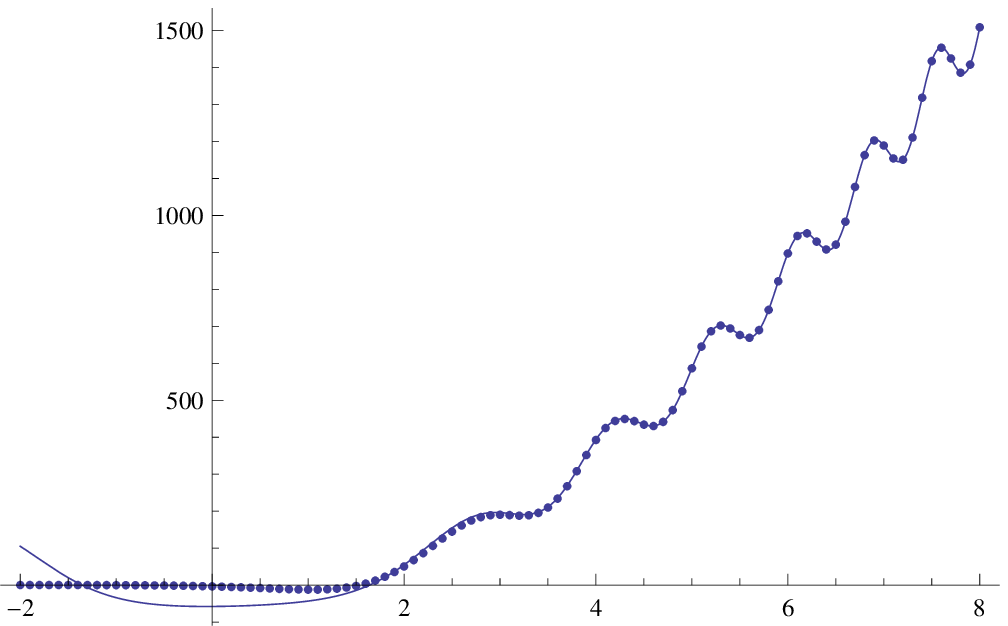}}
\\ (c) & (d)
\end{tabular}
\end{center}
  \caption{(a) The perturbative, (b) 1-instanton, (c) 2-instanton and
    (d) 3-instanton corrections of the grand potential $J_k(\mu)$ at
    $k=2$. The dots represent the grand potential obtained by the exact partition function while the solid lines represent
    the fitted functions.}
  \label{fig:J2}
\end{figure}

\begin{figure}[tb]
\begin{center}
\begin{tabular}{cc}
\resizebox{60mm}{!}{\includegraphics{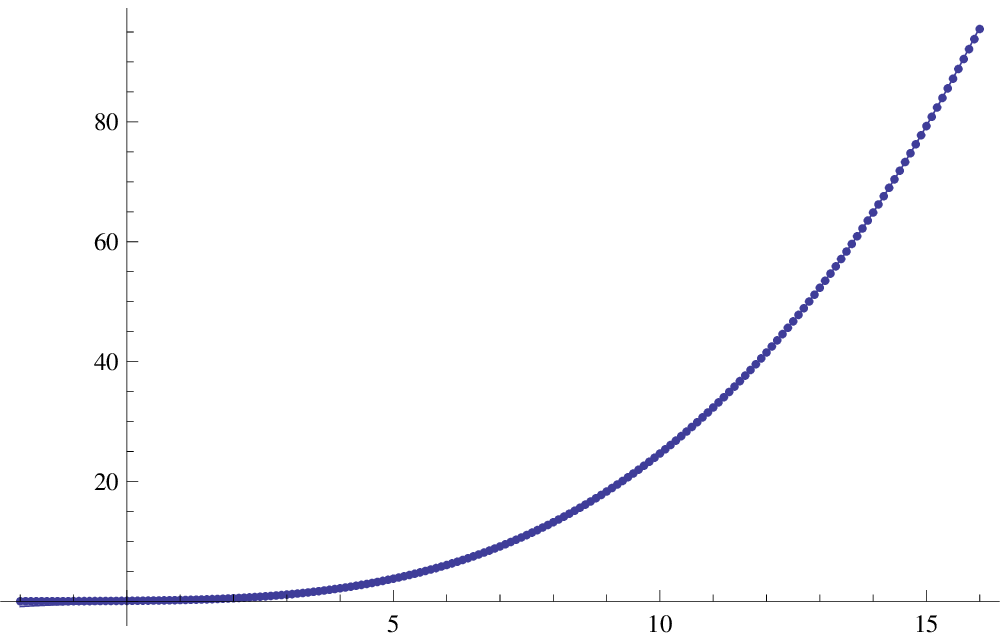}}
&
\resizebox{60mm}{!}{\includegraphics{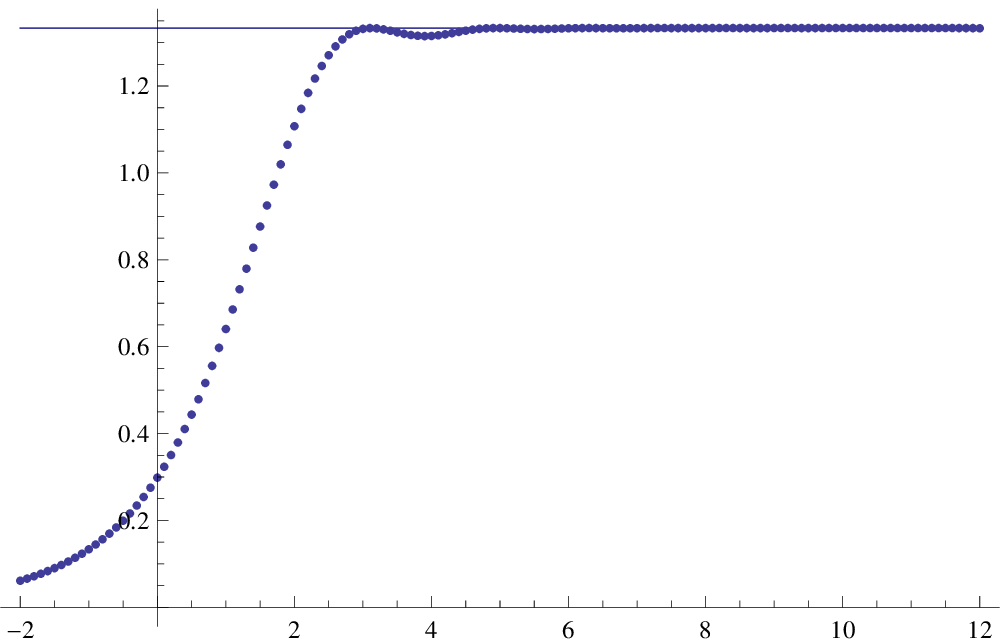}}
\\ (a) & (b)\\[10pt]
\resizebox{60mm}{!}{\includegraphics{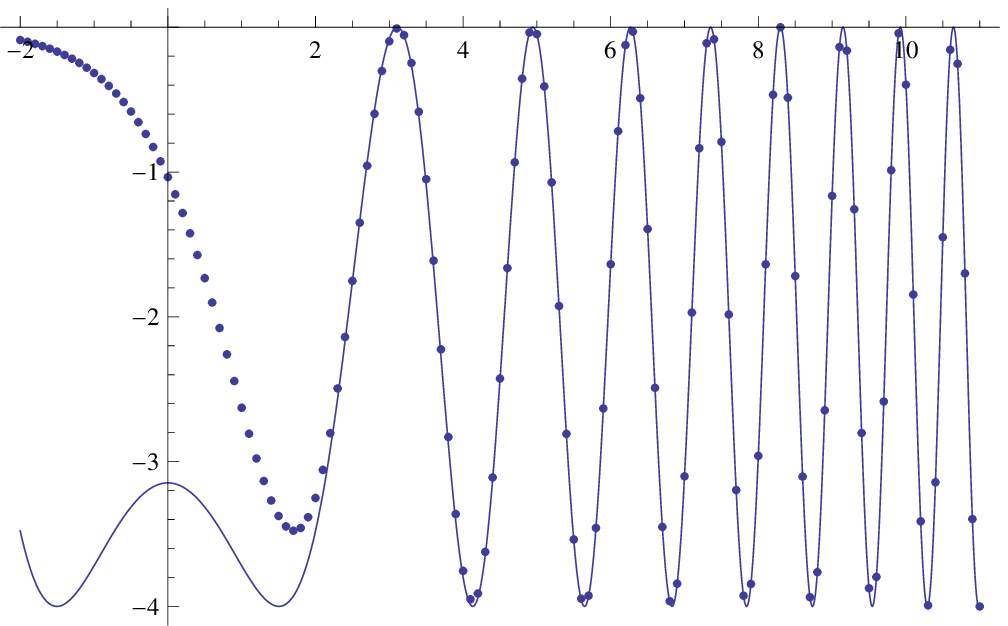}}
&
\resizebox{60mm}{!}{\includegraphics{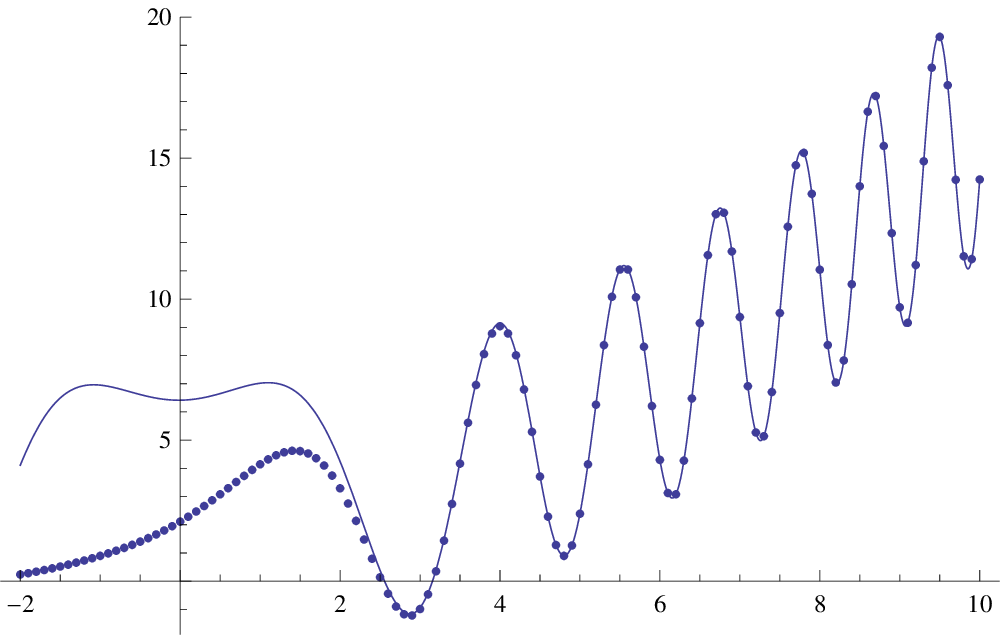}}
\\ (c) & (d)
\end{tabular}
\end{center}
  \caption{(a) The perturbative, (b) 1-instanton, (c) 2-instanton and
    (d) 3-instanton corrections of the grand potential $J_k(\mu)$ at
    $k=3$. The dots represent the grand potential obtained by the exact partition function while the solid lines represent
    the fitted functions.}
  \label{fig:J3}
\end{figure}

\begin{figure}[tb]
\begin{center}
\begin{tabular}{cc}
\resizebox{60mm}{!}{\includegraphics{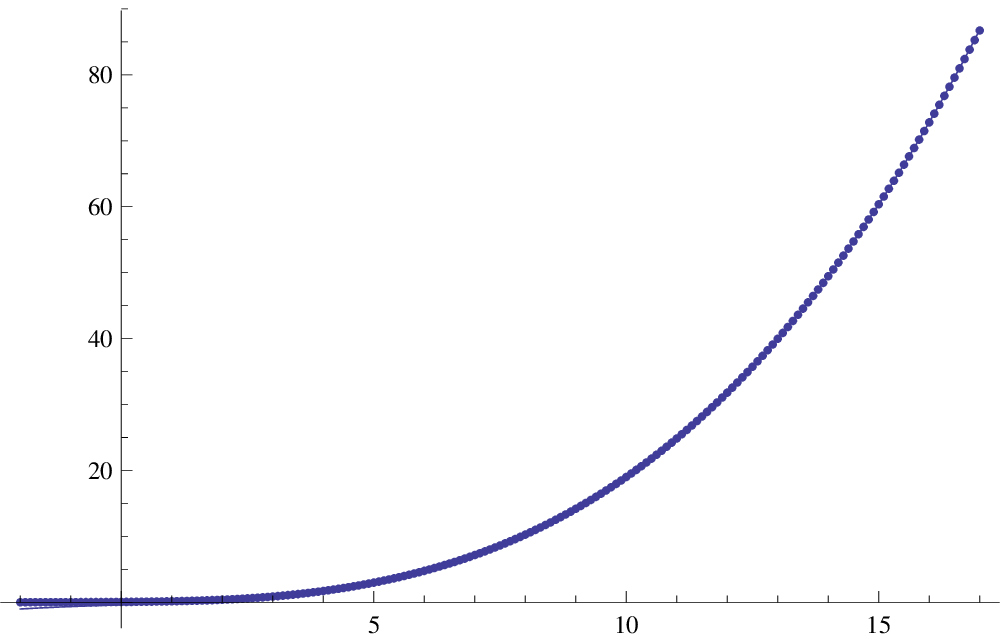}}
&
\resizebox{60mm}{!}{\includegraphics{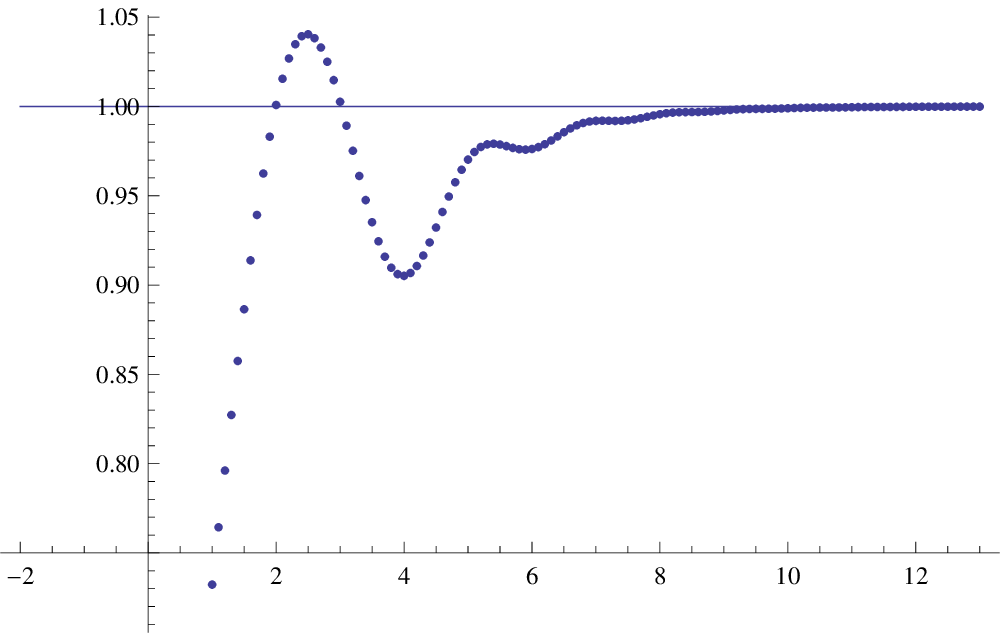}}
\\ (a) & (b)\\[10pt]
\resizebox{60mm}{!}{\includegraphics{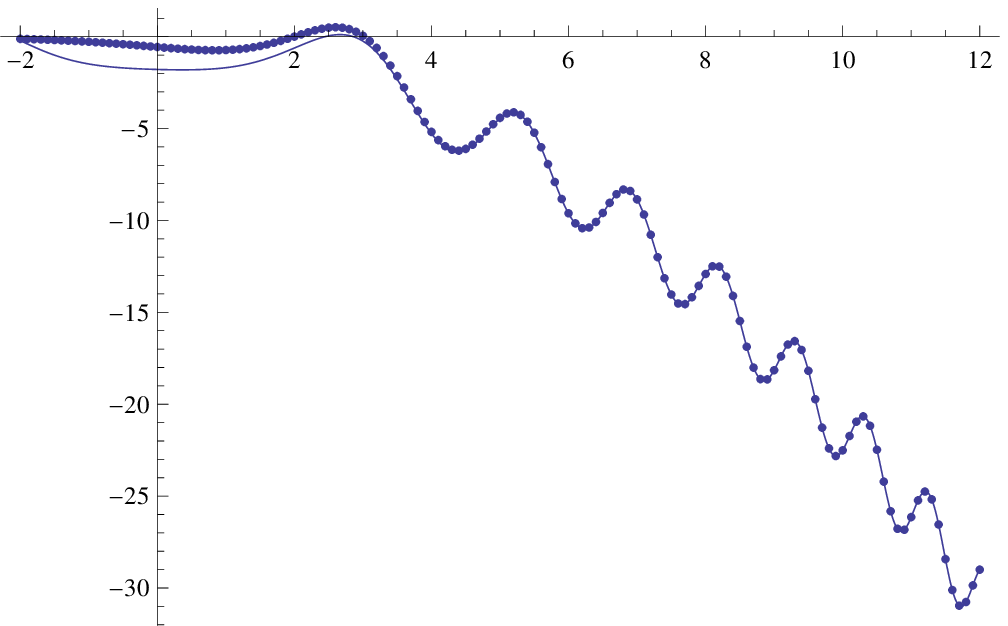}}
&
\resizebox{60mm}{!}{\includegraphics{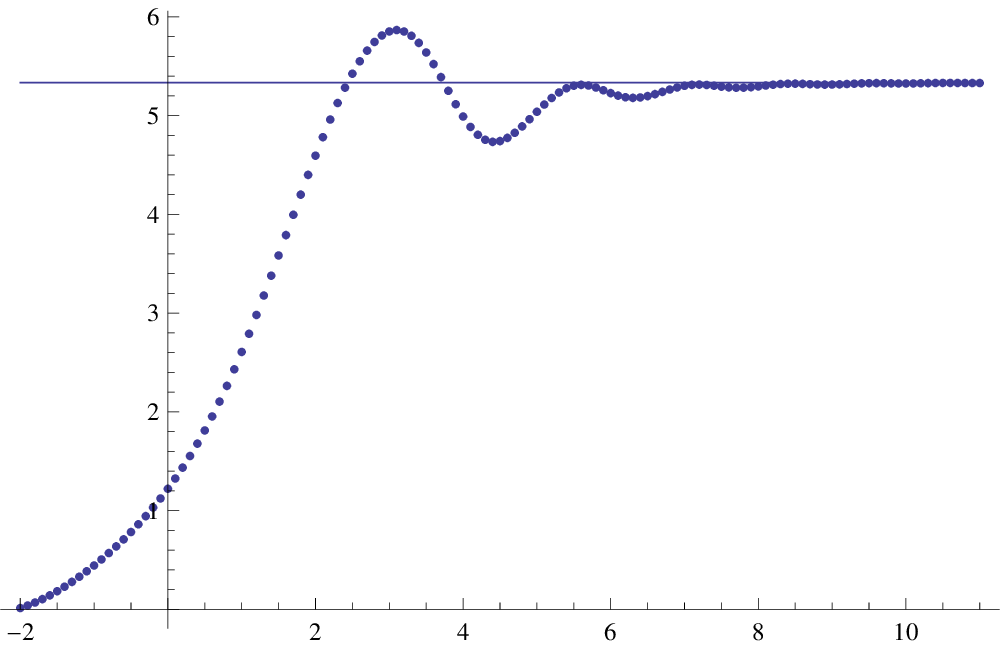}}
\\ (c) & (d)
\end{tabular}
\end{center}
  \caption{(a) The perturbative, (b) 1-instanton, (c) 2-instanton and
    (d) 3-instanton corrections of the grand potential $J_k(\mu)$ at
    $k=4$. The dots represent the grand potential obtained by the exact partition function while the solid lines represent
    the fitted functions.}
  \label{fig:J4}
\end{figure}

\begin{figure}[tb]
\begin{center}
\begin{tabular}{cc}
\resizebox{60mm}{!}{\includegraphics{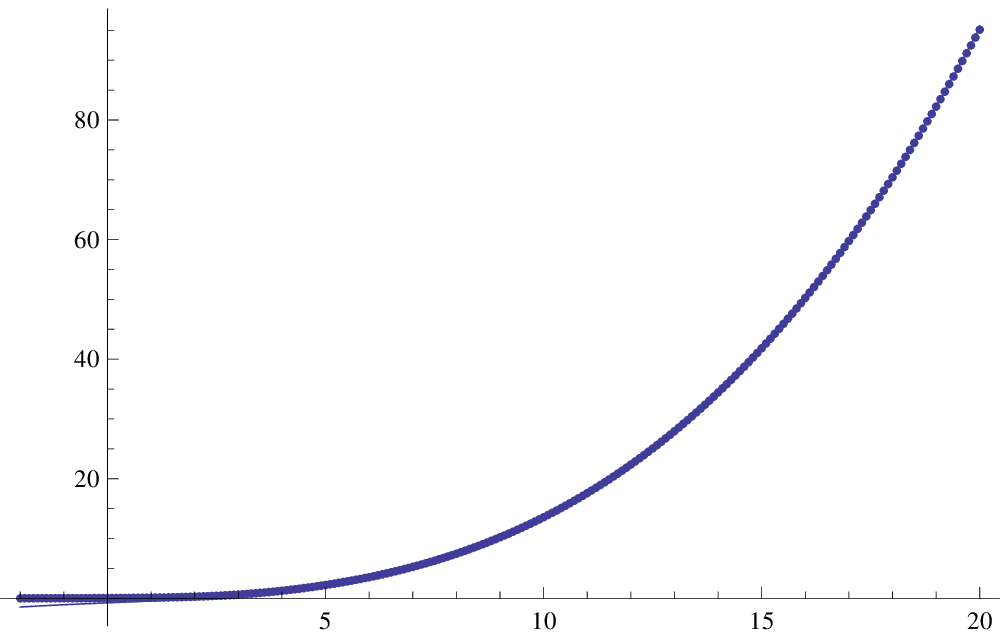}}
&
\resizebox{60mm}{!}{\includegraphics{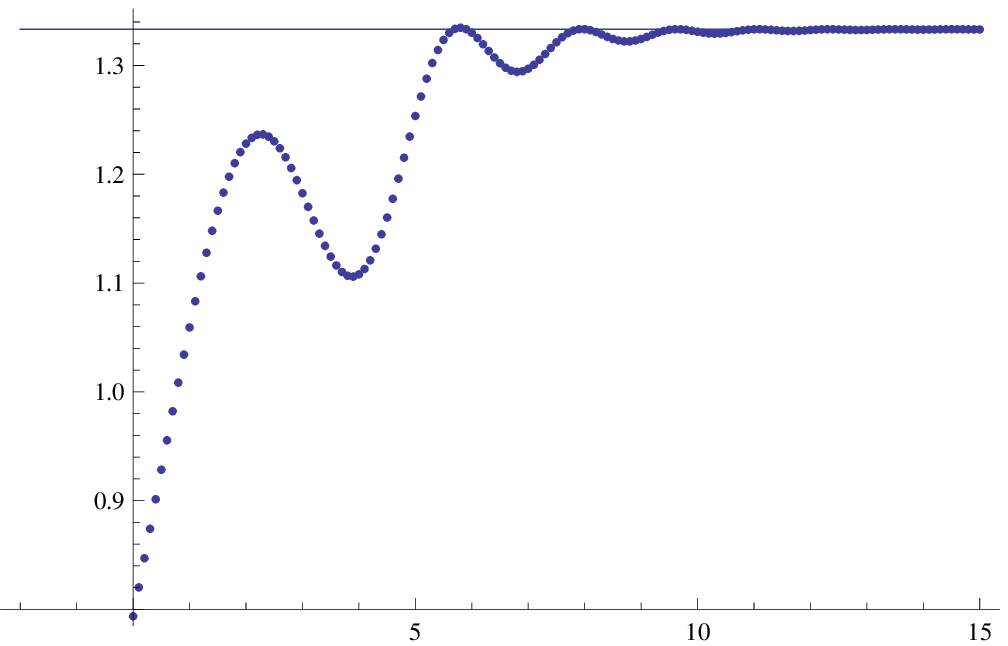}}
\\ (a) & (b)\\[10pt]
\resizebox{60mm}{!}{\includegraphics{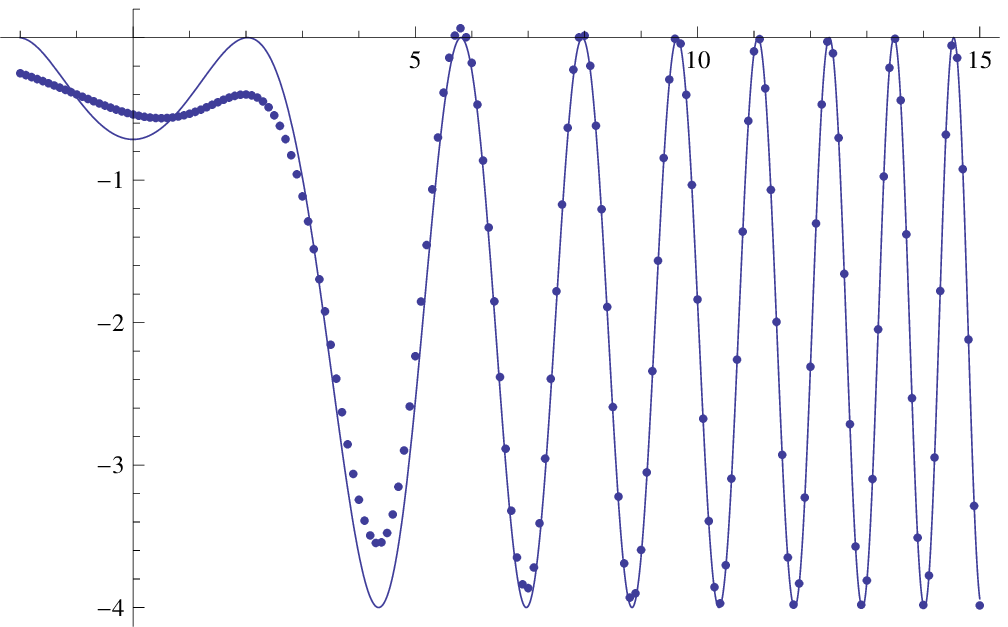}}
&
\resizebox{60mm}{!}{\includegraphics{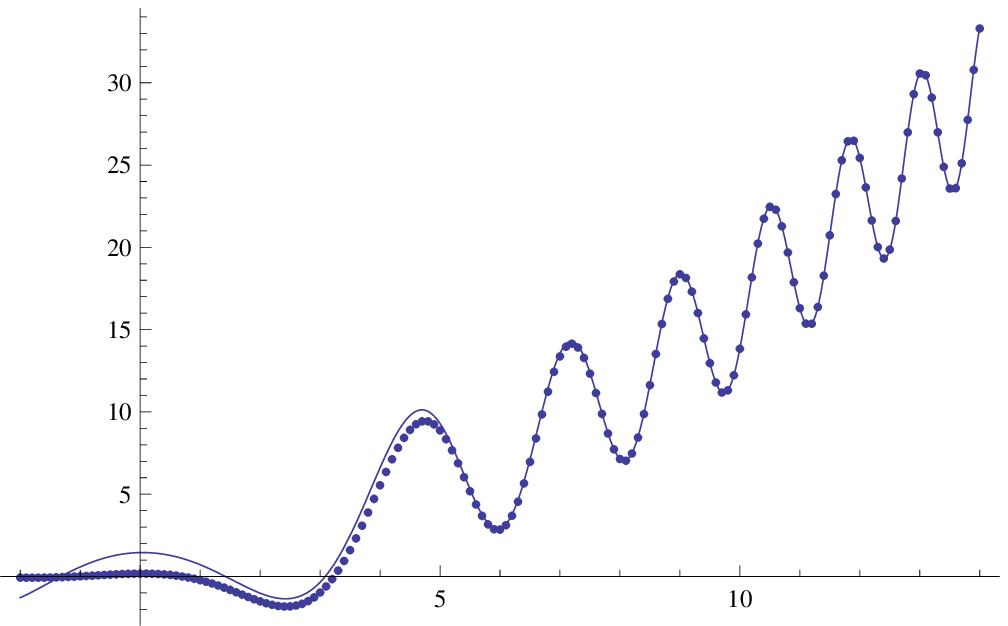}}
\\ (c) & (d)
\end{tabular}
\end{center}
  \caption{(a) The perturbative, (b) 1-instanton, (c) 2-instanton and
    (d) 3-instanton corrections of the grand potential $J_k(\mu)$ at
    $k=6$. The dots represent the grand potential obtained by the exact partition function while the solid lines represent
    the fitted functions.}
  \label{fig:J6}
\end{figure}

We can repeat the same analysis at $k=2, 3, 4, 6$ since we have exact
values.
The results computed from the exact values we find are plotted in
figures~\ref{fig:J2}, \ref{fig:J3}, \ref{fig:J4}, \ref{fig:J6},
respectively together with the fitted functions.
The results are summarized in \eqref{Jnp}.
Using the results in \eqref{Jnp}, we can study various instanton
effects.
This will be done in the subsequent sections.
Before proceeding to it, let us give a physical interpretation for the
oscillating behaviors and present the numerical study of $Z(N)$
without the oscillations.

\subsection{Oscillations in $J_k(\mu)$}\label{sec:osc}

In this subsection, we would like to explain the oscillation in
$J_k(\mu)$ from the periodicity of the grand partition function.
Since the grand partition function $\Xi_k(\mu)=e^{J_k(\mu)}$ is a sum
of $e^{N\mu}$ defined by \eqref{grand}, it must be a periodic function
of $\mu$ with the periodicity $2\pi i$.
However, the naive expression of $\Xi_k(\mu)$ derived from the method of
statistical mechanics in small $k$ perturbation is
not invariant under the shift $\mu\rightarrow\mu+2\pi i$.

To remedy the invariance, we define the $2\pi i$-periodic grand
partition function by
\begin{align}
e^{J_k(\mu)}=\sum_{n=-\infty}^\infty e^{J_k^{\rm naive}(\mu+2\pi i n)}.
\label{periodic}
\end{align}
Here $J_k^{\rm naive}(\mu)$ means the grand potential obtained from
the naive small $k$, large $\mu$ asymptotic expansion, namely
$J_k^{\rm naive}(\mu)=J_k^{\rm (pert)}(\mu)+J_k^{\rm (np)}(\mu)$
with $J_k^{\rm (pert)}(\mu)$ given by \eqref{Jpert}. 
Through this periodic extension \eqref{periodic}, the grand potential
acquires the oscillatory part
\begin{align}
J_k(\mu)=J_k^{\rm naive}(\mu)+J_k^{\rm osc}(\mu),
\end{align}
where $J_k^{\rm osc}(\mu)$ is given by
\begin{align}
 J_k^{\rm osc}(\mu)
=\log\biggl(1+\frac{1}{e^{J^{\rm naive}_k(\mu)}}
\sum_{n\not=0}e^{J^{\rm naive}_k(\mu+2\pi i n)}\biggr)~.
\end{align}

For instance, the oscillatory term in the worldsheet 2-instanton
comes from the $\pm 2\pi i$ shift of the perturbative part
$J_k^{\rm (pert)}(\mu)$.
Since
\begin{align}
\frac{e^{J_k^{\rm (pert)}(\mu+2\pi i)}}{e^{J_k^{\rm (pert)}(\mu)}}
=\exp\biggl[-\frac{8\mu}{k}
+2\pi i\Bigl(C_k\mu^2+B_k-\frac{8}{3k}\Bigr)\biggr],
\end{align}
we find the oscillatory term
\begin{align}
J_k^{\rm osc}(\mu)\approx
\frac{e^{J_k^{\rm (pert)}(\mu+2\pi i)}+e^{J_k^{\rm (pert)}(\mu-2\pi i)}}
{e^{J_k^{\rm (pert)}(\mu)}}
=2\cos\biggl[2\pi\Big(C_k\mu^2+B_k-\frac{8}{3k}\Big)\biggr]
e^{-\frac{8\mu}{k}},
\end{align}
in the worldsheet 2-instanton correction.
Note that this exactly reproduces \eqref{eq:J_2} at $k=1$.

We can proceed further to study the oscillatory term in the
worldsheet 3-instanton for $k>2$ from the worldsheet 1-instanton
contribution
\begin{align}
J_k^{\rm osc}(\mu)&\approx 
\frac{e^{J_k^{\rm (pert)}(\mu+2\pi i)+J_k^{{\rm WS}(1)}(\mu+2\pi i)}
+e^{J_k^{\rm (pert)}(\mu-2\pi i)+J_k^{{\rm WS}(1)}(\mu-2\pi i)}}
{e^{J_k^{\rm (pert)}(\mu)+J_k^{{\rm WS}(1)}(\mu)}}
\nn
&=2\cos\biggl[2\pi\Bigl(C_k\mu^2+B_k-\frac{8}{3k}\Bigr)\biggr]
e^{-\frac{8\mu}{k}}
+4d_k^{(1)}\sin\frac{4\pi}{k}
\sin\biggl[2\pi\Bigl(C_k\mu^2+B_k-\frac{14}{3k}\Bigr)\biggr]
e^{-\frac{12\mu}{k}}.
\label{Josci}
\end{align}
For $k=1,2$, however, since the D2 $m$-instanton
$J_k^{{\rm D2}(m)}\sim e^{-2m\mu}$ ($m=2,1$ for $k=1,2$) has the same
contribution as the worldsheet 1-instanton term
$J_k^{\rm WS(1)}(\mu)\sim e^{-\frac{4\mu}{k}}$, we have to be
careful.
The mixing between these two kinds of instanton effects will be
studied in more details in section 5.
Using $\alpha_k^{(1)}$ and $\beta_k^{(1)}$ in \eqref{Jexpand}, we find
that the oscillatory term in the worldsheet 3-instanton for $k=1,2$ is
given by
\begin{align}
J_k^{(3)\rm osc}(\mu)&=-4\pi(2\alpha_k^{(1)}\mu+\beta_k^{(1)})
\sin\biggl[2\pi\Bigl(C_k\mu^2+B_k-\frac{8}{3k}\Bigr)\biggr]
\nn&\quad
-8\pi^2\alpha_k^{(1)}
\cos\biggl[2\pi\Bigl(C_k\mu^2+B_k-\frac{8}{3k}\Bigr)\biggr].
\label{Josc-ab}
\end{align}
Note that this expression for $k=1$ reproduces \eqref{J13}.

As we have seen in the previous subsection, we actually observed an
oscillatory behavior in the numerical analysis of $J_k(\mu)$.
Since the amplitude and the phase of oscillation perfectly matches
\eqref{Josci} (and \eqref{Josc-ab}), we strongly believe that our
numerically observed oscillation is originated from the
$2\pi i$-periodic extension of the naive grand partition function.

One important consequence of the periodicity of the grand partition
function is that in the integral transformation from $\Xi_k(\mu)$ to
$Z_k(N)$ we can extend the integration range from $[-\pi i,\pi i]$ to 
$(-i\infty,i\infty)$
\begin{align}
Z_k(N)=\int_{-\pi i}^{\pi i} \frac{d\mu}{2\pi i} e^{J_k(\mu)-N\mu}
=\int_{-i\infty}^{i\infty} \frac{d\mu}{2\pi i} e^{J_k^{\rm naive}(\mu)-N\mu}~.
\label{XitoZ}
\end{align}
Since the last expression of \eqref{XitoZ} does not involve
$J_k^{\rm osc}$, there is no oscillation in the canonical partition
function $Z_k(N)$.
The last expression in \eqref{XitoZ} 
also explains the good agreement of the Airy function with the
exact partition function $Z_k(N)$.%
\footnote{
To obtain the Airy function expression, we need to deform the
integration contour from $[-\pi i,\pi i]$ to $(-i\infty,i\infty)$.
Naively, this deformation causes the error
$\cO(e^{-2\mu_*/k})=\cO(e^{-\pi\sqrt{2N/k}})$, where $\mu_*$ is the
value of the saddle-point, as was explained in \cite{MP}.
However, such a correction does not actually appear in the difference
of the exact partition function and the perturbative result.
The procedure here naturally explains this fact.
}

\subsection{Partition Function}\label{sec:Z}
In the previous subsections, we have studied the numerical behavior of
the grand potential $J_k(\mu)$ and observed an interesting oscillating
behavior.
Though interesting from a theoretical viewpoint, the oscillating
behavior causes a larger error in finding the exact values of
instanton prefactors.
As mentioned above, this behavior will disappear after an integral
transformation from the grand potential to the canonical partition
function.
For this reason, as a complementary analysis, let us look at the
partition function as well.
We find exactly the same values from the fitting of $J_k(\mu)$ and the
fitting of $Z_k(N)$.
This agreement gives a strong support for the validity of our method.

\begin{figure}[tb]
\begin{center}
\begin{tabular}{cc}
\hspace{-3mm}
\resizebox{80mm}{!}{\includegraphics{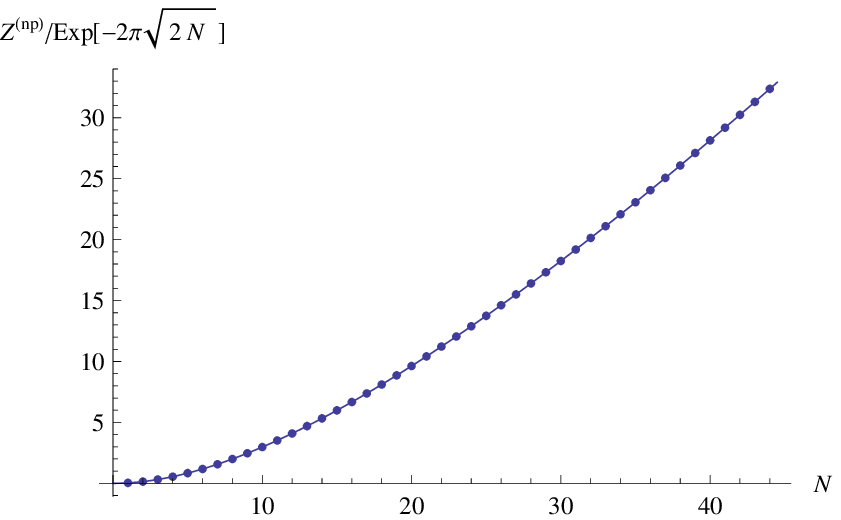}}
\hspace{-4mm}
&
\resizebox{80mm}{!}{\includegraphics{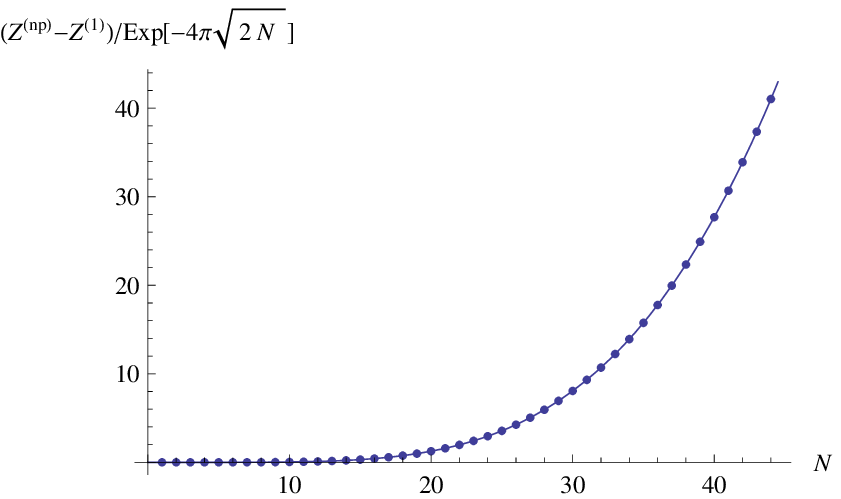}}
\hspace{-5mm}
\\ (a) & (b)
\vspace{-5mm}
\end{tabular}
\end{center}
  \caption{The 1- and 2-instanton corrections to the partition
    function at $k=1$.
    We plot (a) $Z_1^{\rm (np)}/e^{-2\pi\sqrt{2N}}$ and (b)
$(Z_1^{\rm (np)}-Z_1^{(1)})/e^{-4\pi\sqrt{2N}}$ against
    $N$. 
    The instanton corrections in terms of the Airy function (the solid
    lines) excellently agree with the exact values (the dots) even for
    small $N$.}
  \label{fig:Z1-inst}
\end{figure}

Let us consider the case $k=1$.
In this case, the non-perturbative correction to the grand potential
starts from $\cO(e^{-4\mu})$ as found in \cite{HMO}.
The (naive) grand potential is generically given by \eqref{Jexpand}.
We would like to determine the constants $\alpha_1^{(n)}$,
$\beta_1^{(n)}$ and $\gamma_1^{(n)}$ from the exact data.
Since the partition function is computed as 
\begin{align}
Z_1(N)&=\int_{-i\infty}^{i\infty}\frac{d\mu}{2\pi i}
e^{J_1^{\rm naive}(\mu)-N\mu}\nn
&=\int_{-i\infty}^{i\infty}\frac{d\mu}{2\pi i}
e^{J_1^{\rm (pert)}(\mu)-N\mu}
[1+(\alpha_1^{(1)}\mu^2+\beta_1^{(1)}\mu+\gamma_1^{(1)})e^{-4\mu}+\cdots],
\end{align}
the 1-instanton correction normalized by the perturbative
partition function is given by
\begin{align}
Z_1^{(1)}(N)
&=\frac{1}{Z_1^{\rm (pert)}(N)}\int_{-i\infty}^{i\infty}\frac{d\mu}{2\pi i}
e^{J_1^{\rm (pert)}(\mu)-N\mu}
(\alpha_1^{(1)}\mu^2+\beta_1^{(1)}\mu+\gamma_1^{(1)})e^{-4\mu} \nn
&=\frac{\alpha_1^{(1)}\pd_N^2-\beta_1^{(1)}\pd_N+\gamma_1^{(1)}}{Z_1^{\rm (pert)}(N)}
\int_{-i\infty}^{i\infty}\frac{d\mu}{2\pi i}e^{J_1^{\rm (pert)}(\mu)-(N+4)\mu}\nn
&=\Bigl(C_1^{-1}(N+4-B_1)\alpha_1^{(1)}+\gamma_1^{(1)}\Bigr)
\frac{\Ai[C_1^{-1/3}(N+4-B_1)]}{\Ai[C_1^{-1/3}(N-B_1)]}\nn
&\qquad
-C_1^{-1/3}\beta_1^{(1)}\frac{\Ai'[C_1^{-1/3}(N+4-B_1)]}{\Ai[C_1^{-1/3}(N-B_1)]},
\label{instAiry}
\end{align}
where $Z_k^{\rm (pert)}(N)$ is the perturbative contribution, which is
given by \cite{FHM,MP},
\begin{align}
Z_k^{\rm (pert)}(N)=C_k^{-1/3} e^{A_k} \Ai \left[ C_k^{-1/3}(N-B_k) \right].
\label{eq:Zpert}
\end{align}
The expression \eqref{instAiry} is the form expected from the Fermi
gas consideration.
We compare this result with the non-perturbative correction computed from
the exact values of the partition function using our definition of
$Z_1^{\rm (np)}(N)$ in \eqref{eq:Znp}.
To fix the constants $\alpha_1^{(1)}$, $\beta_1^{(1)}$ and $\gamma_1^{(1)}$,
we fit the exact data $Z_1^{\rm (np)}(N)/e^{-2\pi\sqrt{2N}}$ for 
$20\leq N\leq 44$ by the expected form
$Z_1^{(1)}(N)/e^{-2\pi\sqrt{2N}}$, and find that
$\alpha_1^{(1)}$, $\beta_1^{(1)}$ and $\gamma_1^{(1)}$ are very close to (in
about 14-digit agreement)
\begin{align}
\alpha_1^{(1)}=\frac{4}{\pi^2},\qquad
\beta_1^{(1)}=\frac{1}{\pi^2},\qquad
\gamma_1^{(1)}=\frac{1}{4\pi^2}.
\label{eq:alpha1}
\end{align}
This result, of course, is consistent with \eqref{k1Jfit} 
obtained from the fitting of $J_1(\mu)$.
In figure~\ref{fig:Z1-inst}(a), we show the exact data
$Z_1^{\rm (np)}(N)/e^{-2\pi\sqrt{2N}}$ (the dots) and our result
$Z_1^{(1)}(N)/e^{-2\pi\sqrt{2N}}$ with
\eqref{eq:alpha1} (the solid line).
Our Airy function expression of the instanton correction
\eqref{instAiry} is in excellent agreement with the exact data from
$N=1$ to $N=44$.
In the large $N$ limit, $Z_1^{(1)}(N)$ is
asymptotically expanded as
\begin{align}
Z_1^{(1)}(N)
&=e^{-2\pi\sqrt{2N}}
\biggl[2N
-\frac{13\pi^2-2}{2\sqrt{2}\pi}\sqrt{N}
+\frac{169\pi^4+116\pi^2+8}{32\pi^2}\notag\\
&\quad
-\frac{\pi(2197\pi^2+4012)}{384\sqrt{2N}}
+\frac{28561\pi^4+73424\pi^2-32768}{12288N}+\cO(N^{-3/2})\biggr].
\end{align}
The leading term $2Ne^{-2\pi\sqrt{2N}}$ agrees with the result in \cite{PY}.
Note that our result is exact at all orders in $1/N$.

Once we fix $\alpha_1^{(1)}$, $\beta_1^{(1)}$ and $\gamma_1^{(1)}$, we can
proceed to determining $\alpha_1^{(2)}$, $\beta_1^{(2)}$ and
$\gamma_1^{(2)}$.
The 2-instanton correction to the partition function is given by
\begin{align}
Z_1^{(2)}(N)
&=\biggl(2N^2
+\frac{2\pi^4\alpha_1^{(2)}C_1+3C_1+244}{8}N
\notag\\
&\qquad+\frac{32\pi^4\gamma_1^{(2)}+1}{128}C_1^2
+\frac{122\pi^4\alpha_1^{(2)}+119}{64}C_1
+\frac{3721}{32}\biggr)
\frac{\Ai[C_1^{-1/3}(N+8-B_1)]}{\Ai[C_1^{-1/3}(N-B_1)]}\notag\\
&\quad
-C_1^{2/3}\(N
+\frac{4\pi^4\beta_1^{(2)}+1}{16}C_1
+\frac{29}{8}\)
\frac{\Ai'[C_1^{-1/3}(N+8-B_1)]}{\Ai[C_1^{-1/3}(N-B_1)]}.
\label{eq:2-inst}
\end{align}
In the same way as above, from the numerical fitting we find
\begin{align}
\alpha_1^{(2)}=-\frac{26}{\pi^2},\qquad
\beta_1^{(2)}=-\frac{1}{4\pi^2},\qquad
\gamma_1^{(2)}=-\frac{9}{32\pi^2}+2~.
\label{eq:alpha2}
\end{align}
These are in about 14-digit agreement with the fitted values.
In figure~\ref{fig:Z1-inst}(b), we show the exact data
$(Z_1^{\rm (np)}(N)-Z_1^{(1)}(N))/e^{-4\pi\sqrt{2N}}$
and our result
$Z_1^{(2)}(N)/e^{-4\pi\sqrt{2N}}$ with \eqref{eq:alpha2}.

Repeating this method, we find the values of $\alpha_1^{(n)}$,
$\beta_1^{(n)}$ and $\gamma_1^{(n)}$ up to $n=4$.
Furthermore we can apply this procedure to other levels.
Our results at $k=1,2,3,4,6$ are summarized in \eqref{Jnp}.

Using \eqref{Jnp}, one can easily know the instanton corrections to
the partition function.
We have indeed checked that the instanton corrections to the partition
function obtained from \eqref{Jnp} precisely agree with the exact
data.
The results in \eqref{Jnp} show that the prefactors at $k=2,6$ are strongly
related to those at $k=1,3$, respectively.
Currently, we do not understand the physical origin of this relation.
It would be interesting to study what this means.

\section{Worldsheet Instantons}
In the previous section, we have obtained the exact non-perturbative
correction from the numerical study of the grand potential $J_k(\mu)$
and the partition function $Z_k(N)$.
The non-perturbative correction consists of several contributions
written as \eqref{eq:Jnp0}.
In this section, we would like to determine the prefactors of
worldsheet instanton corrections.
The prefactors of the D2-instanton corrections will be considered in
section \ref{sec:D2-inst}.
The prefactor $d_k^{(n)}$ of the worldsheet $n$-instanton is expected to
be independent of $\mu$.
We can determine $d_k^{(n)}$ 
from the topological string on local $\mathbb{F}_0$ (note that $\mathbb{F}_0\cong
\mathbb{P}^1\times \mathbb{P}^1$).

\subsection{Topological String on Local $\mathbb{F}_0$}\label{sec:WS-inst}
As discussed in \cite{Marino:2009jd,DMP1},
the ABJM matrix model \eqref{ABJMmatrix}
(and its generalization to the  ABJ model \cite{ABJ}) 
is related to the lens space matrix model \cite{Marino:2002fk,Aganagic:2002wv}\footnote{The ABJ(M) matrix model and the lens space matrix model are also closely related to the supermatrix model in \cite{DT}.} 
by an analytic continuation
of the rank of the gauge group, and it is further related to the
topological string on local $\mathbb{F}_0$ via a large $N$ duality.
Using this chain of dualities, we can compute the worldsheet instanton corrections 
in the ABJM theory from the knowledge of the topological string on local $\mathbb{F}_0$.

In general, the worldsheet instanton sum in the topological string has
the following structure \cite{Gopakumar:1998jq}
\begin{align}
F^{\mathrm{top}}=\sum_{n=1}^\infty\sum_{g,\vec{d}}
n^g_{\vec{d}}\left(2\sin\frac{n\lambda_s}{2}\right)^{2g-2}
\frac{e^{-n\vec{d}\cdot\vec{T}}}{n},
\label{GVexpansion}
\end{align}
where $\lambda_s$ denotes the string coupling of topological string
theory.

The free energy of ABJM matrix model and the free energy of
topological string on local $\mathbb{F}_0$ are related by a Fourier
transformation \cite{MP} due to the wavefunction nature of
the topological string partition functions \cite{Witten:1993ed}.
Since the canonical partition function $Z_k(N)$ is obtained from the
integral transform of grand partition function $e^{J_k(\mu)}$, it is
natural to identify the grand potential $J_k(\mu)$ with the free energy
$F^{\rm top}(T_1,T_2)$ of topological string on local $\mathbb{F}_0$.
Here $T_1$ and $T_2$ denote the K\"{a}hler parameters of the two
$\mathbb{P}^1$'s in $\mathbb{F}_0$. Since 
the ABJM slice is defined by $T_1=T_2$,
we arrive at the relation
\begin{align}
J_k(\mu)=F^{\rm top}(T_1=T_2\equiv T),
\label{JtoFtop}
\end{align}
with $T$ and $\mu$ related by \cite{MP}
\begin{align}
T=\frac{4\mu}{k}-\pi i~.
\label{Tvsmu}
\end{align}

To compare the topological string and ABJM theory, we should be
careful about the normalization of string coupling.
The string coupling $\lambda_s$ in the topological string theory and
the string coupling $g_s=\frac{2\pi i}{k}$ in the ABJM theory are
related by
\begin{align}
g_s^{2g-2}\til{F}_g^{\mathrm{top}}=\lambda_s^{2g-2}F_g^{\mathrm{top}},
\end{align}
where \cite{MP}
\begin{align}
\til{F}_g^{\mathrm{top}}=(-4)^{g-1}F_g^{\mathrm{top}},\quad(g\geq1).
\end{align} 
Therefore, $\lambda_s$ is given by
\begin{align}
\lambda_s=-2ig_s=\frac{4\pi}{k}~,
\end{align}
and the worldsheet instanton sum \eqref{GVexpansion} becomes
\begin{align}
F^{\mathrm{top}}=\sum_{n=1}^\infty\sum_{g,\vec{d}}
n^g_{\vec{d}}\left(2\sin\frac{2\pi n}{k}\right)^{2g-2}
\frac{e^{-n\vec{d}\cdot\vec{T}}}{n}.
\end{align}

The Gopakumar-Vafa invariants $n^g_{\vec{d}}$ of local $\mathbb{F}_0$
were computed  in \cite{Aganagic:2002qg}.
Up to total degree $d_1+d_2=3$, the GV invariants $n^g_{\vec{d}}$ are
given by
\begin{align}
n^0_{1,0}=n_{0,1}^0=-2,\quad
n^{0}_{1,1}=-4,\quad
n^0_{2,1}=n_{1,2}^0=-6,
\label{GVinv}
\end{align}
and $n^g_{\vec{d}}=0~(g\geq1)$ for the above cases.
Hence $F^{\mathrm{top}}$ of local $\mathbb{F}_0$ is expanded as
\begin{align}
F^{\mathrm{top}}=
-\sum_{n=1}^\infty\left[\frac{e^{-nT_1}+e^{-nT_2}}{2n\sin^2\frac{2\pi n}{k}}
+\frac{e^{-n(T_1+T_2)}}{n\sin^2\frac{2\pi n}{k}}
+\frac{3(e^{-n(2T_1+T_2)}+e^{-n(T_1+2T_2)})}{2n\sin^2\frac{2\pi n}{k}}
+\cdots\right].
\end{align}
Using the relations \eqref{JtoFtop} and \eqref{Tvsmu}, we can read off the coefficient $d_k^{(n)}$ of the
worldsheet $n$-instanton $e^{-\frac{4n\mu}{k}}$.
The first three coefficients are given by \eqref{eq:d_n}. 
More generally, we can write down the worldsheet instanton correction to
the grand potential in the form
\begin{align}
J^{\rm WS}_k(\mu)=\sum_{n,g,d=1}^\infty n_d^g\left(2\sin\frac{2\pi n}{k}\right)^{2g-2}\frac{(-1)^{dn}e^{-\frac{4dn\mu}{k}}}{n}~,
\end{align}
where $n_d^g$ denotes the weighted sum of GV invariants on local $\mathbb{F}_0$  \cite{Katz:1999xq}
\begin{align}
n_d^g=\sum_{d_1+d_2=d}n_{d_1,d_2}^g. 
\end{align}
For instance, using the data
$n_4^0=-48, ~n_4^1=9,~n_4^g=0\,(g\geq2)$ and \eqref{GVinv},
the 4-instanton coefficient $d_k^{(4)}$ is found to be
\begin{align}
d_k^{(4)}=9-\frac{12}{\sin^2\frac{2\pi}{k}}-\frac{1}{2\sin^2\frac{4\pi}{k}}
-\frac{1}{4\sin^2\frac{8\pi}{k}}~.
\label{d4}
\end{align}
In a similar manner, we can determine the higher instanton coefficients 
using the data of GV invariants in \cite{Aganagic:2002qg, Katz:1999xq}.

\subsection{Worldsheet Instanton in the 't Hooft Limit}
In the previous subsection we have determined the worldsheet instanton
correction $J_k^{\rm WS}(\mu)$ from the topological string on local
${\mathbb F}_0$.
As a consistency check, we consider the 't Hooft limit of the
worldsheet instanton corrections derived from the results in the
previous subsection and compare with the known results of the
genus-zero and genus-one free energy.
In the 't Hooft limit, the free energy has the genus expansion
\begin{align}
F(g_s,\lambda)\equiv\log Z(g_s, \lambda)
=\sum_{g=0}^\infty g_s^{2g-2} F_g(\lambda),
\end{align}
where $g_s=\frac{2\pi i}{k}$ is the string coupling and 
$\lambda\equiv\frac{N}{k}$ is the 't Hooft coupling.

One easily finds that the worldsheet instanton corrections to the
partition function up to 3-instanton are given by%
\footnote{We assume that the worldsheet instanton corrections up to
3-instanton are not mixed to the D2-instanton corrections.
This assumption is true for $k>6$, and the 't Hooft limit, of course,
satisfies this condition.}
\begin{align}
Z_k^{{\rm WS}(1)}(N)
&=d_k^{(1)}
\frac{\Ai[C_k^{-1/3}(N+\frac{4}{k}-B_k)]}{\Ai[C_k^{-1/3}(N-B_k)]},\nn
Z_k^{{\rm WS}(2)}(N)
&=\biggl(d_k^{(2)}+\frac{(d_k^{(1)})^2}{2}\biggr)
\frac{\Ai[C_k^{-1/3}(N+\frac{8}{k}-B_k)]}{\Ai[C_k^{-1/3}(N-B_k)]},\nn
Z_k^{{\rm WS}(3)}(N)
&=\biggl(d_k^{(3)}+d_k^{(2)} d_k^{(1)}+\frac{(d_k^{(1)})^3}{6}\biggr)
\frac{\Ai[C_k^{-1/3}(N+\frac{12}{k}-B_k)]}{\Ai[C_k^{-1/3}(N-B_k)]},
\label{eq:WS-3inst}
\end{align}
where $d_k^{(n)}$ $(n=1,2,3)$ are given by \eqref{eq:d_n}.
It is convenient to introduce $\hat{N}=N-\frac{k}{24}$ in the analysis
below.
In the large $\hat{N}$ limit,%
\footnote{Strictly, we consider the large $\hat{\lambda}=\hat{N}/k$
limit (strong coupling limit).}
$Z_k^{{\rm WS}(1)}(N)$ is 
asymptotically expanded as
\begin{align}
Z_k^{{\rm WS}(1)}(N)
=\frac{e^{-2\pi\sqrt{2\hat{N}/k}}}{\sin^2(2\pi/k)}
\biggl[ 1-\frac{5\sqrt{2}\pi}{3k^2} \sqrt{\frac{k}{\hat{N}}}
+\(\frac{25\pi^2}{9k^4}-\frac{1}{k^2}\)\frac{k}{\hat{N}}
+\cO\biggl(\biggl(\frac{k}{\hat{N}}\biggr)^{3/2}\biggr)\biggr].
\end{align}
If we further expand this in $g_s=\frac{2\pi i}{k}$, we obtain
\begin{align}
Z_k^{{\rm WS}(1)}(N) 
=e^{-2\pi\sqrt{2\hat{\lambda}}}
\biggl[-g_s^{-2}+\frac{1}{3}-\frac{5}{6\sqrt{2}\pi \sqrt{\hat{\lambda}}}
-\frac{1}{4\pi^2\hat{\lambda}}+\cO(g_s^2)\biggr],
\end{align}
where $\hat{\lambda}\equiv \frac{\hat{N}}{k}=\lambda-\frac{1}{24}$.
Thus the worldsheet 1-instanton correction to the free energy are
given by
\begin{align}
F_{g=0}^{{\rm WS}(1)}(\hat{\lambda})
=-e^{-2\pi\sqrt{2\hat{\lambda}}},\quad 
F_{g=1}^{{\rm WS}(1)}(\hat{\lambda})
=\biggl[\frac{1}{3}
-\frac{5}{6\sqrt{2}\pi\sqrt{\hat{\lambda}}}
-\frac{1}{4\pi^2\hat{\lambda}}\biggr]
e^{-2\pi\sqrt{2\hat{\lambda}}}.
\label{eq:F^(1-inst)}
\end{align}
In a similar way, we obtain the genus-zero and genus-one part of
the 2- and 3-instanton corrections from \eqref{eq:WS-3inst} as
\begin{align}
F_{g=0}^{{\rm WS}(2)}(\hat{\lambda})
&=\biggl[\frac{9}{8}
+\frac{1}{\pi\sqrt{2\hat{\lambda}}}\biggr]
e^{-4\pi\sqrt{2\hat{\lambda}}},\nn
F_{g=1}^{{\rm WS}(2)}(\hat{\lambda})
&=\biggl[-\frac{1}{2}
+\frac{83}{24\sqrt{2}\pi\sqrt{\hat{\lambda}}}
+\frac{91}{48\pi^2\hat{\lambda}}
+\frac{23}{24\sqrt{2}\pi^3\hat{\lambda}^{3/2}}
+\frac{1}{8\pi^4\hat{\lambda}^2}\biggr]
e^{-4\pi\sqrt{2\hat{\lambda}}}.
\label{eq:F^(2-inst)}
\end{align}
and
\begin{align}
F_{g=0}^{{\rm WS}(3)}(\hat{\lambda})
&=\biggl[-\frac{82}{27}
-\frac{9}{2\sqrt{2}\pi\sqrt{\hat{\lambda}}}
-\frac{1}{\pi^2\hat{\lambda}}
-\frac{1}{6\sqrt{2}\pi^3\hat{\lambda}^{3/2}}\biggr]
e^{-6\pi \sqrt{2\hat{\lambda}}},\nn
F_{g=1}^{{\rm WS}(3)}(\hat{\lambda})
&=\biggl[\frac{10}{9}
-\frac{1205}{54\sqrt{2}\pi\sqrt{\hat{\lambda}}}
-\frac{1145}{72\pi^2\hat{\lambda}}
-\frac{185}{16\sqrt{2}\pi^3\hat{\lambda}^{3/2}}
\notag \\ &\quad
-\frac{47}{16\pi^4\hat{\lambda}^2}
-\frac{47}{48\sqrt{2}\pi^5\hat{\lambda}^{5/2}}
-\frac{1}{12\pi^6\hat{\lambda}^3}
\biggr]e^{-6\pi \sqrt{2\hat{\lambda}}}~,
\label{eq:F^(3-inst)}
\end{align}
where we have used the relation between the instanton corrections to
the partition function and to the free energy:
\begin{align}
F_k^{{\rm WS}(1)}(N)
&=Z_k^{{\rm WS}(1)}(N),\quad
F_k^{{\rm WS}(2)}(N)
=Z_k^{{\rm WS}(2)}(N)-\frac{1}{2}\bigl(Z_k^{{\rm WS}(1)}(N)\bigr)^2,\nn
F_k^{{\rm WS}(3)}(N)
&=Z_k^{{\rm WS}(3)}(N)
-Z_k^{{\rm WS}(2)}(N)Z_k^{{\rm WS}(1)}(N)
+\frac{1}{3}\bigl(Z_k^{{\rm WS}(1)}(N)\bigr)^3.
\end{align}

\begin{figure}[tb]
\begin{center}
\begin{tabular}{cc}
\hspace{-3mm}
\resizebox{80mm}{!}{\includegraphics{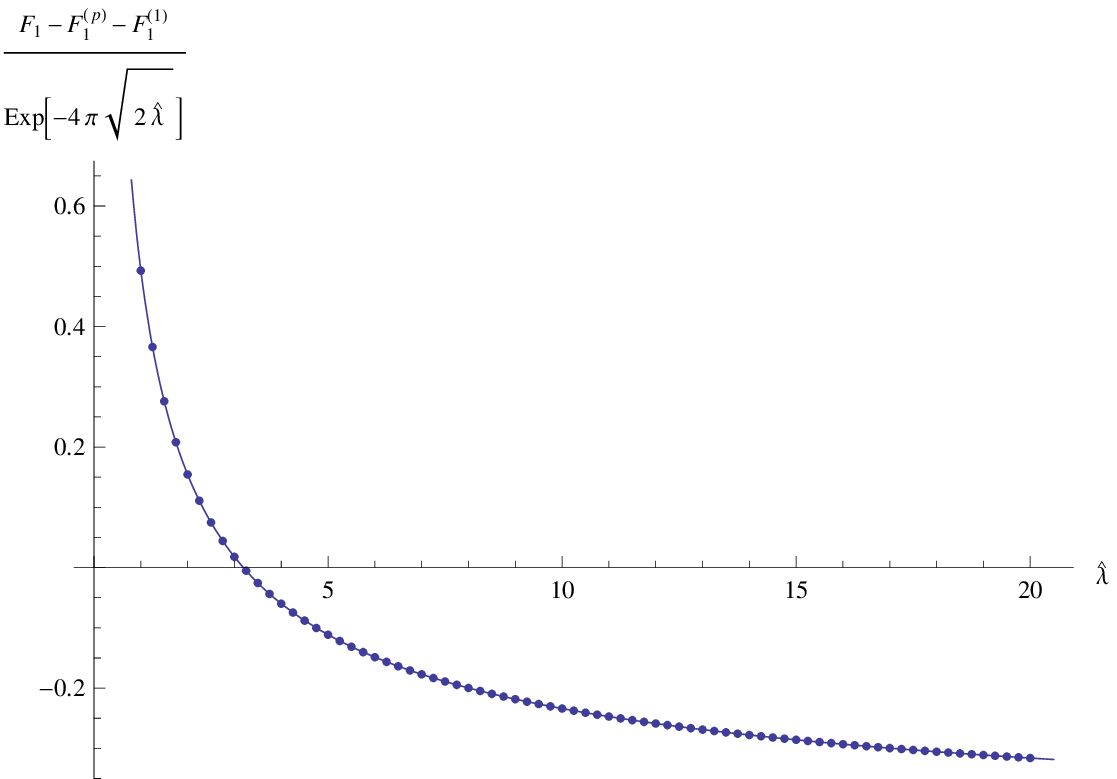}}
\hspace{-4mm}
&
\resizebox{80mm}{!}{\includegraphics{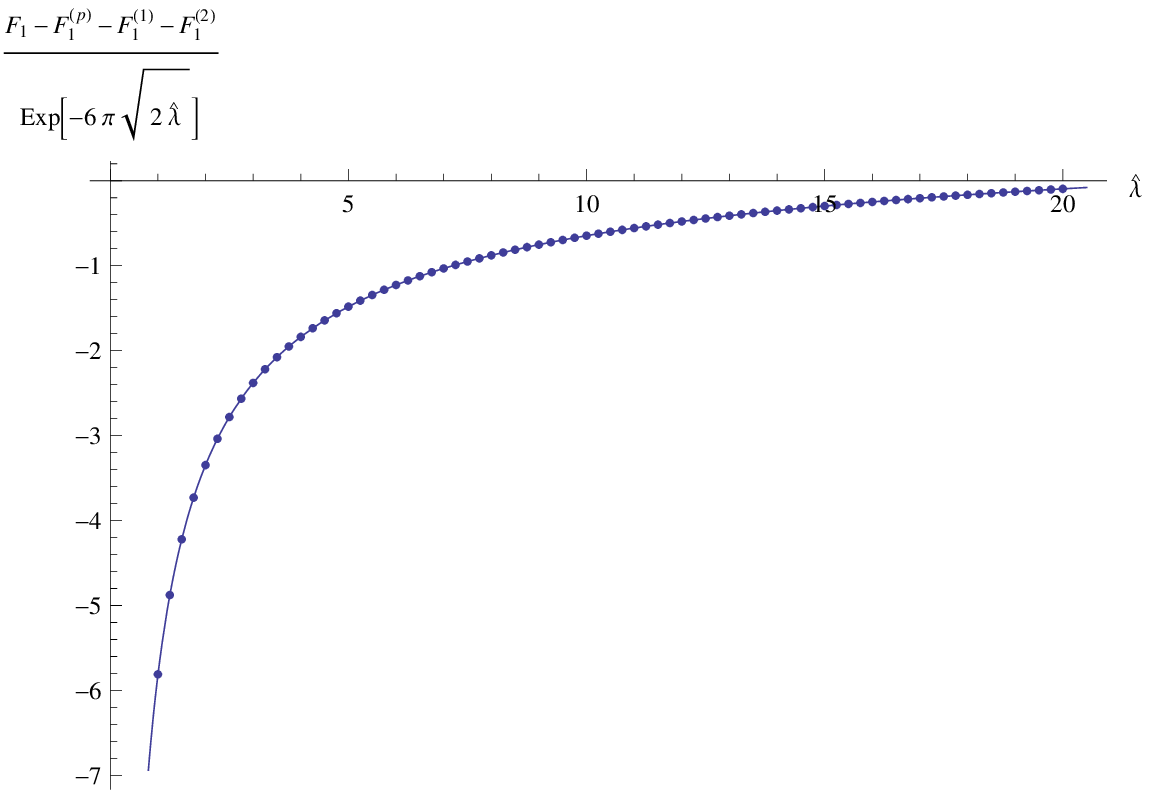}}
\hspace{-5mm}
\\ (a) & (b)
\vspace{-5mm}
\end{tabular}
\end{center}
  \caption{(a) The 2- and (b) 3-instanton corrections to the genus one free energy.}
  \label{fig:F1-inst}
\end{figure}

Let us compare these results with the result in \cite{DMP1}.
One immediately finds that the instanton corrections to the genus-zero
free energy exactly agree with the result in \cite{DMP1} (see also
\cite{KEK}).
Let us consider the genus-one free energy.
The genus-one free energy is given by
\begin{align}
F_{g=1}=-\log\eta(\tau),
\label{eq:F_g=1}
\end{align}
where $\eta(\tau)$ is the Dedekind eta function, and the modulus
$\tau$ is related to the 't Hooft coupling through
\begin{align}
\tau=i\frac{K'(\frac{i\kappa}{4})}{K(\frac{i\kappa}{4})},\quad
\lambda=\frac{\kappa}{8\pi}\,
{}_3F_2\(\frac{1}{2},\frac{1}{2},\frac{1}{2}
;1,\frac{3}{2};-\frac{\kappa^2}{16}\).
\end{align}
In the strong coupling limit, the genus-one free energy behaves as
\begin{align}
F_{g=1}=\frac{\pi\sqrt{2\hat{\lambda}}}{6}
-\frac{1}{2}\log(2\sqrt{2\hat{\lambda}})
+\biggl[\frac{1}{3}
-\frac{5}{6\sqrt{2}\pi\sqrt{\hat{\lambda}}}
-\frac{1}{4\pi^2\hat{\lambda}}\biggr]
e^{-2\pi \sqrt{2\hat{\lambda}}}
+\cO(e^{-4\pi \sqrt{2\hat{\lambda}}}).
\end{align}
The 1-instanton correction exactly agrees with \eqref{eq:F^(1-inst)}.
It becomes messy to compute the higher instanton corrections.
Instead, we numerically evaluate \eqref{eq:F_g=1}, and extract the 2-
and 3-instanton corrections from it.
In figure~\ref{fig:F1-inst}(a), we plot
$(F_{g=1}-F_{g=1}^\text{(pert)}-F_{g=1}^{{\rm WS}(1)})/e^{-4\pi\sqrt{2\hat{\lambda}}}$
against $\hat{\lambda}$.
The dots represent the numerical data obtained from \eqref{eq:F_g=1}
while the solid line represents our prediction \eqref{eq:F^(2-inst)}.
Our results show a perfect agreement with the numerical evaluation.
Similarly, in figure~\ref{fig:F1-inst}(b), we plot
$(F_{g=1}-F_{g=1}^\text{(pert)}-F_{g=1}^{{\rm WS}(1)}
-F_{g=1}^{{\rm WS}(2)})/e^{-6\pi\sqrt{2\hat{\lambda}}}$
against $\hat{\lambda}$.
Our 3-instanton prediction \eqref{eq:F^(3-inst)} again agrees with
the numerical ones.

\section{Conjecture of D2-instanton Corrections}\label{sec:D2-inst}
The coefficients $d_k^{(n)}$ of the worldsheet $n$-instanton corrections diverge
when $k\in 2n/\mathbb{N}$.
However, we confirmed numerically that the grand potential itself is
finite at these values of $k$. 
Furthermore, at $k=2n/m$ ($m=1,2,\dots$), the worldsheet $n$-instanton correction
is just the same order as the D2 $m$-instanton correction because $e^{-4n\mu/k}=e^{-2m\mu}$ .
Therefore it is natural to expect that D2-instanton corrections are
also divergent at these values of $k$, and the divergence coming from
the worldsheet instanton is canceled by the divergence from
D2-instanton.
The sum of these two contributions leads to a finite result.
This story imposes a strong constraint on the pole structure of the D2-instanton corrections.

\subsection{D2 1-instanton correction}
Let us consider the $e^{-2\mu}$ term in $J^{\rm (np)}_k(\mu)$.
At $k=2n$ with $n\in\mathbb{Z}$, the $e^{-2\mu}$ term receives two
contributions: one from the D2 1-instanton and the other from
worldsheet $n$-instanton.
The coefficient of worldsheet $n$-instanton is divergent in
the limit $k\rightarrow 2n$.
This divergence has the general structure
\begin{align}
\lim_{k\rightarrow 2n}d_k^{(n)}e^{-\frac{4n\mu}{k}}
=(-1)^{n-1}\left[\frac{4n}{\pi^2(k-2n)^2}
+\frac{4(\mu+1)}{\pi^2(k-2n)}
+\frac{2\mu^2+2\mu+1}{n\pi^2}+w^{(n)}
\right]e^{-2\mu},
\label{WSknpole}
\end{align}
with some constant $w^{(n)}$.
We have checked this behavior up to $n=7$ using the Gopakumar-Vafa 
invariant of local $\mathbb{F}_0$ in \cite{Aganagic:2002qg} and found
\begin{align}
&w^{(1)}=\frac{1}{3},\quad
w^{(2)}=\frac{7}{6},\quad
w^{(3)}=\frac{37}{9},\quad
w^{(4)}=\frac{187}{12},\quad
w^{(5)}=\frac{661}{15}+12\sqrt{5},\nn
&w^{(6)}=\frac{6919}{18},\quad
w^{(7)}=\frac{19783-16380\cos\frac{\pi}{7}
+29903\cos\frac{2\pi}{7}
-22428\cos\frac{3\pi}{7}
}{42\sin^2\frac{\pi}{7}}.
\end{align}

Next, let us consider the D2 1-instanton correction 
\begin{align}
J^{\rm D2(1)}_k=\Big[a_k^{(1)}\mu^2+b_k^{(1)}\mu+c_k^{(1)}\Big]e^{-2\mu}~.
\end{align}
The small $k$ expansion of the coefficients $a_k^{(1)}, b_k^{(1)}$ and $c_k^{(1)}$
was computed in \cite{MP} up to
$k^3$:
\begin{align}
a_k^{(1)}&=
-\frac{4}{\pi^2 k}+\frac{k}{2}
-\frac{\pi^2k^3}{96}+\cdots,\nn
b_k^{(1)}&=
\frac{4}{\pi^2 k}-\frac{5k}{6}
+\frac{67\pi^2k^3}{1440}
+\cdots,\nn
c_k^{(1)}&=
\frac{4}{\pi^2 k}-\frac{2}{3k}
+\frac{\pi^2-1}{12}k-\frac{\pi^2(104+5\pi^2)}{2880}k^3+\cdots.
\end{align}
From our results at $k=1,3$, we expect that the coefficients
of $e^{-2\mu}$ vanish for odd $k\in\mathbb{Z}$.%
\footnote{For odd $k$, it is plausible that there are no D2-branes wrapping $\mathbb{RP}^3$ as in the case of $k=1$ discussed in \cite{PY}. 
Thus it is expected that the D2-instanton correction starts from $\cO(e^{-2\pi\sqrt{2kN}})$. 
Further, we have numerically checked that this expectation is
true at $k=5,7$ by using the TBA-like equations in \cite{HMO}.}
Also, in order to cancel the poles coming from the worldsheet instanton 
\eqref{WSknpole}, the D2 1-instanton coefficients should behave as
\begin{align}
\lim_{k\rightarrow 2n}a_k^{(1)}&={\cal O}(1),\nn
\lim_{k\rightarrow 2n}b_k^{(1)}&=-\frac{4(-1)^{n-1}}{\pi^2(k-2n)}+{\cal O}(1),\nn
\lim_{k\rightarrow 2n}c_k^{(1)}&=(-1)^{n-1}\left[
-\frac{4n}{\pi^2(k-2n)^2}-\frac{4}{\pi^2(k-2n)}
\right]+{\cal O}(1).
\end{align}
From the above conditions, we can fix the forms of
$a_k^{(1)},b_k^{(1)}$ and $c_k^{(1)}$.
The result is given by \eqref{eq:a1b1c1}.
Summing the contributions from the worldsheet $n$-instanton and
the D2 1-instanton, the $e^{-2\mu}$ term at $k=2n$ gives the finite answer
\begin{align}
J^{{\rm D2}(1)}_{2n}(\mu)+J^{{\rm WS}(n)}_{2n}(\mu)
=(-1)^{n-1}\left[\frac{4\mu^2+2\mu+1}{n\pi^2}+s^{(n)}
\right]e^{-2\mu},
\label{eq:D2+WS}
\end{align}
where
\begin{align}
s^{(n)}=w^{(n)}+\frac{1}{3n}-\frac{2n}{3}.
\end{align}
For lower $n$, we find
\begin{align}
&s^{(1)}=0,\quad
s^{(2)}=0,\quad
s^{(3)}=\frac{20}{9},\quad
s^{(4)}=13,
\end{align}
and \eqref{eq:D2+WS} for $n=1,2,3$ correctly reproduce our results  at
$k=2,4,6$ in \eqref{Jnp}!

\begin{figure}[tb]
\begin{center}
\begin{tabular}{ccc}
\resizebox{75mm}{!}{\includegraphics{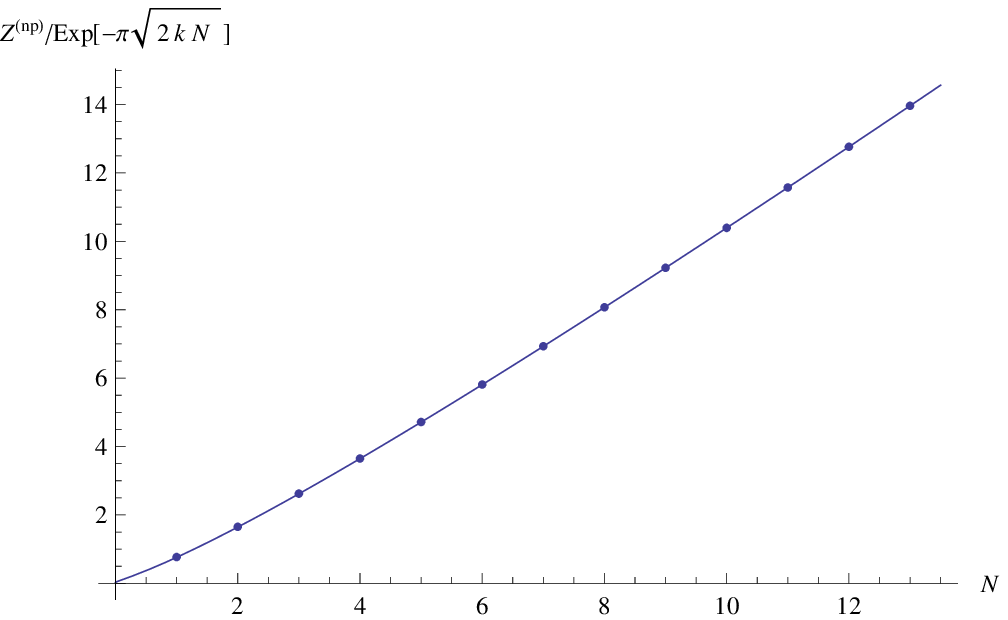}}
&
\resizebox{75mm}{!}{\includegraphics{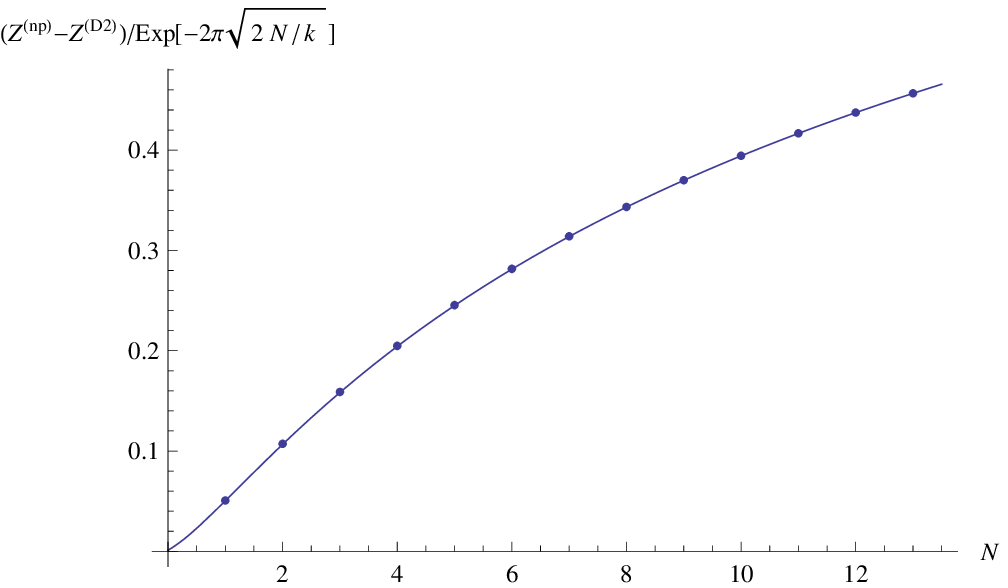}}
\\ (a) & (b)
\vspace{0.5cm}
\end{tabular}
\end{center}
 \caption{In these figures, we plot (a) the leading and (b) the next-to-leading contributions of the instanton corrections at $k=\frac{3}{2}$.
 The dots represent the numerical values from the TBA-like equations while the solid lines represent our theoretical predictions.
 The leading correction is captured by the D2 1-instanton, and the next-to-leading correction by the worldsheet 1-instanton.} 
\label{fig:k=3/2}
\end{figure}

As a further check of our conjecture \eqref{eq:a1b1c1}, we compare the instanton corrections to the partition function
at various $k$ with the numerical results obtained by using the TBA-like equations.
As explained in \cite{HMO} (see also \cite{PY}), the ABJM partition
function can be also computed by solving the TBA-type integral equations.%
\footnote{One advantage to use the TBA-like equations
is that we can numerically compute the partition function at any $k$. However, we cannot neglect numerical errors as $N$ grows. }
In figure~\ref{fig:k=3/2}, for instance, we show the instanton corrections at $k=\frac{3}{2}$.
At $k=\frac{3}{2}$, the leading contribution is the D2 1-instanton correction.
In figure~\ref{fig:k=3/2}(a), the dots represent $Z_{3/2}^{\rm (np)}(N)/e^{-\pi\sqrt{2k N}}$ computed by the TBA-like equations
while the solid line represents the D2 1-instanton correction $Z^{\rm D2}_{3/2}(N)/e^{-\pi\sqrt{2k N}}$ given by \eqref{eq:Zinst_D2} 
with our conjecture \eqref{eq:a1b1c1}.
In figure~\ref{fig:k=3/2}(b), we plot the next-to-leading contribution.
At this order, the worldsheet 1-instanton correction is dominant.
The dots represent $\bigl(Z_{3/2}^{\rm (np)}(N)-Z^{\rm D2}_{3/2}(N)\bigr)/e^{-2\pi\sqrt{2N/k}}$ computed by the TBA-like equations
(and \eqref{eq:Zinst_D2})
while the solid line represents the worldsheet 1-instanton correction $Z^{\rm WS}_{3/2}(N)/e^{-2\pi\sqrt{2N/k}}$ given by \eqref{eq:Zinst_ws}.
In both figures, our theoretical predictions \eqref{eq:Zinst_D2} and \eqref{eq:Zinst_ws} perfectly explain the numerical results from TBA.
We have checked the validity of our conjecture \eqref{eq:a1b1c1} for other $k$'s in the same way.

\subsection{Comment on D2 2-instanton correction}
By the similar reasoning, it seems to be possible to guess the forms of D2
2-instanton coefficients $a_k^{(2)},b_k^{(2)}$ and $c_k^{(2)}$.
However, the constraint is much weaker than the 1-instanton case,
hence we can not determine the forms of 2-instanton coefficients
completely.
We need more information to fix them.
 
The $k\rightarrow n$ limit of worldsheet $n$-instanton has the general
structure
\begin{align}
\lim_{k\rightarrow n}d_k^{(n)}e^{-\frac{4n\mu}{k}}
=-(4+5(-1)^n)\left[\frac{n}{4\pi^2(k-n)^2}+\frac{2\mu+1}{2\pi^2(k-n)}
+\frac{2\mu^2+\mu+\frac{1}{4}}{n\pi^2}+v^{(n)}\right]e^{-4\mu}.
\end{align}
We have checked this behavior up to $n=7$ and found
\begin{align}
&v^{(1)}=\frac{1}{3},\quad
v^{(2)}=\frac{1}{18},\quad
v^{(3)}=\frac{37}{9},\quad
v^{(4)}=\frac{13}{36},\quad
v^{(5)}=\frac{661}{15}-12\sqrt{5},\nn
&v^{(6)}=\frac{193}{54},\quad
v^{(7)}=\frac{19783-22428\cos\frac{\pi}{7}+16380\cos\frac{2\pi}{7}
-29903\cos\frac{3\pi}{7}}{42\sin^2\frac{2\pi}{7}}.
\end{align}

Naively, these poles should be canceled by the D2 2-instanton.
At this order, however, there might be another contribution that we will discuss in the next section,
i.e., the contribution from a kind of bound states of the D2-instanton and the worldsheet instanton.
At present, we cannot conclude whether such a contribution exists or not.
If exists, we need to consider its pole structure. 
We leave this problem as a future work.

\section{Discussions}
In this paper, by generalizing our previous method 
we continued the exact computation of the
partition function in the ABJM theory using the Fermi gas formalism.
Using these exact values of the partition functions,
we further studied the instanton effects.
We have found that the grand potential contains an interesting
oscillatory behavior and it can be explained by the periodicity of
the grand partition function.
The worldsheet instanton effects are understood from the results in the
topological string on local ${\mathbb F}_0$.
After subtracting the worldsheet instanton, 
we studied the D2-instanton effects,
only little of which has been known so far.
We proposed an analytic expression of
the D2 1-instanton correction which is consistent with all of the known
properties so far.
We would like to discuss some open questions in our analysis.

In our previous work \cite{HMO}, we have found that the partition
functions $Z_1(N)$ at $k=1$ are written as a polynomial of $\frac{1}{\pi}$
with rational coefficients.
Here we have studied the partition functions for few other $k$'s.
We find that at $k=2$ the partition function are polynomials of
$\frac{1}{\pi^2}$, at $k=3$, polynomials of $\frac{1}{\pi}$ and $\frac{1}{\sqrt{3}}$, at
$k=4$, polynomials of $\frac{1}{\pi}$ and at $k=6$, polynomials of
$\frac{1}{\sqrt{3}\pi}$.
We note in passing a curious observation about the zeros of
these polynomials.
For example, 
if we replace $\frac{1}{\pi}$ by a variable $x$
in the partition function $Z_1(N)$ at $k=1$, 
it becomes a polynomial of $x$ of degree $[\frac{N}{2}]$.
If we order the zeros of this polynomial by their absolute values,
the first few zeros are very close to $\frac{1}{\pi},-\frac{3}{\pi},\frac{5}{\pi}$, and so on.
We observed a similar pattern of zeros for other $k$'s.
We would like to understand the general structure of $Z_k(N)$ better.

We have determined the analytic expression of the leading D2-instanton
correction.
We would like to study this effect from the worldvolume theory of
the Euclidean D2-brane wrapping on $\mathbb{RP}^3$.
We expect that the prefactor of D2-instanton correction
is coming from the 1-loop determinant of the fluctuation around
the classical D2-brane action.
To analyze such effects,
the recent result of reproducing the 1-loop log term from the gravitational
computation in \cite{BGMS} might be a helpful guide.

Although the worldsheet instanton coefficients $d_k^{(n)}$ obtained
from the topological string correctly reproduce the 't Hooft limit of
free energy studied in \cite{DMP1}, we find some discrepancy between
the prediction of $d_k^{(n)}$ and our results \eqref{Jnp} obtained by
the numerical analysis.
For example, let us take a closer look at the $k=3,6$ cases in
\eqref{Jnp}.
The coefficients of the first two terms in $J_3^{\rm (np)}$ and
$J_6^{\rm (np)}$ agree with the value of $d_k^{(1)}$ and $d_k^{(2)}$
in \eqref{eq:d_n}.
However, the fourth term of $J_3^{\rm (np)}$ and $J_6^{\rm (np)}$ is
different from $d_3^{(4)}=d_6^{(4)}=-8$ predicted from the topological
string \eqref{d4}.
Also, for the $k=4$ case, the third term of $J_4^{\rm (np)}$ is different from the
prediction $d_4^{(3)}=\frac{10}{3}$, while the first term agrees with
$d_4^{(1)}=1$.
We believe that the predictions of worldsheet instanton corrections
from the topological string \eqref{eq:d_n} and \eqref{d4} are correct at
generic $k$, and we conjecture that the reason of this discrepancy is
due to a kind of bound states of the D2-instanton and the
worldsheet instanton.
This conjecture is based on the observation that after a few
worldsheet instanton terms which agree with $d_k^{(n)}$, the
discrepancy arises just after the first D2-instanton term appeared in
$J_k^{\rm (np)}$.
We leave more detailed study of the origin of this discrepancy as a
future problem.

Recently, a very interesting relation between the sphere partition
functions and the non-perturbative completion of topological string
partition functions is proposed in \cite{Lockhart:2012vp}.
This proposal is reminiscent of the relation between the canonical and
the grand canonical partition functions \eqref{XitoZ}. As discussed in
\cite{MP}, it is
natural to identify the grand partition function $\Xi(\mu)$ of ABJM
theory with the non-perturbative completion of the topological string
partition function on local $\mathbb{F}_0$.
It would be interesting to study this relation further.

\vskip5mm
\centerline{\bf Acknowledgements}
\vskip3mm
\noindent
We are grateful to Edna Cheung, Yang-Hui He, Shinji Hirano, Hiroaki
Kanno, Nakwoo Kim, Gregory Korchemsky, Kimyeong Lee, Tomoki Nosaka,
Jaemo Park, Zhao Peng, Ryu Sasaki, Rak-Kyeong Seong, Masaki Shigemori,
Fumihiko Sugino, Shigeki Sugimoto, Masahito Yamazaki for useful
discussions.
The work of Y.H. is supported in part by the JSPS Research Fellowship
for Young Scientists, while the work of K.O. is supported in part by
JSPS Grant-in-Aid for Young Scientists (B) \#23740178.

\appendix

\section{Proof of \eqref{Hankelidentity} for General Hankel Matrices}

In this appendix, we will prove the identity \eqref{Hankelidentity}
assuming that a Hankel matrix has the following integral
representation
\begin{align}
H_{m,n}=\frac{1}{4\pi}\int_{-1}^1dt\,w(t)\,t^{m+n}~.
\end{align}
We also assume that the weight $w(t)$ is a positive even function of $t$
on the interval $(-1,1)$
\begin{align}
w(-t)=w(t)>0,\quad t\in(-1,1)~.
\end{align}
By the change of variable
\begin{align}
t=t_q\equiv\tanh\frac{q}{2},
\end{align}
the matrix element $H_{m,n}$ is rewritten as
\begin{align}
H_{m,n}=\int_{-\infty}^\infty\frac{dq}{2\pi}\rho_m(q)\rho_n(q),\quad
\rho_n(q)=\frac{\sqrt{w(t_q)}}{\sqrt{2}\cosh\frac{q}{2}}t_q^n~,
\end{align}
and the trace becomes
\begin{align}
\Tr H^\ell=\int_{-\infty}^\infty\prod_{i=1}^\ell \frac{dq_i}{2\pi}
\rho(q_1,q_2)\cdots\rho(q_\ell,q_1)=\Tr\rho^\ell,
\end{align}
where
\begin{align}
\rho(q,q')=\sum_{n=0}^\infty\rho_n(q)\rho_n(q')=
\frac{\sqrt{w(t_q)w(t_{q'})}}{2\cosh\frac{q-q'}{2}}~.
\end{align}

Let us prove the identity \eqref{Hankelidentity} for this class of matrix
\begin{align}
\frac{\det(1+zH_-)}{\det(1-zH_+)}=H(z),
\label{Hgeneralid}
\end{align}
where $H(z)$ is defined by
\begin{align}
H(z)=\sum_{m=0}^\infty (1-zH_+)^{-1}_{0,m}=\bra e_0|\frac{1}{1-zH_+}|v\ket~.
\label{Hzgeneral}
\end{align}
with $|e_0\ket$ and $|v\ket$ given in \eqref{evdef}.
\subsection{Proof of the Even Part}
First we consider the even part of \eqref{Hgeneralid}
\begin{align}
\frac{\det(1-z^2H_-^2)}{\det(1-z^2H_+^2)}=
H(z)H(-z)~.
\label{Hid-even}
\end{align}
As discussed in our previous paper \cite{HMO}, using the relation
$H_-^2=H_+(1-|e_0\ket\bra e_0|)H_+$ the left hand side
of \eqref{Hid-even}
is written as
\begin{align}
\frac{\det(1-z^2H_-^2)}{\det(1-z^2H_+^2)}=
\bra e_0|\frac{1}{1-z^2H_+^2}|e_0\ket~.
\end{align}

By applying the lemma of Tracy-Widom to the kernel
\begin{align}
\rho_+(q,q')=\frac{\rho(q,q')+\rho(q,-q')}{2}=\frac{E(q)E(q')}{\cosh q+\cosh q'},\quad
E(q)=\sqrt{w(t_q)}\cosh\frac{q}{2}~,
\end{align}
we find
\begin{align}
\rho_+^{2\ell+1}(q,q')=\frac{E(q)E(q')}{\cosh q+\cosh q'}
\sum_{k=0}^{2\ell}(-1)^k\phi^k(q)\phi^{2\ell-k}(q')~.
\label{TWid}
\end{align}
The left hand side of \eqref{TWid} is rewritten as
\begin{align}
\rho_+^{2\ell+1}(q,q')&=\int\frac{dq_1}{2\pi}\cdots\int\frac{dq_{2\ell}}{2\pi}
\rho_+(q,q_1)\cdots\rho_+(q_{2\ell},q')\nn
&=\sum_{m=0}^\infty\cdots\sum_{n=0}^\infty
\int\frac{dq_1}{2\pi}\cdots\int\frac{dq_{2\ell}}{2\pi}\rho_{2m}(q)\rho_{2m}(q_1)\cdots
\rho_{2n}(q_{2\ell})\rho_{2n}(q')\nn
&=\sum_{m,n=0}^\infty\rho_{2m}(q)(H_+^{2\ell})_{m,n}\rho_{2n}(q')~,
\end{align}
and $\phi^k(q)$ in \eqref{TWid} is defined recursively
\begin{align}
\phi^k(q)=\frac{1}{E(q)}\int\frac{dq'}{2\pi}\rho_+(q,q')E(q')\phi^{k-1}(q')~.
\end{align}
If we expand $\phi^k(q)$ in terms of $\rho_{2n}(q)$
\begin{align}
\phi^k(q)=\frac{\sqrt{2}}{E(q)}\sum_{n=0}^\infty \rho_{2n}(q)\alpha_n^{(k)}
=\frac{1}{\cosh^2\frac{q}{2}}\sum_{n=0}^\infty t_q^{2n}\alpha_n^{(k)}~,
\label{phivsal}
\end{align}
then $\alpha_n^{(k)}$ satisfies
\begin{align}
\alpha_n^{(k)}=\sum_{m=0}^\infty (H_+)_{n,m}\alpha_m^{(k-1)}~.
\label{alrecur}
\end{align}
The initial condition $\phi^0(q)=1$ corresponds to
\begin{align}
\alpha_m^{(0)}=1\quad(\forall m)~.
\end{align}
In other words, $\alpha_m^{(0)}=v_m$.
From \eqref{phivsal} and \eqref{alrecur}, $\phi^k(q)$ is found to be
\begin{align}
\phi^k(q)=\frac{1}{\cosh^2\frac{q}{2}}\sum_{n,m=0}^\infty t_q^{2n}(H_+^k)_{n,m}
~.
\label{phikq}
\end{align}
Setting $q=0$ in \eqref{phikq}, we find
\begin{align}
\phi^k(0)=\sum_{m=0}^\infty (H_+^k)_{0,m}=\bra e_0|H_+^k|v\ket~.
\end{align}
This implies that $H(z)$ in \eqref{Hzgeneral} can also be written as
\begin{align}
H(z)=\sum_{n=0}^\infty \phi^n(0)z^n~.
\end{align}

Finally, taking the limit $q,q'\rightarrow0$ in \eqref{TWid} 
we find
\begin{align}
(H_+^{2\ell})_{0,0}=\bra e_0|H_+^{2\ell}|e_0\ket
=\sum_{k=0}^{2\ell}(-1)^k\phi^k(0)\phi^{2\ell-k}(0)~.
\end{align}
This is the desired relation \eqref{Hid-even}.

\subsection{Proof of the Odd Part}
Let us prove the odd part of \eqref{Hzgeneral}.
The kernel $\rho(q,q')$ can be written as a matrix element in
the ``auxiliary'' quantum mechanical system
\begin{align}
\rho(q,q')=\bra q|\sqrt{w(t_q)}c_p\sqrt{w(t_q)}|q'\ket,\quad
[q,p]=2\pi i~.
\end{align}
In this subsection, we use the notation
\begin{align}
c_q=\frac{1}{2\cosh\frac{q}{2}},\quad
s_q=\frac{1}{2\sinh\frac{q}{2}},\quad
t_q=\tanh\frac{q}{2}~.
\end{align}

Let us consider the following function
\begin{align}
\Xi_1(z)&=\det(1+zR_q\Pi c_qc_p),\qquad R_q=w(t_q)c_q^{-1}~.
\end{align}
Using the commutation relation found in our previous paper \cite{HMO},
\begin{align}
c_pc_q=-is_q\Pi s_p,\quad \Pi=1-|0_p\ket\bra 0_q|,
\end{align}
we can see that $\Xi_1(z)$ is an even function of $z$, as follows.
We first rewrite $\Xi_1(z)$ as
\begin{align}
\Xi_1(z)&=\det(1+izR_q\Pi s_p\Pi^\dag s_q)\nn
&=\det(1+izR_qs_q\Pi s_p\Pi^\dag )\nn
&=\det(1-zR_qc_pc_q\Pi^\dag )~.
\end{align}
Since $\Xi_1(z)$ is a real function for $z\in\mathbb{R}$, we
find
\begin{align}
\Xi_1(z)=\Xi_1(z)^\dag
=\det(1-z(R_qc_pc_q\Pi^\dag)^\dag )=\Xi_1(-z)~.
\end{align}

On the other hand, $\Xi_1(z)$ can also be written as
\begin{align}
\Xi_1(z)&=\det(1+zw(t_q)c_p)\cdot\bra 0_q|R_q^{-1}\frac{1}{1+zw(t_q)c_p}R_q|0_p\ket\nn
&=\det(1+z\rho)\cdot\bra 0_q|E_q^{-1}\frac{1}{1+z\rho}E_q|0_p\ket~,
\label{Xi-one}
\end{align}
where
\begin{align}
E_q=\sqrt{w(t_q)}\cosh\frac{q}{2},\quad
\rho=\sqrt{w(t_q)}c_p\sqrt{w(t_q)}~.
\end{align}
One can easily see that the second factor in the last line of \eqref{Xi-one} is
\begin{align}
\bra 0_q|E_q^{-1}\frac{1}{1+z\rho}E_q|0_p\ket=\sum_{n=0}^\infty 
\phi^n(0)(-z)^n=H(-z)~.
\end{align}
Then $\Xi_1(-z)=\Xi_1(z)$ implies the following relation
\begin{align}
\frac{H(z)}{H(-z)}=\frac{\det(1+zH_+)\det(1+zH_-)}{\det(1-zH_+)\det(1-zH_-)}~.
\end{align}
This completes the proof of the odd part of \eqref{Hgeneralid}.

\section{Exact Values of $Z_k(N)$}\label{sec:exactvalue}
Here let us summarize the exact values of $Z_k(N)$ ($k=1,2,3,4,6$). 
Our main task is to solve the integral equation \eqref{eq:int_eq_phi}.
Here we use the method in \cite{PY}. This method allows us to rewrite the convolution in  \eqref{eq:int_eq_phi} as the sum of the residues.

\subsection{Recursive Relations at Even $k$}
Let us first consider the even $k$ case.
Introducing the variables $u=e^{q/k}$ and $v=e^{q'/k}$,
then the integral equation \eqref{eq:int_eq_phi} becomes
\begin{align}
\varphi^\l(u)=\frac{1}{2\pi} \frac{u}{u+1} \int_0^\infty dv \frac{v^{(k-2)/2}(v+1)}{(u+v)(v^k+1)} \varphi^{\l-1}(v),
\label{eq:int_eq_varphi_k}
\end{align}
where $\varphi^\l(u)=\phi^\l(q)$.
Note that $v^{(k-2)/2}$ is rational when $k$ is even.
From the solutions for lower $\l$, 
we observe that the solution $\varphi^\l(u)$ have the following form
\begin{align}
\varphi^\l(u)=\sum_{j=0}^\l A_\l^{(j)}(u) \log^j u,
\end{align}
where $A_\l^{(j)}(u)$ is a rational function.
We assume that $A_\l^{(j)}(u)$ has the poles at the roots of $(u+1)(u^k+1)=0$.
We can rewrite the equation \eqref{eq:int_eq_varphi_k} by using the
technique in \cite{PY} as follows,
\begin{align}
\varphi^{\l+1}(u)&=\frac{1}{2\pi}\frac{u}{u+1}
\int_0^\infty dv \frac{v^{k/2-1}(v+1)}{(u+v)(v^k+1)} \sum_{j=0}^\l
A_\l^{(j)}(v) \log^j v \notag \\
&=-\frac{1}{2\pi}\frac{u}{u+1} \sum_{j=0}^\l \frac{(2\pi i)^j}{j+1}
\oint_\gamma dv \frac{v^{k/2-1}(v+1)}{(u+v)(v^k+1)}
A_\l^{(j)}(v) B_{j+1}\( \frac{\log v}{2\pi i} \) \notag \\
&=-\frac{1}{2\pi}\frac{u}{u+1}
\sum_{j=0}^\l \frac{(2\pi i)^{j+1}}{j+1}
\sum_{\text{all poles}} \Res_v \frac{v^{(k-2)/2}(v+1)}{(u+v)(v^k+1)}
A_\l^{(j)}(v) B_{j+1}\( \frac{\log v}{2\pi i} \) \notag \\
&=-\frac{(-u)^{k/2}(u-1)}{2\pi(u+1)(u^k+1)}
\sum_{j=0}^\l \frac{(2\pi i)^{j+1}}{j+1} A_\l ^{(j)}(-u)
B_{j+1}\( \frac{\log u}{2\pi i}+\frac{1}{2}\) +f_\l(u),
 \label{eq:varphi^l+1} 
\end{align}
where $B_n(x)$ is the Bernoulli polynomial, and $f_\l(u)$ is a
rational function.
In the second line in \eqref{eq:varphi^l+1}, we have used the integral
formula noted in \cite{PY},
\begin{align}
\int_0^\infty dv\, C(v) \log^j v = -\frac{(2\pi i)^j}{j+1}
\oint_\gamma dv\, C(v) B_{j+1}\( \frac{\log v}{2\pi i} \),
\end{align}
where we choose the branch cut of $\log$ as the positive real axis,
and integration contour $\gamma$ is chosen as the closed path from
$+\infty$ to $0$ infinitesimally below the cut and then from $0$ to
$+\infty$ infinitesimally above the cut.
Using the identity:
\begin{align}
B_n(x+a)=\sum_{p=0}^n \binom{n}{p}B_{n-p}(a) x^p,
\end{align}
the first term of the right hand side in \eqref{eq:varphi^l+1} becomes
\begin{align}
&-\frac{(-u)^{k/2}(u-1)}{2\pi(u+1)(u^k+1)}
\sum_{j=0}^\l \sum_{p=0}^{j+1} \frac{(2\pi i)^{j+1}}{j+1}\binom{j+1}{p} B_{j+1-p}\( \frac{1}{2} \) A_\l^{(j)}(-u) 
\(\frac{\log u}{2\pi i}\)^p \notag \\
&=-\frac{(-u)^{k/2}(u-1)}{2\pi(u+1)(u^k+1)}
\sum_{p=0}^\l \sum_{j=0}^{p+1} \frac{(2\pi i)^{p+1}}{p+1}\binom{p+1}{j} B_{p+1-j}\( \frac{1}{2} \) A_\l^{(p)}(-u) 
\(\frac{\log u}{2\pi i}\)^j \notag \\
&=-\frac{(-u)^{k/2}(u-1)}{2\pi(u+1)(u^k+1)}
\sum_{j=0}^{\l+1} \sum_{\stackrel{\scriptstyle p=j-1}{\scriptstyle p\geq 0}}^{\l} \frac{(2\pi i)^{p+1-j}}{p+1}\binom{p+1}{j} B_{p+1-j}\( \frac{1}{2} \) A_\l^{(p)}(-u) \log^j u .
\end{align}
We note that $B_{p+1-j}(1/2)$ vanishes if $p+1-j$ is odd.
Thus setting $p+1-j=2m$, we obtain 
\begin{align*}
-\frac{(-u)^{k/2}(u-1)}{(u+1)(u^k+1)}
\sum_{j=0}^{\l+1} \sum_{\stackrel{\scriptstyle m=0}{\scriptstyle j+2m-1\geq 0}}^{[\frac{\l-j+1}{2}]} \frac{(-1)^m (2\pi)^{2m-1}}{j+2m}
\binom{j+2m}{j} B_{2m}\( \frac{1}{2} \) A_\l^{(j+2m-1)}(-u) \log^j u .
\end{align*}
Comparing the coefficients of $\log^j u$, we obtain the recursive
relations for $A_\l^{(j)}(u)$.

For $1 \leq j \leq \l+1$, we find
\begin{align}
A_{\l+1}^{(j)}(u)=-\frac{(-u)^{k/2}(u-1)}{(u+1)(u^k+1)} \sum_{m=0}^{[\frac{\l-j+1}{2}]} \frac{(-1)^m (2\pi)^{2m-1}}{j+2m}
\binom{j+2m}{j} B_{2m}\( \frac{1}{2} \) A_\l^{(j+2m-1)}(-u),
\end{align}
while for $j=0$ we find
\begin{align}
A_{\l+1}^{(0)}(u)=-\frac{(-u)^{k/2}(u-1)}{(u+1)(u^k+1)} \sum_{m=1}^{[\frac{\l+1}{2}]} \frac{(-1)^m (2\pi)^{2m-1}}{2m}
B_{2m}\( \frac{1}{2} \) A_\l^{(2m-1)}(-u)+f_\l(u),
\end{align}
where the rational function $f_\l(u)$ is given by
\begin{align}
f_\l(u)=-\frac{u}{2\pi(u+1)} \sum_{j=0}^\l \frac{(2\pi i)^{j+1}}{j+1} \sum_{v \ne -u} \Res_v \frac{v^{k/2-1}(v+1)}{(u+v)(v^k+1)}A_\l^{(j)}(v)
B_{j+1}\( \frac{\log v}{2\pi i} \) .
\end{align}
where $\sum_{v \ne -u}$ means the sum over all the poles except for $v=-u$.
It is convenient to redefine
\begin{align}
A_\l^{(j)}(u)=\frac{1}{(2\pi)^j j!} \( \frac{u^{k/2}}{u^k+1} \)^j \hat{A}_\l^{(j)}(u).
\end{align}
Then we can rewrite the recursive relations as
\begin{align}
\hat{A}_{\l+1}^{(j)}(u)&=(-1)^{kj/2+1} \frac{u-1}{u+1}
\sum_{m=0}^{[\frac{\l-j+1}{2}]} c_m 
\(\frac{u^{k/2}}{u^k+1} \)^{2m} \hat{A}_\l^{(j+2m-1)}(-u),
\quad (1\leq j \leq \l+1), \label{eq:rec_hatA_k}\\
\hat{A}_{\l+1}^{(0)}(u)&=-\frac{u-1}{u+1} 
\sum_{m=1}^{[\frac{\l+1}{2}]} c_m 
\(\frac{u^{k/2}}{u^k+1} \)^{2m} \hat{A}_\l^{(2m-1)}(-u)+f_\l(u),
\end{align}
where
\begin{align}
c_m&=\frac{(-1)^m}{(2m)!}B_{2m}\( \frac{1}{2} \)= \frac{(-1)^m}{(2m)!} \( \frac{1}{2^{2m-1}}-1\)B_{2m} ,\\
f_\l(u)&=-\frac{u}{u+1} \sum_{j=0}^\l \frac{i^{j+1}}{(j+1)!} \sum_{v \ne -u} \Res_v \frac{v+1}{v(u+v)} \( \frac{v^{k/2}}{v^k+1} \)^{j+1}
\hat{A}_\l^{(j)}(v) B_{j+1}\( \frac{\log v}{2\pi i} \). \label{eq:f_l(u)}
\end{align}
The initial condition is trivially given by
\begin{align}
\hat{A}_0^{(0)}(u)=1. \label{eq:hatA_0^0}
\end{align}
The equations \eqref{eq:rec_hatA_k}-\eqref{eq:hatA_0^0} are the
recursive relations to be solved.

Once these equations are solved, we obtain $\varphi^\l(u)$, and thus 
can compute $\phi^\l(0)=\varphi^\l(1)$ and $\Tr \rho_+^{2n}$.
The matrix element $\rho_+^{2n}(u_1,u_2)$ is given by
\begin{align}
\rho_+^{2n}(u_1,u_2)=F_k(u_1,u_2)
\sum_{\l=0}^{2n-1} (-1)^\l \varphi^{\l}(u_1) \varphi^{2n-1-\l}(u_2),
\end{align}
where
\begin{align}
F_k(u_1,u_2)=\frac{(u_1 u_2)^{\frac{k}{4}+\frac{1}{2}}(u_1+1)(u_2+1)}
{2k(u_1-u_2)(u_1 u_2-1)(u_1^k+1)^{1/2}(u_2^k+1)^{1/2}}.
\end{align}
The trace of $\rho_+^{2n}$ is then computed by
\begin{align}
\Tr \rho_+^{2n}=\int \frac{dq}{2\pi} \rho_+^{2n}(q,q)
=\frac{k}{2\pi}\int_0^\infty du \frac{\rho_+^{2n}(u,u)}{u}.
\end{align}
It is easy to see that $\rho_+^{2n}(u,u)/u$ has the following form
\begin{align}
\frac{\rho_+^{2n}(u,u)}{u}=\sum_{j=0}^{2n-1} R_{2n}^{(j)}(u) \log^j u .
\end{align}
where $R_{2n}^{(j)}(u)$ is a rational function.
From the computation for low $n$'s, we observe that $R_{2n}^{(j)}(u)$
has the poles at the roots of $u^{2k}-1=0$.
We assume that this observation is valid for arbitrary $n$.
Let $u_m$ be the root of $u^{2k}-1=0$,
\begin{align}
u_m=\cos \( \frac{\pi m}{k} \)+i \sin\( \frac{\pi m}{k}\),
\qquad (m=0,\pm 1,\pm 2, \dots).
\end{align}
Then,  $\Tr \rho_+^{2n}$ is computed as
\begin{align}
\Tr \rho_+^{2n}&=\frac{k}{2\pi} \int_0^\infty du \sum_{j=0}^{2n-1} R_{2n}^{(j)}(u) \log^j u \notag \\
&= -\frac{k}{2\pi} \sum_{j=0}^{2n-1} \frac{(2\pi i)^{j+1}}{j+1} \sum_{m=0}^{2k-1} \Res_{u=u_m} R_{2n}^{(j)}(u) B_{j+1}\( \frac{\log u}{2\pi i} \).
\label{eq:Trrho_+^2n-evenk}
\end{align} 

\subsection{Recursive Relations at Odd $k$}
Next let us consider the odd $k$ case.
Changing the variables to $t=e^{q/2k}$ and $s=e^{q'/2k}$, we obtain
\begin{align}
\psi^\l(t)=\frac{1}{\pi} \frac{t^2}{t^2+1} \int_0^\infty ds \frac{s^{k-1}(s^2+1)}{(t^2+s^2)(s^{2k}+1)} \psi^{\l-1}(s),
\label{eq:int_eq_varphi_k0}
\end{align}
where $\psi^\l(t)=\phi^\l(q)$.
We observe that the solution takes the form
\begin{align}
\psi^\l(t)=\sum_{j=0}^{[\frac{\l}{2}]} A_\l^{(j)}(t) \log^j t.
\end{align}
where a rational function $A_{\l}^{(j)}(t)$ has poles at the roots of $t^{4k}-1=0$.
Let $\nu_m$ be the roots of $t^{4k}-1=0$,
\begin{align}
\nu_m=\cos\(\frac{m\pi}{2k}\)+i\sin\(\frac{m\pi}{2k}\),\quad
(m=0\pm1,\pm2,\dots).
\end{align}
In the same computation in the previous subsection, we find the recursive relations for $A_\l^{(j)}(t)$,
\begin{align}
A_{\l+1}^{(j)}(t)&=(-1)^{\frac{k+1}{2}}\frac{t^k(t^2-1)}{(t^2+1)(t^{2k}-1)}
\sum_{p=j-1}^{[\frac{\l}{2}]} \frac{(2\pi i)^{p-j}}{p+1} 
\binom{p+1}{j} B_{p+1-j}\(\frac{1}{4}\)\notag \\
&\quad\times \bigl[A_\l^{(p)}(it)+(-1)^{p-j}A_\l^{(p)}(-it) \bigr],
\qquad (1\leq j \leq [\l/2]+1),
\label{eq:rec_A_oddk}\\
A_{\l+1}^{(0)}(t)&=(-1)^{\frac{k+1}{2}}\frac{t^k(t^2-1)}{(t^2+1)(t^{2k}-1)}
\sum_{p=0}^{[\frac{\l}{2}]} \frac{(2\pi i)^{p}}{p+1}B_{p+1}\(\frac{1}{4}\)
\bigl[A_\l^{(p)}(it)+(-1)^{p}A_\l^{(p)}(-it) \bigr]+f_\l(t) \notag,
\end{align}
where
\begin{align}
f_\l(t)=-\frac{1}{\pi} \frac{t^2}{t^2+1} \sum_{j=0}^{[\frac{\l}{2}]}
\frac{(2\pi i)^{j+1}}{j+1}
\sum_{m=0}^{4k-1} \Res_{s=\nu_m} 
\frac{s^{k-1}(s^2+1)}{(t^2+s^2)(s^{2k}+1)}A_\l^{(j)}(s)B_{j+1}
\(\frac{\log s}{2\pi i}\).
\label{eq:f_oddk}
\end{align}
The matrix element $\rho_+^{2n}(t_1,t_2)$ is given by
\begin{align}
\rho_+^{2n}(t_1,t_2)=\tilde{F}_k(t_1,t_2)
\sum_{\l=0}^{2n-1}(-1)^\l \psi^\l(t_1) \psi^{2n-1-\l}(t_2),
\end{align}
where
\begin{align}
\tilde{F}_k(t_1,t_2)
=\frac{(t_1t_2)^{k/2+1}(t_1^2+1)(t_2^2+1)}
{2k(t_1^2-t_2^2)(t_1^2t_2^2-1)(t_1^{2k}+1)^{1/2}(t_2^{2k}+1)^{1/2}}.
\end{align}
We further assume that $\rho_+^{2n}(t_1,t_2)/t$ takes the form
\begin{align}
\frac{\rho_+^{2n}(t,t)}{t}=\sum_{j=0}^{n-1} R_{2n}^{(j)}(t) \log^j t,
\end{align}
where a rational function $R_{2n}^{(j)}(t)$ has poles at the roots of
$t^{4k}-1=0$.
Thus $\Tr \rho_+^{2n}$ is computed by
\begin{align}
\Tr \rho_+^{2n}&=\frac{k}{\pi}\int_0^\infty dt
\sum_{j=0}^{n-1} R_{2n}^{(j)}(t) \log^j t\notag \\
&=-\frac{k}{\pi}\sum_{j=0}^{n-1} \frac{(2\pi i)^{j+1}}{j+1}
\sum_{m=0}^{4k-1} \Res_{t=\nu_m} R_{2n}^{(j)}(t)
B_{j+1}\(\frac{\log t}{2\pi i} \).
\label{eq:Trrho_+^2n-oddk}
\end{align}

Let us summarize our procedure here.
For even $k$, we solve the recursive relations
\eqref{eq:rec_hatA_k}-\eqref{eq:hatA_0^0}, then compute
$\Tr\rho_+^{2n}$ by \eqref{eq:Trrho_+^2n-evenk}.
For odd $k$, the recursive relations to be solved are
\eqref{eq:rec_A_oddk}-\eqref{eq:f_oddk}.
We can compute $\Tr\rho_+^{2n}$ by \eqref{eq:Trrho_+^2n-oddk}.
These equations do not contain any integrals, and we can simply solve
by using {\tt Mathematica}.
We have actually solved these equations, and have computed $Z_k(N)$ up
to $N=44,20,18,16,14$ at $k=1,2,3,4,6$ respectively.
These values are summarized in figures~\ref{fig:Z_1-1}-\ref{fig:Z_6}.

\begin{figure}[htb]
\begin{center}
\rotatebox{-90}{
\resizebox{20.5cm}{!}{
\vbox{
\begin{align*}
Z_1(1)&=1/4,\qquad
Z_1(2)=1/(16 \pi),\qquad
Z_1(3)=(-3 + \pi)/(64 \pi),\qquad
Z_1(4)=(10 - \pi^2)/(1024 \pi^2),\qquad
Z_1(5)=(26 + 20 \pi - 9 \pi^2)/(4096 \pi^2),\qquad
Z_1(6)=(78 - 121 \pi^2 + 36 \pi^3)/(147456 \pi^3),\qquad
Z_1(7)=(-126 + 174 \pi + 193 \pi^2 - 75 \pi^3)/(196608 \pi^3)
\nn
Z_1(8)&=(876 - 4148 \pi^2 - 2016 \pi^3 + 1053 \pi^4)/(18874368 \pi^4),\qquad
Z_1(9)=(4140 + 8880 \pi - 15348 \pi^2 - 13480 \pi^3 + 5517 \pi^4)/(75497472 \pi^4),\qquad
Z_1(10)=(16860 - 136700 \pi^2 + 190800 \pi^3 + 207413 \pi^4 - 
 81000 \pi^5)/(7549747200 \pi^5)
 \nn
Z_1(11)&=(-122580 + 381900 \pi + 837300 \pi^2 - 1289300 \pi^3 - 1091439 \pi^4 + 
 447525 \pi^5)/(30198988800 \pi^5),\qquad
Z_1(12)=(626760 - 8856300 \pi^2 - 18446400 \pi^3 + 35287138 \pi^4 + 30204000 \pi^5 - 
 12504375 \pi^6)/(4348654387200 \pi^6)
 \nn
Z_1(13)&=(1563480 + 6714000 \pi - 17252100 \pi^2 - 40746000 \pi^3 + 49141894 \pi^4 + 
 45780780 \pi^5 - 18083925 \pi^6)/(5798205849600 \pi^6)
 \nn
Z_1(14)&=(21382200 - 421152060 \pi^2 + 
  1918350000 \pi^3 + 2614227910 \pi^4 - 5654854800 \pi^5 - 3965159223 \pi^6 + 
  1732468500 \pi^7)/(3409345039564800 \pi^7)
\nn
Z_1(15)&=(-222059880 + 1271579400 \pi + 
  3613033620 \pi^2 - 12266517900 \pi^3 - 17757814914 \pi^4 + 
  28941378130 \pi^5 + 21727092861 \pi^6 - 9162734175 \pi^7)/(13637380158259200 \pi^7)
\nn
Z_1(16)&=(288454320 - 8196414240 \pi^2 - 54540622080 \pi^3 + 83379537976 \pi^4 + 
   337956998400 \pi^5 - 310977507352 \pi^6 - 354450849984 \pi^7 + 
   132764935275 \pi^8)/(872792330128588800 \pi^8)
\nn
Z_1(17)&=(3171011760 + 23555952000 \pi - 71723746080 \pi^2 - 333199608000 \pi^3 + 
   542885550648 \pi^4 + 1355261623520 \pi^5 - 1384280129304 \pi^6 - 
   1337978574000 \pi^7 + 518021476875 \pi^8)/(3491169320514355200 \pi^8)
\nn
Z_1(18)&=(4970745360 - 180631896480 \pi^2 + 2270514395520 \pi^3 + 
   2444801550408 \pi^4 - 18251132155200 \pi^5 - 13590443330584 \pi^6 + 
   35949047139936 \pi^7 + 20671882502409 \pi^8 - 
   9607077219600 \pi^9)/(377046286615550361600 \pi^9)
\nn
Z_1(19)&=(-2636096400 + 24895105200 \pi + 79219113120 \pi^2 - 487774106400 \pi^3 - 
   852843285000 \pi^4 + 3053792290360 \pi^5 + 3630439618136 \pi^6 - 
   6122444513560 \pi^7 - 4288974330849 \pi^8 + 
   1840384320075 \pi^9)/(55858709128229683200 \pi^9)
\nn
Z_1(20)&=(361238119200 - 17220213738000 \pi^2 - 311199827904000 \pi^3 + 
   324620205099120 \pi^4 + 4438412804448000 \pi^5 - 
   2865149087601400 \pi^6 - 16634573387107200 \pi^7 + 
   10309541088246858 \pi^8 + 14094700187198400 \pi^9 
   \nn &\hspace{0.5cm}- 
   4979889726328125 \pi^{10})/(603274058584880578560000 \pi^{10})
\nn
Z_1(21)&=(5559930784800 + 65609800200000 \pi - 214704822906000 \pi^2 - 
   1706798293200000 \pi^3 + 3122732798020080 \pi^4 + 
   15272968981284000 \pi^5 - 19948322354095800 \pi^6 - 
   51815653479308000 \pi^7 + 47747674792873242 \pi^8
   \nn &\hspace{0.5cm} + 
   47396132831956500 \pi^9 - 
   18101156239333125 \pi^{10})/(2413096234339522314240000 \pi^{10})
\nn
Z_1(22)&=(25844898895200 - 1500424524522000 \pi^2 + 46684893872232000 \pi^3 + 
   35164010074651920 \pi^4 - 740786057283888000 \pi^5 - 
   402614116951888600 \pi^6 + 3793710085975101600 \pi^7 + 
   2103302735093392558 \pi^8 - 6256890344362442400 \pi^9
\nn 
   &\hspace{0.5cm} - 
   3210299116242733125 \pi^{10}+ 1551114253328062500 \pi^{11})/(1167938577420328800092160000 \pi^{11})
\nn
Z_1(23)&=(-496645968218400 + 7222685259636000 \pi + 23988125937006000 \pi^2 - 
   241443239284170000 \pi^3 - 452965450178712240 \pi^4 + 
   2930864234437062000 \pi^5 + 4031113133564569800 \pi^6 - 
   14930904869365243000 \pi^7 - 15786905528345919066 \pi^8 
   \nn 
   &\hspace{0.5cm}+ 
   27222797625410117610 \pi^9 + 18233204850331964775 \pi^{10} - 
   7915669863827158125 \pi^{11})/(4671754309681315200368640000 \pi^{11})
\nn
Z_1(24)&=(1204104891528000 - 86634289079553600 \pi^2 - 3819319790489280000 \pi^3 + 
   2602976790752830800 \pi^4 + 100433912061555072000 \pi^5 - 
   40665871963026060960 \pi^6 - 833015249980407648000 \pi^7 + 
   330789435742955662220 \pi^8 
   \nn 
   &\hspace{0.5cm}+ 2433643974077824569600 \pi^9 -1163340151816826557764 \pi^{10} - 1809616125406822207200 \pi^{11} + 
   614736070878601546875 \pi^{12})/(1345465241188218777706168320000 \pi^{12})
\nn
Z_1(25)&=(2787296269665600 + 49358959705440000 \pi - 164816144642318400 \pi^2 - 
   2065776940458000000 \pi^3 + 3923291222794334160 \pi^4 + 
   32721154659114768000 \pi^5 - 46361197897271310240 \pi^6 - 
   236251940186723032000 \pi^7 
   \nn
   &\hspace{0.5cm} + 268046321968484065644 \pi^8 + 
   718649496677642926000 \pi^9  - 615263237145329762916 \pi^{10} - 
   623636559509281432200 \pi^{11} + 
   235728636221685481875 \pi^{12})/(597984551639208345647185920000 \pi^{12})
\nn
Z_1(26)&=(112416665053492800 - 9539770920889809600 \pi^2 + 
   679050246669809414400 \pi^3 + 342466609375043833680 \pi^4 - 
   17962043874947027856000 \pi^5 - 6538704303611999307360 \pi^6 + 
   175693381923498780781440 \pi^7 + 67906500509898555099452 \pi^8 
   \nn
   &\hspace{0.5cm} - 
   732942357598839937348800 \pi^9 - 340632835309262972353644 \pi^{10} + 
   1094527855640549701449936 \pi^{11} + 521717090330960815763475 \pi^{12} - 
   258856873289859609765000 \pi^{13})/(3638138012172943574917479137280000 \
\pi^{13})
\nn
Z_1(27)&=(-2837783495662315200 + 60576452126406216000 \pi + 
   201778747825504449600 \pi^2 - 3107566895728375080000 \pi^3 - 
   5911562189814162401520 \pi^4 + 62328030245503696674000 \pi^5 + 
   89382763214933595881760 \pi^6
   \nn
   &\hspace{0.5cm} - 596840165742039299652000 \pi^7 - 
   710285728070533874332788 \pi^8 + 2688501047332900049538860 \pi^9 + 
   2631613640469642289362564 \pi^{10} - 4607029942628752806862500 \pi^{11} - 
   2990942157677468764775625 \pi^{12} 
   \nn &\hspace{0.5cm} + 
   1309268426599376039671875 \
\pi^{13})/(14552552048691774299669916549120000 \pi^{13})
\nn
Z_1(28)&=(12969482004169891200 - 1313678612062041825600 \pi^2 - 
   130700139399240559948800 \pi^3 + 57505552398883116047520 \pi^4 + 
   5584407193839421531584000 \pi^5 - 1387144249864003881410160 \pi^6 - 
   82911547437043021895278080 \pi^7  
   \nn
   &\hspace{0.5cm}+ 19393348493721346914697928 \pi^8 + 
   522086937010309473445843200 \pi^9 - 
   149459897799711476890866044 \pi^{10} - 
   1306120034852713622360250432 \pi^{11} + 
   517277319128577543305082654 \pi^{12} + 
   887470472209191527843877600 \pi^{13}
   \nn
   &\hspace{0.5cm} - 
   292941866060855176428388125 \
\pi^{14})/(11409200806174351050941214574510080000 \pi^{14})
\nn
Z_1(29)&=(356597653882854384000 + 9094131200774299680000 \pi - 
   30038899805342890862400 \pi^2 - 561876995802308266080000 \pi^3 + 
   1062648884610900631044000 \pi^4 + 13932441993654741781224000 \pi^5 - 
   20002293314753598671462640 \pi^6
   \nn
   &\hspace{0.5cm} - 174087341108384502887088000 \pi^7 + 
   208625406705598959725168520 \pi^8 + 1102194201726340597528654000 \pi^9 - 
   1127922427503232413314046876 \pi^{10} - 
   3106694796352524493370982000 \pi^{11} + 
   2512129123834667330105351214 \pi^{12} 
   \nn
   &\hspace{0.5cm} + 
   2590737342046361982559266300 \pi^{13} - 
   971173742724208860654943125 \
\pi^{14})/(45636803224697404203764858298040320000 \pi^{14})
\nn
Z_1(30)&=(1516562211091623696000 - 177090468308691897960000 \pi^2 + 
   27136431922196521884960000 \pi^3 + 9017847591500109108146400 \pi^4 - 
   1077113128060105913256480000 \pi^5 - 
   256681199098892348426646000 \pi^6 + 
   17131486119850520142660936000 \pi^7
   \nn
   &\hspace{0.5cm} + 
   4335785790236401386882916920 \pi^8 - 133790555452597949153456208000 \pi^9 - 
   42194568890414867791467229900 \pi^{10} + 
   497101282096648260408501524400 \pi^{11} + 
   205605092600586074416500476442 \pi^{12} - 
   697543537772745100897507429200 \pi^{13} 
   \nn
   &\hspace{0.5cm}- 315952706311600959133249618125 \pi^{14} + 
   159786318876990035803260187500 \
\pi^{15})/(41073122902227663783388372468236288000000 \pi^{15})
\nn
Z_1(31)&=(-5491025370413030448000 + 166004632983882092400000 \pi + 
   541029306660436634040000 \pi^2 - 12171024630433182313800000 \pi^3 - 
   22742123903644972954327200 \pi^4 + 365896598220831803825700000 \pi^5 + 
   521165244159524214764514000 \pi^6 
   \nn
   &\hspace{0.5cm} - 
   5649100224613410554739750000 \pi^7 -6885817029748362284443224360 \pi^8 + 
   46819207040545598624314349000 \pi^9 + 
   50706666062879865301800728100 \pi^{10} - 
   194513286041291875665429285500 \pi^{11} - 
   180463482287010623903232340206 \pi^{12}
   \nn
   &\hspace{0.5cm} + 
   319234142124859370913055662750 \pi^{13} + 
   202614683135946238835538564375 \pi^{14} - 89233886735707696636857328125 \
\pi^{15})/(18254721289878961681505943319216128000000 \pi^{15})
\nn
Z_1(32)&=(26378192029689051552000 - 3580049562649923242880000 \pi^2 - 
   757132674251549925288960000 \pi^3 + 
   214985863491932279351404800 \pi^4 + 
   48450137344329251746575360000 \pi^5 - 
   7380031300608320195089248000 \pi^6
   \nn
   &\hspace{0.5cm} - 
   1141207573627181061863237376000 \pi^7 + 
   155907850426836134859071679840 \pi^8 +12598650074482753279364061696000 \pi^9 - 
   2023063868341111494879077115200 \pi^{10} - 
   66659187395884098831853323686400 \pi^{11}
   \nn
   &\hspace{0.5cm} + 
   14997054793962534623552524193424 \pi^{12} + 
   149677247547917349534167645606400 \pi^{13} - 
   51289784809585816523329448010000 \pi^{14} - 
   95114261347756370696738538000000 \pi^{15}
   \nn
   &\hspace{0.5cm} + 30707052223734699797941597265625 \pi^{16})/(21029438925940563857094846703736979456000000 \pi^{16})
\nn
Z_1(33)&=(938391146991441535392000 + 33397176671939336256000000 \pi - 
   106969904782051711720320000 \pi^2 - 
   2869098805616518579104000000 \pi^3 + 
   5271786469822631334556780800 \pi^4 + 
   102890141125717999328795520000 \pi^5
   \nn
   &\hspace{0.5cm} - 
   144488682885133549916265312000 \pi^6 - 
   1981982524489386054179544000000 \pi^7 +2355690689320882868185315724640 \pi^8 + 
   21374876708906309003768152800000 \pi^9 - 
   22562204399707743197488161172800 \pi^{10}
   \nn
   &\hspace{0.5cm} - 
   123272970849585219046190213680000 \pi^{11} + 
   116183975669541868107557485762704 \pi^{12} + 
   327775305305306832840663235099200 \pi^{13} - 
   252942132305492678291871933104400 \pi^{14} - 264752932078174137625582781940000 \pi^{15}
   \nn
   &\hspace{0.5cm} + 
   98556910328894483095041052265625 \
\pi^{16})/(84117755703762255428379386814947917824000000 \pi^{16})
\end{align*}
}}
}
\end{center}
\caption{The exact values of $Z_1(N)$ up to $N=44$.}
\label{fig:Z_1-1}
\end{figure}

\begin{figure}[htb]
\begin{center}
\rotatebox{-90}{
\resizebox{20cm}{!}{
\vbox{
\begin{align*}
Z_1(34)&=(3748468635840087038880000 - 576507161916513586991232000 \pi^2 + 
   181107877834055659744857600000 \pi^3 + 
   39495907538962450446622752000 \pi^4 - 
   10032226168135957514354545920000 \pi^5 - 
   1562219208018840347512324588800 \pi^6 
   \nn
   &\hspace{0.5cm} + 
   234599266467263376168123553920000 \pi^7 + 
   38603239231337392309701577173600 \pi^8 - 
   2932684274647031133516687271872000 \pi^9 - 
   600271248924227540020652100904640 \pi^{10} + 
   20159955677466022545588253272480000 \pi^{11}
   \nn
   &\hspace{0.5cm}  + 
   5573654587796424883190789264494160 \pi^{12} - 
   69709092298931690593778835499708800 \pi^{13} - 
   26569384887189154463277158157743664 \pi^{14} + 
   93832677823825548373924023227006400 \pi^{15} + 
   40956483693699951777807871375445625 \pi^{16}
   \nn
   &\hspace{0.5cm}  - 
   21010234864640717487636413068500000 \
\pi^{17})/(97240125593549167275206571158079793004544000000 \pi^{17})
\nn
Z_1(35)&=(-155394189350732306598624000 + 6470406921313139884279200000 \pi + 
   20298820779309274250944896000 \pi^2 - 
   644237716951136200599907200000 \pi^3 - 
   1159426487470456162673933817600 \pi^4 + 
   27184334880173756913462875040000 \pi^5
   \nn
   &\hspace{0.5cm} + 
   37441712501857940589355673030400 \pi^6 - 
   615920597406312013511953377120000 \pi^7 - 
   737394576503418581858954301538080 \pi^8 + 
   8083939286151916595206710203292000 \pi^9 + 
   8891745399143918867683322258656320 \pi^{10}
   \nn &\hspace{0.5cm} - 
   60913223839020959285920727787464000 \pi^{11} - 
   61938023356886006851232342745704688 \pi^{12} + 
   239331994588192630847481782926880400 \pi^{13} + 
   213684392930403426579039783004190352 \pi^{14} - 
   380703883988580770014135383951190800 \pi^{15}
   \nn
   &\hspace{0.5cm}  - 
   237579775604713372691749420127176875 \pi^{16} + 
   105110725506673319593388326238765625 \
\pi^{17})/(388960502374196669100826284632319172018176000000 \pi^{17})
\nn
Z_1(36)&=(205449951732963783797952000 - 36007191928401346014826464000 \pi^2 - 
   15429603717166905589305894912000 \pi^3 + 
   2840446363912624200459178252800 \pi^4 + 
   1397363237706856558307600471040000 \pi^5 - 
   131357413180004913764646329145600 \pi^6
   \nn
   &\hspace{0.5cm} - 
   48296798398041665685367034661580800 \pi^7 + 
   3883644257708774705265766583576640 \pi^8 + 
   831449949125205134950635749923584000 \pi^9 - 
   74966355578150879844550753874386080 \pi^{10} - 
   7594849087488363684192700143463050240 \pi^{11}
   \nn &\hspace{0.5cm} + 
   921494334533293208733930116271590944 \pi^{12} + 
   35556136643862328691978571274984281600 \pi^{13} - 
   6631878927263733277152930079345687248 \pi^{14} - 
   73649263521729777257048019164426375424 \pi^{15} + 
   22472441260989832462435020189734075550 \pi^{16}
   \nn &\hspace{0.5cm} + 
   44419572428849001959102055192929040000 \pi^{17} - 
   14087391698607910119184937538270796875 \
\pi^{18})/(168030937025652961051556954961161882311852032000000 \pi^{18})
\nn
Z_1(37)&=(344658385327622963268672000 + 16696275491606487560784000000 \pi - 
   51162197245362439004504736000 \pi^2 - 
   1908396918432031940168256000000 \pi^3 + 
   3353443144313022700702847884800 \pi^4 + 
   93646325534078783973438771840000 \pi^5
   \nn
   &\hspace{0.5cm} - 
   126013527688531282433233414214400 \pi^6 - 
   2584302296129015527191445038720000 \pi^7 + 
   2947512886694682727712026276981440 \pi^8 + 
   42496143753135725779909660878000000 \pi^9 - 
   43492244872629368821077100733533920 \pi^{10}
   \nn &\hspace{0.5cm} - 
   411952846480704125142412609087072000 \pi^{11} + 
   391369473271880097081811721638702944 \pi^{12} + 
   2208912587183296832097721173734094400 \pi^{13} - 
   1942475465461536631413106579427584752 \pi^{14} - 
   5601595538289187295174864184144292800 \pi^{15}
   \nn &\hspace{0.5cm} + 
   4154644618393775505851430554713346250 \pi^{16} + 
   4405698003009276640285707242239312500 \pi^{17} - 
   1630215031442982020424151720597265625 \
\pi^{18})/(24893472151948586822452882216468427009163264000000 \pi^{18})
\nn
Z_1(38)&=(34670979645901746307288128000 - 
   6793886204041650800608345824000 \pi^2 + 
   4202171522328588667044126804864000 \pi^3 + 
   602363513763875692829011006195200 \pi^4 - 
   307858874840519708881339429224960000 \pi^5
   \nn
   &\hspace{0.5cm} - 
   31540825837298445178949118762681600 \pi^6 + 
   9864116085598242470628358653522201600 \pi^7 + 
   1067017024880265019830906086415725760 \pi^8 - 
   178863182236055742780317653014420096000 \pi^9 
   \nn &\hspace{0.5cm} - 
   23936003744539769855200093232751864480 \pi^{10} + 
   1950971531415274034868089224826497687680 \pi^{11} + 
   350412682409633888332406915496775780576 \pi^{12} - 
   12409259364678554294486401637878253126400 \pi^{13}
   \nn &\hspace{0.5cm} - 
   3139232417115718702513052765143276159248 \pi^{14} + 
   40954558592321625344372057487446221111488 \pi^{15} + 
   14707382078185621167538839421215582064938 \pi^{16} - 
   53558352942981242055276392144650651147200 \pi^{17}
   \nn &\hspace{0.5cm} - 
   22737208019032436560743755455032885721875 \pi^{18} + 
   11792146368175750691275947360407520187500 \
\pi^{19})/(970546692260171503033792971855671032233257336832000000 \pi^{19})
\nn
Z_1(39)&=(-1805904761210022513031992000000 + 
   101264771459781609028217668800000 \pi + 
   302381839841825904791539997088000 \pi^2 - 
   13176351246073953964062105669600000 \pi^3 - 
   22547043029826270566695173241152000 \pi^4
   \nn &\hspace{0.5cm} + 
   744314423610103051803959582360640000 \pi^5 + 
   975340325878260593619959519459731200 \pi^6 - 
   23199798857345391433844810282804640000 \pi^7 - 
   26711620734978466698070836639363700800 \pi^8
   \nn &\hspace{0.5cm} + 
   439479520445053419416040646131248808000 \pi^9 + 
   474622609600170794900723628199437876960 \pi^{10} - 
   5182443088226552783132606566883784996000 \pi^{11} - 
   5352004748469079784950507079841404737440 \pi^{12}
   \nn &\hspace{0.5cm} + 
   36609605305304644024267199595719579514400 \pi^{13} + 
   35696346279919871803666484520944563156656 \pi^{14} - 
   138288504259608600641365483955310789896400 \pi^{15} - 
   120070444515693036962048824855438711760814 \pi^{16}
    \nn &\hspace{0.5cm} + 
   214933454616669430963676986801739145664350 \pi^{17} + 
   132413932097779200697842535294520418050625 \pi^{18} - 
   58786281703205223493628273568619944796875 \
\pi^{19})/(3882186769040686012135171887422684128933029347328000000 \pi^{19})
\nn
Z_1(40)&=(3313407099483014501115308160000 - 
   728706417258906691069678012800000 \pi^2 - 
   608182922633989792238430586905600000 \pi^3 + 
   73078733921342337274895668091232000 \pi^4 + 
   74789865796932340434322044015513600000 \pi^5
   \nn &\hspace{0.5cm} - 
   4375786226122492542181263106984320000 \pi^6  - 
   3591647407788274468425117170821616640000 \pi^7 + 
   171966169868878863391432759716498422400 \pi^8 + 
   89396439569269955177081324398616355840000 \pi^9
   \nn &\hspace{0.5cm} - 
   4587910535468435970070816103837918736000 \pi^{10} - 
   1256990379521422138044925687677477286272000 \pi^{11} + 
   82902295060566290599656834282057236202080 \pi^{12} + 
   10039970400288518423746300585677739736832000 \pi^{13}
   \nn &\hspace{0.5cm} - 
   977622393658912874385390765869430737777600 \pi^{14} - 
   42879173273044008727492193823473545336435200 \pi^{15} + 
   6872848003775367964689883607589382366122276 \pi^{16} + 
   83385731667764285467035752727475736616691200 \pi^{17}
   \nn &\hspace{0.5cm} - 
   23118022037577776612867744615810440354417500 \pi^{18} - 
   48213251336088521517706522595147840710500000 \pi^{19} + 
   15067070464843057084805405574377241085546875 \
\pi^{20})/(3105749415232548809708137509938147303146423477862400000000 \
\pi^{20})
\nn
Z_1(41)&=(188552554107834316310149599360000 + 
   12181876275428018759588788800000000 \pi - 
   35377424389814520926191021142400000 \pi^2 - 
   1791039072418016434645144312800000000 \pi^3 + 
   2978072491010546421716012885128032000 \pi^4
   \nn &\hspace{0.5cm} + 
   115438425598273892671366932298176000000 \pi^5 - 
   146933315003754853108030592402866560000 \pi^6 - 
   4341672644396621320587818863381536000000 \pi^7 + 
   4655916455401019181633684383830965494400 \pi^8 + 
   101590883601155325905883170798566584000000 \pi^9
   \nn &\hspace{0.5cm} - 
   97328965015814936106947092222276335888000 \pi^{10} - 
   1485822441477880712510802305222954666400000 \pi^{11} + 
   1333986362874605156945144268101045806838880 \pi^{12} + 
   13224665210527084220043346689487904097760000 \pi^{13}
   \nn &\hspace{0.5cm} - 
   11453230318564880605521757863123786256420800 \pi^{14} - 
   66752596672330883653121453078480975195280000 \pi^{15} + 
   55252876428294224301856488042579575422831716 \pi^{16} + 
   162631581966724743013205035185677953194330000 \pi^{17}
   \nn &\hspace{0.5cm} - 
   116528745702624231727209092112137641932697500 \pi^{18} - 
   125022202563610959798844577657382305614875000 \pi^{19} + 
   46015866656808256643855211590685881916796875 \
\pi^{20})/(12422997660930195238832550039752589212585693911449600000000 \
\pi^{20})
\nn
Z_1(42)&=(652054007140292642485862021760000 - 
   158609532299761374501366467971200000 \pi^2 + 
   186665933069369587242955731029644800000 \pi^3 + 
   17667040439233978395848715632434272000 \pi^4 - 
   17341808515612802831731351146454137600000 \pi^5
   \nn &\hspace{0.5cm} - 
   1181703323131690599008517421707914880000 \pi^6 + 
   721507120701706831784587777403347345920000 \pi^7 + 
   52283990298487438398501982107140915894400 \pi^8 - 
   17688551061484841941422127936622960605440000 \pi^9 - 
   1587625873080343611845622796580563481904000 \pi^{10}
   \nn &\hspace{0.5cm} + 
   277866469275880390352640356421283819649856000 \pi^{11} + 
   33172764898064489324223178149704271334493280 \pi^{12} - 
   2797062373428974139738431202487364917367712000 \pi^{13} - 
   463731667775739696657987276161848827861454400 \pi^{14}
   \nn &\hspace{0.5cm} + 
   16963673930573775986898603325483482902501433600 \pi^{15} + 
   4037967583397126356247643211196064937066409076 \pi^{16} - 
   54301142172300990346205630377022886585223939200 \pi^{17} - 
   18651852806565966140235166219665124918106872500 \pi^{18}
   \nn &\hspace{0.5cm} + 
   69572684576100005245492562401033725156003750000 \pi^{19} + 
   28912911736598093723356727818160605656338671875 \pi^{20} - 
   15123842776704225142080715041329580376171875000 \
\pi^{21})/(21914167873880864401300618270123567371001164059797094400000000 \
\pi^{21})
\nn
Z_1(43)&=(-4690890099680215571830765595520000 + 
   347706936435165183500269293648000000 \pi + 
   980408078432865164250109198857600000 \pi^2 - 
   57381500752645218341404421729520000000 \pi^3 - 
   92542081171004968477059413691146784000 \pi^4
   \nn &\hspace{0.5cm} + 
   4186995271371558735485685491694492000000 \pi^5 + 
   5165487961081754007275196622046087040000 \pi^6 - 
   171069646711081801655850927133868784000000 \pi^7 - 
   187451035104173854607005695220950617404800 \pi^8 + 
   4381198954546784211621461197678965762960000 \pi^9
   \nn &\hspace{0.5cm} + 
   4593784730914452699343606984329562512912000 \pi^{10} - 
   73759420627788934223619084008448252598800000 \pi^{11} - 
   75685608695625981059554748404625147226147360 \pi^{12} + 
   810639408697052935993089371122080494389820000 \pi^{13}
   \nn &\hspace{0.5cm}+ 
   809226449024092218247897464124723440885547200 \pi^{14}  - 
   5483597517663405354963411593361462541397720000 \pi^{15} - 
   5206012282960668679004955749048286463474524732 \pi^{16} + 
   20144697614905921149634983527723386257006267300 \pi^{17}
   \nn &\hspace{0.5cm} + 
   17138251758578416646182596839689019341863263100 \pi^{18} - 
   30768265195311795048444961610941146078115057500 \pi^{19} - 
   18767865800397767795244152180070051534142640625 \pi^{20}
    \nn &\hspace{0.5cm} + 
   8354300037621534380360695550081236149835546875\pi^{21})
   /(9739630166169273067244719231166029942667184026576486400000000 \pi^{21})
\nn
Z_1(44)&=(142931890873823264893971843552000000 - 
   38553378409104080305189241348472960000 \pi^2 - 
   60614534957567443707225912078650112000000 \pi^3 + 
   4790543294359345843403437778178955200000 \pi^4 + 
   9808736383860604407333289076500166169600000 \pi^5
   \nn &\hspace{0.5cm} - 
   360383888599616080396448567562027573792000 \pi^6 - 
   628666211083068466325488711249908753254400000 \pi^7 + 
   18137355507059295783542804084757443032800000 \pi^8 + 
   21457010641852687290301334499478957149834240000 \pi^9
   \nn &\hspace{0.5cm} - 
   636620294962971485971408211824996384138934400 \pi^{10} - 
   431332117611131270207589343066614841133748480000 \pi^{11} + 
   15745638355751087943361823829820040497121592000 \pi^{12} + 
   5249902023807224367631030778442404101568775552000 \pi^{13}
   \nn &\hspace{0.5cm} - 
   270373302125035780750199426666347000631816216480 \pi^{14} - 
   37895915015999999636329927324860042910775239296000 \pi^{15} + 
   3084963302019749474899609295576812641891628219800 \pi^{16} + 
   150594281662017253081339601367290144798174955763200 \pi^{17} 
   \nn &\hspace{0.5cm}- 
   21278839362544516171943770372568304263208365874756 \pi^{18} - 
   278275878517177410683024330346304640100600126187200 \pi^{19} + 
   71147751494386429152328551217019038071603976848750 \pi^{20} + 
   155358605341892695289654301851159713038128800500000 \pi^{21}
   \nn &\hspace{0.5cm} - 
   47950596725606860210962244245014538457634628515625 \
\pi^{22})/(169703316015333413923671987883836905721033014479068699033600000000 \pi^{22})
\end{align*}
}}
}
\end{center}
\caption{The exact values of $Z_1(N)$ up to $N=44$ (continued).}
\label{fig:Z_1-2}
\end{figure}

\begin{figure}[htb]
\begin{center}
\rotatebox{-90}{
\resizebox{20cm}{!}{
\vbox{
\begin{align*}
Z_2(1)&=1/8, \qquad
Z_2(2)=1/(32 \pi^2) ,\qquad
Z_2(3)=(10 - \pi^2)/(512 \pi^2) \qquad
Z_2(4)=(24 - 32 \pi^2 + 3 \pi^4)/(49152 \pi^4),\qquad
Z_2(5)=(1032 - 904 \pi^2 + 81 \pi^4)/(1179648 \pi^4),\qquad
Z_2(6)=(360 - 1440 \pi^2 + 2807 \pi^4 - 270 \pi^6)/(70778880 \pi^6)
\nn
Z_2(7)&=(359280 - 918600 \pi^2 + 589034 \pi^4 - 50625 \pi^6)/(16986931200 \pi^6),\qquad
Z_2(8)=(20160 - 161280 \pi^2 + 1366288 \pi^4 - 1228032 \pi^6 + 
 110565 \pi^8)/(507343011840 \pi^8)
 \nn
Z_2(9)&=(145010880 - 757626240 \pi^2 + 1292565904 \pi^4 - 723655312 \pi^6 + 
 60824925 \pi^8)/(426168129945600 \pi^8),\qquad
Z_2(10)=(302400 - 4032000 \pi^2 + 130880400 \pi^4 - 291792800 \pi^6 + 
 176521079 \pi^8 - 15025500 \pi^{10})/(1217623228416000 \pi^{10})
 \nn
Z_2(11)&=(74456323200 - 664676712000 \pi^2 + 2192481033120 \pi^4 - 
   3105786871600 \pi^6 + 1607656687398 \pi^8 - 
   133217004375 \pi^{10})/(18410463213649920000 \pi^{10})
\nn
Z_2(12)&=(558835200 - 11176704000 \pi^2 + 1296197179200 \pi^4 - 
   5590427673600 \pi^6 + 7745173132312 \pi^8 - 3722093964000 \pi^{10} + 
   303293615625 \pi^{12})/(432032203413651456000 \pi^{12})
\nn
Z_2(13)&=(194023988928000 - 2657411755891200 \pi^2 + 14431398548212800 \pi^4 - 
   38219526536737920 \pi^6 + 48507686982371320 \pi^8 - 
   23823195742030968 \pi^{10} + 
   1954077051898125 \pi^{12})/(5132542576554179297280000 \pi^{12})
\nn
Z_2(14)&=(108972864000 - 3051240192000 \pi^2 + 1166600822587200 \pi^4 - 
   8454347869968000 \pi^6 + 21741780886785960 \pi^8 - 
   23339271580526720 \pi^{10} + 9641958314010093 \pi^{12} - 
   759070490928750 \pi^{14})/(18871166645108295598080000 \pi^{14})
\nn
Z_2(15)&=(787945104032486400 - 15409012442388364800 \pi^2 + 
   124996926434092352640 \pi^4 - 531883490421787183680 \pi^6 + 
   1234759142546311157776 \pi^8 - 1454710320020335668472 \pi^{10} + 
   687260380130920046898 \pi^{12}
   \nn &\hspace{0.5cm} - 
   55929616500716083125 \pi^{14})/(2720165444892490160689643520000 \pi^{14})
\nn
Z_2(16)&=(86306508288000 - 3222109642752000 \pi^2 + 3749940341731584000 \pi^4 - 
   41498126318134886400 \pi^6 + 174585692946050305920 \pi^8 - 
   342711141392257397760 \pi^{10} + 312740486692520191712 \pi^{12} - 
   115997092985213588736 \pi^{14}
   \nn &\hspace{0.5cm} + 
   8880888816551626875 \pi^{16})/(3826166779628997149101916160000 \pi^{16})
\nn
Z_2(17)&=(1174520269429258752000 - 31153310208889989120000 \pi^2 + 
   353604351346282465228800 \pi^4 - 2206170741163938861312000 \pi^6 + 
   8099218379243719626138240 \pi^8 - 17182132282133938856172800 \pi^{10} 
   \nn &\hspace{0.5cm} + 
   19149751631150222464481184 \pi^{12} - 8774440277838741548100000 \pi^{14} + 
   709487455486528859765625 \
\pi^{16})/(626726118503229733022893867008000000 \pi^{16})
\nn
Z_2(18)&=(44505389440512000 - 2136258693144576000 \pi^2 + 
   7045481116676075673600 \pi^4 - 111270721326427856793600 \pi^6 + 
   701552429702000576219520 \pi^8 - 2223697665491636216762880 \pi^{10} + 
   3677115584554262596762016 \pi^{12} 
   \nn &\hspace{0.5cm} - 2994137064571764051874752 \pi^{14} + 
   1023827859827918055824283 \pi^{16} - 
   76595501984076690615000 \
\pi^{18})/(568231680776261624607423772753920000 \pi^{18})
\nn
Z_2(19)&=(3275305728023106505728000 - 113378060870190461689344000 \pi^2 + 
   1717995073916883638151475200 \pi^4 - 
   14779491212904380459162265600 \pi^6 + 
   78664265877719398398796834560 \pi^8 - 
   260899435751714064910595652480 \pi^{10} 
   \nn &\hspace{0.5cm} + 
   516998276680469455334417352256 \pi^{12}- 
   551137960521620544983472700128 \pi^{14} + 
   246212931459333034614107699550 \pi^{16} - 
   19799702711969287082272171875 \
\pi^{18})/(312982009771564902833769014032859136000000 \pi^{18})
\nn
Z_2(20)&=(101472287924367360000 - 6088337275462041600000 \pi^2 + 
   53490193257256743955968000 \pi^4 - 
   1147279812673273334415360000 \pi^6 + 
   10180480198229895204569318400 \pi^8 - 
   47824540468978617999502848000 \pi^{10} 
   \nn &\hspace{0.5cm} + 
   126522983630835272734199506560 \pi^{12} - 
   185451912547633642015127859200 \pi^{14} + 
   138679030761041176293852038232 \pi^{16} - 
   44485762576443496864931210400 \pi^{18} 
   \nn &\hspace{0.5cm} + 
   3263739072714937244162109375 \
\pi^{20})/(414581834294360481313576384601260032000000 \pi^{20})
\nn
&
\nn
Z_3(1)&=1/12, \qquad
Z_3(2)=(-3 + \pi)/(48 \pi) ,\qquad
Z_3(3)=(9 + 108 \pi - 64 \sqrt{3} \pi)/(5184 \pi),\qquad
Z_3(4)=(378 + 180 \pi - 539 \pi^2 + 256 \sqrt{3} \pi^2)/(82944 \pi^2),\qquad
Z_3(5)=(702 + 6480 \pi - 768 \sqrt{3} \pi - 5701 \pi^2 + 
 2304 \sqrt{3} \pi^2)/(2985984 \pi^2)
 \nn
Z_3(6)&=(-2430 + 1998 \pi + 13923 \pi^2 + 3840 \sqrt{3} \pi^2 - 
 6673 \pi^3)/(11943936 \pi^3), \qquad
Z_3(7)=(17982 + 1487160 \pi - 134784 \sqrt{3} \pi - 235575 \pi^2 - 
 435456 \sqrt{3} \pi^2 - 2586816 \pi^3 + 
 1601728 \sqrt{3} \pi^3)/(3869835264 \pi^3)
 \nn
Z_3(8)&=(1128492 + 1185840 \pi - 16264476 \pi^2 + 
  3068928 \sqrt{3} \pi^2 - 13831992 \pi^3 + 35414397 \pi^4 - 
  17292800 \sqrt{3} \pi^4)/(123834728448 \pi^4)
\nn
Z_3(9)&=(1472580 + 89719488 \pi - 3068928 \sqrt{3} \pi - 
  46127556 \pi^2 + 45536256 \sqrt{3} \pi^2 - 378069120 \pi^3 + 
  52376064 \sqrt{3} \pi^3 + 378253327 \pi^4 - 169072128 \sqrt{3} \pi^4)/(4458050224128 \pi^4),
  \nn
Z_3(10)&=(-145260540 + 208275300 \pi + 3644198100 \pi^2 + 
  632448000 \sqrt{3} \pi^2 - 2972675700 \pi^3 - 19775216637 \pi^4 - 
  6435033600 \sqrt{3} \pi^4 + 10570974175 \pi^5 - 335923200 \sqrt{3} \pi^5)/(445805022412800 \pi^5)
\nn
Z_3(11)&=(299312820 + 88250698800 \pi - 
  2356128000 \sqrt{3} \pi - 12526577100 \pi^2 - 33312384000 \sqrt{3} \pi^2 - 
  1023089097600 \pi^3 + 91604563200 \sqrt{3} \pi^3 + 78245930511 \pi^4 + 
  332129548800 \sqrt{3} \pi^4 + 1443913778700 \pi^5
  \nn &\hspace{0.5cm} - 
  902368702400 \sqrt{3} \pi^5)/(48146942420582400 \pi^5)
\nn
Z_3(12)&=(8708546520 + 15182154000 \pi - 355130527500 \pi^2 + 
   53318476800 \sqrt{3} \pi^2 - 643639834800 \pi^3 + 4434569325126 \pi^4 - 
   1499789952000 \sqrt{3} \pi^4 + 4618381766220 \pi^5 - 
   157212057600 \sqrt{3} \pi^5 - 10986823994175 \pi^6
   \nn &\hspace{0.5cm} + 
   5450595219200 \sqrt{3} \pi^6)/(770351078729318400 \pi^6)
\nn
Z_3(13)&=(25651672920 + 6020447083200 \pi - 76624081920 \sqrt{3} \pi - 
   1788699404700 \pi^2 + 3205211212800 \sqrt{3} \pi^2 - 
   106194939321600 \pi^3 + 3814758374400 \sqrt{3} \pi^3 + 
   59135521614966 \pi^4 - 61844039654400 \sqrt{3} \pi^4 + 
   326960550693360 \pi^5 
   \nn &\hspace{0.5cm} - 49162115356416 \sqrt{3} \pi^5 - 
   353685210676475 \pi^6 + 
   164389189305600 \sqrt{3} \pi^6)/(83197916502766387200 \pi^6)
\nn
Z_3(14)&=(-5599326148920 + 12192598920600 \pi + 327391390155060 \pi^2 + 
   47611234944000 \sqrt{3} \pi^2 - 331983697632300 \pi^3 - 
   5928588773974566 \pi^4 - 1240653059251200 \sqrt{3} \pi^4 + 
   4694982635549430 \pi^5 - 482647063449600 \sqrt{3} \pi^5 
   \nn &\hspace{0.5cm} + 
   33032941562527173 \pi^6 + 11297648468171520 \sqrt{3} \pi^6 - 
   18749272956026375 \pi^7+ 
   1083923053516800 \sqrt{3} \pi^7)/(16306791634542211891200 \pi^7)
\nn
Z_3(15)&=(9762153103080 + 8002729127423520 \pi - 80443646277120 \sqrt{3} \pi - 
   842105862572940 \pi^2 - 3248179182904320 \sqrt{3} \pi^2 - 
   251373308727744000 \pi^3 + 6524189288467200 \sqrt{3} \pi^3 - 
   14425113463704846 \pi^4 + 102523904471116800 \sqrt{3} \pi^4
   \nn &\hspace{0.5cm}  + 
   2042292408062643576 \pi^5 - 181471161819298176 \sqrt{3} \pi^5- 
   62639788677842547 \pi^6 - 699544133395704576 \sqrt{3} \pi^6 - 
   2646161456861676600 \pi^7 + 
   1659255242570878400 \sqrt{3} \pi^7)/(1761133496530558884249600 \pi^7)
\nn
Z_3(16)&=(1131904919350800 + 2906885902032000 \pi - 91901054842198560 \pi^2 + 
   12485221294694400 \sqrt{3} \pi^2 - 296821079176870080 \pi^3 + 
   2868185584451494440 \pi^4 - 930153171012710400 \sqrt{3} \pi^4 + 
   6507854398692773280 \pi^5 
   \nn &\hspace{0.5cm} - 341106145615872000 \sqrt{3} \pi^5 - 
   32541301215263383128 \pi^6 + 14410703617461688320 \sqrt{3} \pi^6- 
   37430599676027727024 \pi^7 + 2308644602346700800 \sqrt{3} \pi^7 + 
   85096547210538648225 \pi^8 
   \nn &\hspace{0.5cm} - 
   42660063670951193600 \sqrt{3} \pi^8)/(112712543777955768591974400 \pi^8)
\nn
Z_3(17)&=(873064837174320 + 610268956142952960 \pi - 
   3332148259184640 \sqrt{3} \pi - 107766663838898400 \pi^2 + 
   317593856147742720 \sqrt{3} \pi^2 - 25458746188524364800 \pi^3 + 
   327107190466990080 \sqrt{3} \pi^3 + 13328971254570106296 \pi^4 
   \nn &\hspace{0.5cm} - 
   14907554349424128000 \sqrt{3} \pi^4 + 299477269154064774528 \pi^5- 
   11111084392271173632 \sqrt{3} \pi^5 - 189225618460189692840 \pi^6 + 
   189805167698060630016 \sqrt{3} \pi^6 - 802724981961779683200 \pi^7 + 
   128239157457940337664 \sqrt{3} \pi^7
   \nn &\hspace{0.5cm}  + 914684272980604229075 \pi^8 - 
   434650602648511872000 \sqrt{3}
     \pi^8)/(4057651576006407669311078400 \pi^8)
\nn
Z_3(18)&=(-4357116781128720 + 13351215166858800 \pi + 462502127490651360 \pi^2 + 
   62013565910016000 \sqrt{3} \pi^2 - 406665151540875360 \pi^3 - 
   16710670410456818376 \pi^4 - 2425393899408936960 \sqrt{3} \pi^4 + 
   10314727603578551160 \pi^5 
   \nn &\hspace{0.5cm} - 3951295513985433600 \sqrt{3} \pi^5 + 
   278995588617672585288 \pi^6+ 56862983172665456640 \sqrt{3} \pi^6 - 
   213566979579646867848 \pi^7 + 53942665950893260800 \sqrt{3} \pi^7 - 
   1634570622021407629545 \pi^8 - 569046977693061731328 \sqrt{3} \pi^8 
   \nn &\hspace{0.5cm}+ 
   965826228941841503075 \pi^9 - 
   76073671439016652800 \sqrt{3}
     \pi^9)/(16230606304025630677244313600 \pi^9)
\end{align*}
}}
}
\end{center}
\caption{The exact values of $Z_2(N)$ up to $N=20$ and $Z_3(N)$ up to $N=18$.}
\label{fig:Z_2&3}
\end{figure}

\begin{figure}[htb]
\begin{center}
\rotatebox{-90}{
\resizebox{20cm}{!}{
\vbox{
\begin{align*}
Z_4(1)&=1/16
\nn
Z_4(2)&=(-8 + \pi^2)/(512 \pi^2)
\nn
Z_4(3)&=(-8 - 32 \pi + 11 \pi^2)/(8192 \pi^2)
\nn
Z_4(4)&=(192 - 560 \pi^2 - 384 \pi^3 + 177 \pi^4)/(1572864 \pi^4)
\nn 
Z_4(5)&=(576 + 16896 \pi - 3120 \pi^2 - 13376 \pi^3 + 4023 \pi^4)/(75497472 \pi^4)
\nn
Z_4(6)&=(-23040 + 192960 \pi^2 + 506880 \pi^3 - 494312 \pi^4 - 574080 \pi^5 + 
 214515 \pi^6)/(36238786560 \pi^6)
 \nn
Z_4(7)&=(-115200 - 14653440 \pi + 1396800 \pi^2 + 44582400 \pi^3 - 5718760 \pi^4 - 
 25262112 \pi^5 + 7216425 \pi^6)/(2899102924800 \pi^6), 
 \nn
Z_4(8)&=(6451200 - 106444800 \pi^2 - 820592640 \pi^3 + 
  655571840 \pi^4 + 3206246400 \pi^5 - 1609472800 \pi^6 - 2421603072 \pi^7 + 
  826548975 \pi^8)/(2597596220620800 \pi^8)
\nn
Z_4(9)&=(135475200 + 57823395840 \pi - 2912716800 \pi^2 - 403938877440 \pi^3 + 
   24113927040 \pi^4 + 807646292992 \pi^5 - 87822722400 \pi^6 - 
   394780516992 \pi^7 
   \nn &\hspace{0.5cm} + 109569305475 \pi^8)/(872792330128588800 \pi^8)
\nn
Z_4(10)&=(-1083801600 + 29578752000 \pi^2 + 578233958400 \pi^3 - 
   328147276800 \pi^4 - 4901011046400 \pi^5 + 1776733212800 \pi^6 + 
   12068917073920 \pi^7 - 4240588979256 \pi^8 
   \nn &- 7552952651520 \pi^9 + 
   2442522103875 \pi^{10})/(139646772820574208000 \pi^{10})
\nn
Z_4(11)&=(-29262643200 - 35762356224000 \pi + 981517824000 \pi^2 + 
   456163113369600 \pi^3 - 13560288537600 \pi^4 - 1989404766720000 \pi^5 + 
   94612817577600 \pi^6 + 3271732109716480 \pi^7
   \nn &\hspace{0.5cm} - 321584570195112 \pi^8 - 
   1466895830516640 \pi^9 + 
   399386173165875 \pi^{10})/(60327405858488057856000 \pi^{10})
\nn
Z_4(12)&=(858370867200 - 34978612838400 \pi^2 - 1573543673856000 \pi^3 + 
   607388292403200 \pi^4 + 23505886701158400 \pi^5 - 
   5634782902425600 \pi^6 - 120944965839667200 \pi^7
   \nn &\hspace{0.5cm} + 
   27397474939488192 \pi^8 + 238604318535249920 \pi^9 - 
   65936598439191984 \pi^{10} - 133391624609358720 \pi^{11} + 
   41557316802568875 \pi^{12})/(42470493724375592730624000 \pi^{12})
\nn
Z_4(13)&=(47210397696000 + 147960574613913600 \pi - 2277901688832000 \pi^2 - 
   3024553612050432000 \pi^3 + 47141631161856000 \pi^4 + 
   23265188986777436160 \pi^5
   \nn &\hspace{0.5cm} - 529361495486208000 \pi^6 - 
   80848090651157913600 \pi^7  + 3288078092186986560 \pi^8 + 
   118276011145953165824 \pi^9 - 10811378575671070320 \pi^{10} - 
   50048648198187105600 \pi^{11}
   \nn &\hspace{0.5cm} + 
   13435199121085520625 \pi^{12})/(37374034477450521602949120000 \pi^{12})
\nn
Z_4(14)&=(-1636627120128000 + 93083167457280000 \pi^2 + 
   8976274859910758400 \pi^3 - 2324234978537472000 \pi^4 - 
   209741539423223808000 \pi^5 + 32619991967506944000 \pi^6
   \nn &\hspace{0.5cm} + 
   1851712030245348311040 \pi^7 - 237088654639272916480 \pi^8 - 
   7449382173461106278400 \pi^9 + 1163142218382587643200 \pi^{10} + 
   12836543833770867065856 \pi^{11}
   \nn &\hspace{0.5cm} - 2958230156652192323880 \pi^{12} - 
   6645748001721802646400 \pi^{13} + 
   2013179571256608129375 \pi^{14})/(36277729466111972969262612480000 \
\pi^{14})
\nn
Z_4(15)&=(-574456119164928000 - 4252705107415872307200 \pi + 
   37698682820198400000 \pi^2 + 128051242036867576627200 \pi^3 - 
   1090001563845562368000 \pi^4 
   \nn &\hspace{0.5cm} - 1539456133268861915627520 \pi^5 + 
   17858925579392561664000 \pi^6 + 9220725502522143368478720 \pi^7 - 
   164677338491547053652480 \pi^8 - 27893283955626979109828608 \pi^9
   \nn &\hspace{0.5cm} + 
   988986117012433330190400 \pi^{10} + 37690738162404499115693568 \pi^{11} - 
   3251910491479391104317240 \pi^{12} - 15292312100377632828597600 \pi^{13}
   \nn &\hspace{0.5cm}  + 
   4059760856437163341839375 \
\pi^{14})/(203735728681684840195378831687680000 \pi^{14})
\nn
Z_4(16)&=(4595648953319424000 - 347737437467836416000 \pi^2 - 
   68043281718653956915200 \pi^3 + 11791165027909710643200 \pi^4 + 
   2300873670656175321907200 \pi^5
   \nn &\hspace{0.5cm} - 232102577529072746496000 \pi^6 - 
   31140050639861252169400320 \pi^7 + 1606592053498785797283840 \pi^8 + 
   211017356476758125020446720 \pi^9 - 
   12215592801570528920309760 \pi^{10}
   \nn &\hspace{0.5cm} - 
   728911775434945936571465728 \pi^{11} + 
   80055727755935442206522112 \pi^{12} + 
   1144566042396957200696352768 \pi^{13} - 
   227611843867739033211263040 \pi^{14} 
   \nn &\hspace{0.5cm} - 
   560055899821872309979737600 \pi^{15} + 
   165922990425511426822089375 \
\pi^{16})/(52156346542511319090016980912046080000 \pi^{16})
\end{align*}
}}
}
\end{center}
\caption{The exact values of $Z_4(N)$ up to $N=16$.}
\label{fig:Z_4}
\end{figure}

\begin{figure}[htb]
\begin{center}
\rotatebox{-90}{
\resizebox{20cm}{!}{
\vbox{
\begin{align*}
Z_6(1)&=1/24
\nn
Z_6(2)&=(54 - 5 \pi^2)/(5184 \pi^2)
\nn
Z_6(3)&=(189 + 192 \sqrt{3} \pi - 125 \pi^2)/(186624 \pi^2)
\nn
Z_6(4)&=(5832 - 8856 \pi^2 - 13824 \sqrt{3} \pi^3 + 8459 \pi^4)/(107495424 \pi^4)
\nn
Z_6(5)&=(670680 + 1078272 \sqrt{3} \pi + 1035504 \pi^2 - 3623424 \sqrt{3} \pi^3 + 
 1825661 \pi^4)/(69657034752 \pi^4)
 \nn
Z_6(6)&=(1574640 + 4461480 \pi^2 - 31104000 \sqrt{3} \pi^3 - 10628658 \pi^4 + 
 71147520 \sqrt{3} \pi^5 - 36458705 \pi^6)/(8358844170240 \pi^6)
\nn
Z_6(7)&=(2211581880 + 4551759360 \sqrt{3} \pi + 
  50163219000 \pi^2 - 45568396800 \sqrt{3} \pi^3 - 90128238381 \pi^4 + 
  111021346368 \sqrt{3} \pi^5 - 50074904525 \pi^6)/(40623982667366400 \pi^6)
\nn
Z_6(8)&=(17856417600 + 539043361920 \pi^2 - 1638499000320 \sqrt{3} \pi^3 - 
   5410591285680 \pi^4 + 12087329587200 \sqrt{3} \pi^5 + 
   8110453018896 \pi^6 - 18418078780416 \sqrt{3} \pi^7 
   \nn &\hspace{0.5cm} + 
   8721700685075 \pi^8)/(36399088469960294400 \pi^8)
\nn
Z_6(9)&=(106825423078080 + 260289957150720 \sqrt{3} \pi + 
   10951867524084480 \pi^2 - 5226616796405760 \sqrt{3} \pi^3 - 
   71942666862535056 \pi^4 + 34782176721954816 \sqrt{3} \pi^5
   \nn &\hspace{0.5cm} + 
   81965619896795424 \pi^6 - 73627337521026048 \sqrt{3} \pi^7 + 
   31101639766610525 \pi^8)/(495318795899219686195200 \pi^8)
\nn
Z_6(10)&=(20249177558400 + 2379575970072000 \pi^2 - 
   6463906773811200 \sqrt{3} \pi^3 - 46213099086722400 \pi^4 + 
   99717782505062400 \sqrt{3} \pi^5 + 264787782487023600 \pi^6
   \nn &\hspace{0.5cm} - 
   459434908272936960 \sqrt{3} \pi^7 - 356403859223598126 \pi^8 + 
   585393319191736320 \sqrt{3} \pi^9 - 
   264199539324466375 \pi^{10})/(19812751835968787447808000 \pi^{10})
\nn
Z_6(11)&=(565659007308561600 + 1566936357235200000 \sqrt{3} \pi + 
   184947764670274536000 \pi^2 - 52720374262219776000 \sqrt{3} \pi^3 - 
   2526481305792155280240 \pi^4 
   \nn &\hspace{0.5cm} + 667345559627553408000 \sqrt{3} \pi^5 + 
   9958443555502553986800 \pi^6 - 3679821240531691852800 \sqrt{3} \pi^7 - 
   9435237128848062914391 \pi^8 + 7176870243498043838400 \sqrt{3} \pi^9
   \nn &\hspace{0.5cm} - 
   2898638708018262453125 \pi^{10})/(866609765305274762967121920000 \pi^{10})
\nn
Z_6(12)&=(1082521032272064000 + 365904737966443737600 \pi^2 - 
   1005513034318442496000 \sqrt{3} \pi^3 - 11574416716220175048000 \pi^4 + 
   26694180539585298432000 \sqrt{3} \pi^5 
   \nn &\hspace{0.5cm} + 
   133123683331032238373760 \pi^6 - 
   245089914171050924236800 \sqrt{3} \pi^7 - 
   608208196884211255269960 \pi^8 + 
   912176891685213084057600 \sqrt{3} \pi^9
   \nn &\hspace{0.5cm} + 
   764080450644042337595544 \pi^{10} - 
   1051590384346712928115200 \sqrt{3} \pi^{11} + 
   458877788639987997809375 \pi^{12})/(610093274774913433128853831680000 \
\pi^{12})
\nn
Z_6(13)&=(62293477299179336179200 + 191373715777382807961600 \sqrt{3} \pi + 
   53986778775880541529523200 \pi^2 - 
   9714396039974354792448000 \sqrt{3} \pi^3
   \nn &\hspace{0.5cm} - 
   1257498933415736749070921280 \pi^4 + 
   198742049521922643898613760 \sqrt{3} \pi^5 + 
   10095503690125153171346238720 \pi^6 - 
   2018130164120379746906726400 \sqrt{3} \pi^7
   \nn &\hspace{0.5cm} - 
   31512503968726555514761885512 \pi^8 + 
   9944682708691937722724803584 \sqrt{3} \pi^9 + 
   26854902567483870461075407728 \pi^{10} - 
   18369730532357431360325337600 \sqrt{3} \pi^{11}
   \nn &\hspace{0.5cm} + 
   7175716737492220019983601875 \
\pi^{12})/(39138703763360246562082231009935360000 \pi^{12})
\nn
Z_6(14)&=(75233046700843903872000 + 62316129128911639177996800 \pi^2 - 
   179741171859553222100582400 \sqrt{3} \pi^3 - 
   2920908275659757270602915200 \pi^4 
   \nn &\hspace{0.5cm} + 
   7337476756897728787906560000 \sqrt{3} \pi^5 + 
   55771703247774513136798140480 \pi^6 - 
   111172680157979357109753200640 \sqrt{3} \pi^7 - 
   497846711834332956259427062800 \pi^8 
   \nn &\hspace{0.5cm} + 
   803056003567155007256530944000 \sqrt{3} \pi^9 + 
   1971238373946201476967266648232 \pi^{10} - 
   2654911265312663505530589182976 \sqrt{3} \pi^{11}- 
   2349718100000660190385829901078 \pi^{12}
   \nn &\hspace{0.5cm} + 
   2874604570520265111460706611200 \sqrt{3} \pi^{13} - 
   1222676176528703161321453271875 \
\pi^{14})/(28492976339726259497195864175232942080000 \pi^{14})
\end{align*}
}}
}
\end{center}
\caption{The exact values of $Z_6(N)$ up to $N=14$.}
\label{fig:Z_6}
\end{figure}

\subsection{A Guess}
The exact values of the partition functions obtained here are all polynomials of $1/\pi$.
At $k=1$, in particular, the partition function has the following form
\begin{align}
Z_1(N)=\sum_{\l=0}^{[\frac{N}{2}]} \frac{Z_1^{(\l)}(N)}{\pi^\l},
\end{align}
where the coefficients $Z_1^{(\l)}(N)$ are rational numbers.
It is, of course, difficult to conjecture all the coefficients $Z_1^{(\l)}(N)$ for general $N$.
However, we find that the leading and the next-to-leading coefficients are simply given by
\begin{align}
Z_1^{(M)}(2M)&=\frac{1}{M!}\( \frac{1}{8\sqrt{2}i} \)^M H_M\(\frac{i}{2\sqrt{2}} \),\qquad
Z_1^{(M-1)}(2M)=0 ,\nn
Z_1^{(M)}(2M+1)&=\frac{1}{4M!}\( -\frac{1}{8\sqrt{2}i} \)^M H_M\(\frac{3i}{2\sqrt{2}} \),\nn
Z_1^{(M-1)}(2M+1)&=\frac{1}{64(M-1)!}\( \frac{1}{8\sqrt{2}i} \)^{M-1} H_{M-1}\(\frac{5i}{2\sqrt{2}} \),
\end{align}
where $H_n(x)$ is the $n$-th Hermite polynomial.
One can check that this guess is correct at least up to $N=44$ by
using the values showed in figures~\ref{fig:Z_1-1} and
\ref{fig:Z_1-2}.

\end{document}